\documentclass[11pt,a4paper,oneside,reqno]{amsart}
\usepackage{epsfig}
\usepackage{amssymb}
\usepackage{url}
\usepackage{pstricks}
\theoremstyle{plain}
\newtheorem{thm}{Theorem}[section]
\newtheorem{lem}[thm]{Lemma}
\newtheorem{prop}[thm]{Proposition}
\newtheorem{cor}[thm]{Corollary}

\newtheorem*{OpQu*}{Open question}
\theoremstyle{definition}
\newtheorem{defn}[thm]{Definition}
\theoremstyle{remark}
\newtheorem{note}[thm]{Note}
\newtheorem*{note*}{Note}
\newtheorem{exmp}[thm]{Example}
\newtheorem*{exmp*}{Example}
\newtheorem{exmps}[thm]{Examples}
\newtheorem*{exmps*}{Examples}
\newtheorem*{opq*}{Open question}

\numberwithin{equation}{thm}

\DeclareMathOperator{\ord}{ord}
\DeclareMathOperator{\wt}{wt}

\DeclareMathOperator{\Coef}{Coef}

\DeclareMathOperator{\XOR}{\scriptstyle{\mathsf{XOR}}}
\DeclareMathOperator{\OR}{\scriptstyle{\mathsf {OR}}}
\DeclareMathOperator{\AND}{\scriptstyle{\mathsf {AND}}}

\DeclareMathOperator{\NOT}{\scriptstyle{\mathsf {NOT}}}

\newcommand{\Z}{\mathbb Z}
\newcommand{\Q}{\mathbb Q}
\newcommand{\N}{\mathbb N}
\newcommand{\R}{\mathbb R}

\renewcommand{\>}{\rightarrow}

\usepackage[backref,% %% При включении пакета showkeys отключить hyperref!!!
pagebackref,%
bookmarks=true,%
colorlinks=true%
]%
{hyperref}
\begin{document}

%+Title
\title{Non-Archimedean analysis, $T$-functions,\endgraf \ and cryptography}
\author{Vladimir Anashin}
\thanks%
{Vladimir Anashin is a Professor and Dean of the Faculty of Information
Security at the Russian State University for the Humanities}
\email{vs-anashin@yandex.ru, anashin@rsuh.ru,  vladimir@anashin.msk.su}
\date{}
%\maketitle
%-Title

%+Abstract
\begin{abstract}
These are lecture notes of a 20-hour course at the International  Summer
School \emph{Mathematical Methods and Technologies in Computer Security}
at Lomonosov Moscow State University, July 9--23, 2006. 

Loosely speaking, a $T$-function is a map of $n$-bit words into $n$-bit words such that each $i$-th bit of image depends only on 
low-order bits $0,\ldots, i$ of the pre-image. For example, all arithmetic operations (addition, multiplication) are $T$-functions, 
all bitwise logical operations ($\XOR$, $\AND$, etc.) are $T$-functions. Any composition of $T$-functions is a $T$-function as well. Thus 
$T$-functions are natural computer word-oriented functions. 

It turns out that $T$-functions are continuous (and often differentiable!) functions with respect to the so-called 2-adic 
distance. This observation gives a powerful tool to apply 2-adic analysis to construct wide classes of 
$T$-functions with provable cryptographic properties (long period, balance, uniform distribution, high linear complexity, etc.); these functions 
currently are being used in new generation of fast stream ciphers.
We consider these ciphers as specific automata that could be associated to
dynamical systems on the space of 2-adic integers. From this view the lectures
could be considered as a course in cryptographic applications of the non-Archimedean
dynamics; the latter has recently attracted significant attention in connection
with applications to physics, biology and cognitive sciences.

During the course listeners  study non-Archimedean 
machinery and its applications to stream cipher design.
    
%There is abstract text that you should replace with your own. 
\end{abstract}
%-Abstract

\maketitle

%+Contents
\tableofcontents
%-Contents
\newpage
\section{Introduction}
\subsection{Goals}
% This template provides a sample layout of a AMS-\LaTeX{} Article.
% 
% The front matter has a number of sample entries that you should replace
% with your own.

Imagine we are a team of cryptographers, and we are going to design a software-oriented cipher. That is, we are going to combine
basic microchip instructions to make a very specific transformation of machine words. On the one hand, this transformation must be fast; that is, the corresponding computer
program must achieve high performance. On the other hand, this transformation must be secure:
Having both an output (that is, encrypted text) {\it and the program}, it must
be infeasible to obtain {\it illegally} the corresponding input (i.e., plain text). 

At this
point, we should understand the following issues:
\begin{itemize}
\item What are these basic instructions? What are reasonable compositions
of these instructions?
\item Could we give an evidence that certain transformation of this kind is secure?  
\end{itemize}

Actually, a goal of the course is to clarify these issues. Moreover,
in order to make our considerations not too general, and to conclude with some practical applications, we restrict ourselves with
a certain specific kind of ciphers, the so-called \emph{stream ciphers}.
 
\subsection{What are stream ciphers?} In contemporary  digital computers
information is represented in a binary form, as a sequence of zeros and ones.
So a plaintext is a sequence $\alpha_0,\alpha_1,\alpha_2,\ldots$, where $\alpha_j\in\mathbb
B=\{0,1\}$.
Let $\Gamma =\gamma_0,\gamma_1,\gamma_2,\ldots$ be another sequence of zeros and ones, which is known both to Alice and Bob, and which is known to no
third party. The sequence $\Gamma $ is called a \emph{keystream}. To encrypt a plaintext, Alice just XORes it with the key:
$$
\begin{array}{cll}
~&\alpha_0,\alpha_1,\alpha_2,\ldots, \alpha_i,\ldots &\text{(plaintext)}\\
\bigoplus &~&\text{(bitwise addition modulo 2)}\\
~&\gamma_0,\gamma_1,\gamma_2,\ldots, \gamma_i,\ldots &\text{(keystream)}\\
~& \hfill\hbox to 3.5cm{\hrulefill} &~\\
~&\zeta_0,\zeta_1,\zeta_2,\ldots, \zeta_i,\ldots &\text{(encrypted text)}\\

\end{array}
$$
To decrypt, Bob acts in the opposite order:
$$
\begin{array}{cll}
~&\zeta_0,\zeta_1,\zeta_2,\ldots, \zeta_i,\ldots &\text{(encrypted text)}\\
\bigoplus &~&\text{(bitwise addition modulo 2)}\\
~&\gamma_0,\gamma_1,\gamma_2,\ldots, \gamma_i,\ldots &\text{(keystream)}\\
~& \hfill\hbox to 3.5cm{\hrulefill} &~\\
~&\alpha_0,\alpha_1,\alpha_2,\ldots, \alpha_i,\ldots &\text{(plaintext)}\\
\end{array}
$$
Loosely speaking, Shannon's Theorem states that this  encryption is secure providing the keystream
$\Gamma$ is picked at random for each plaintext. In real life settings we very
rarely could fulfil conditions of Shannon's Theorem, and usually we use a
\emph{pseudorandom} keystream
$\Gamma$ rather than a random one. That is, usually in real life ciphers $\Gamma$ is produced by a certain algorithm, and $\Gamma$ only looks
like random (e.g., passes certain statistical tests).   A \emph{pseudorandom
generator}, or a \emph{pseudorandom number generator} (PRNG) is an algorithm that takes a short random string (which is called
\emph{a key}, or \emph{a seed}) and stretches it 
into a much longer sequence, a \emph{keystream}.  Actually, within the scope
of the course we speak about \emph{stream cipher} meaning the latter is a
pseudorandom generator which is used for encryption according to the protocol
described above. 

Not every PRNG is suitable for stream encryption. Stream ciphers are \emph{cryptographically
secure} PRNG's; that is, they must not only produce statisticlly good sequences,
but also they must withstand \emph{cryptoanalyst's attacks}. %From this point the course
%concerns cryptographically secure PRNG's.
\section{Preliminaries}
\label{sec:Prelm}
Now we will try to state some of the above mentioned notions more formally.
We start with our main notion, a PRNG.

\subsection{Pseudorandom generators}
\label{sub:PRNG} 
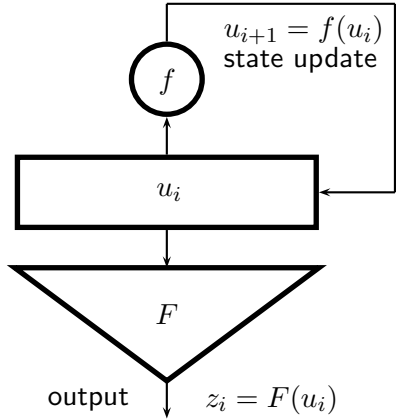
\begin{figure}
%\psblurbox{\parbox[c]{10.7cm}{
\begin{quote}\psset{unit=0.5cm}
 \begin{pspicture}(2,0)(24,12)
%\pscustom[fillstyle=slopes,
%slopecolors=0 1 1 .9  17 .5 1 .5  23 0 0.5 0.5  3]{
%\psccurve(0,2.5)(12,3.5)(20,4)(23,2)(17,2.5)}
%\psaxes(0,0)(0,12)(25,0)
\pscircle[linewidth=2pt](12,10){1}
  \psline(12,11)(12,12)
  \psline{->}(18,7)(16,7)
  \psline(18,12)(18,7)
  \psline(12,12)(18,12)
  \psline{->}(12,8)(12,9)
  \psline{->}(12,2)(12,1)
  \psline{<-}(12,5)(12,6)
  \psframe[linewidth=2pt](8,6)(16,8)
  \pspolygon[linewidth=2pt](8,5)(16,5)(12,2)
  \uput{0}[180](12.4,7){$u_i$}
  \uput{0}[90](12,9.6){$f$}
  \uput{1}[90](12,2.5){$F$}
  \uput{1}[0](12.5,11.3){$u_{i+1}=f(u_i)$}
  \uput{1}[0](12.5,10.5){{\sf state update}}
  \uput{1}[0](12,1.5){$z_i=F(u_i)$}
  \uput{1}[180](12,1.5){{\sf output}}
%\psgrid(0,0)(-1,-1)(3,2)
 \end{pspicture}
\end{quote}
%}
%}
\caption{Ordinary PRNG}
\label{fig:PRNG}
\end{figure}

Basically, a generator we consider during the course 
is a finite automaton
${\mathfrak A}=\langle N,M,f,F,u_0\rangle $ with a finite state set $N$, state
transition (or, state update) function
$f:N\rightarrow N$, finite output alphabet $M$, output function  $F:N\rightarrow M$
and an initial state (seed) $u_0\in N$. Thus, this generator (see Figure
\ref{fig:PRNG}) produces a sequence
$$\mathcal S=\{F(u_0), F(f(u_0)), F(f^{(2)}(u_0)),\ldots, F(f^{(j)}(u_0)),\ldots\}$$ 
over
the set $M$, where 
$$f^{(j)}(u_0)=\underbrace{f(\ldots f(}_{j
\;\text{times}}u_0)\ldots)\ \ (j=1,2,\ldots);\quad f^{(0)}(u_0)=u_0.$$ 

\begin{figure}
\begin{quote}\psset{unit=0.5cm}
 \begin{pspicture}(2,0)(24,12)
%\pscustom[fillstyle=slopes,
%slopecolors=0 1 1 .9  17 .5 1 .5  23 0 0.5 0.5  3]{
%\psccurve(0,2.5)(12,3.5)(20,4)(23,2)(17,2.5)}
%\psaxes(0,0)(0,12)(25,0)
\pscircle[linewidth=2pt](12,10){1}
  \psline(12,11)(12,12)
  \psline{->}(18,7)(16,7)
  \psline(18,12)(18,7)
  \psline(12,12)(18,12)
  \psline{->}(12,8)(12,9)
  \psline{->}(12,2)(12,1)
  \psline{<-}(12,5)(12,6)
  \psframe[linewidth=2pt](8,6)(16,8)
  \pspolygon[linewidth=2pt](8,5)(16,5)(12,2)
  %\fromSlide{2}{%
%   \psline[linecolor=red]{->}(5,6.6)(12,9.8)
%   \pscircle[linecolor=red](12.3,9.9){0.4}
%   \psline[linecolor=red]{->}(5,6.8)(8.3,10.9)
%   \pscircle[linecolor=red](8.6,11.1){0.4}
%   %\uput{1}[0](-1,6.7){{\red Заметьте}}
  %}
%   \pscircle[linecolor=white](12.5,3.8){0.4}
%   %\fromSlide{3}{%
%   \pscircle[linecolor=red](12.5,3.6){0.4}
%   \psline[linecolor=red]{->}(7,3.2)(12.1,3.7)
%   \pscircle[linecolor=red](16.7,1.3){0.4}
%   \psline[linecolor=red]{->}(7,3)(16.4,1.2)
%   %\uput{1}[0](1,3.1){{\red Заметьте}}
  %}
  \uput{0}[180](12.4,7){$u_i$}
  %\fromSlide{2}{%
  \uput{0}[90](12,9.7){$f_i$}
  %}
  %\fromSlide{3}{%
  \uput{1}[90](12.1,2.3){$F_i$}
  %}
  %\fromSlide{2}{%
  \uput{1}[0](12.5,11.3){$u_{i+1}=f_i(u_i)$}
  %}
  %\fromSlide{2}{%
  \uput{1}[0](12.5,10.5){{\sf state update}}
  %}
  %\fromSlide{3}{%
  \uput{1}[0](12,1.5){$z_i=F_i(x_i)$}
  %}
  %\fromSlide{3}{%
  \uput{1}[180](12,1.5){{\sf output}}
  %}
%\psgrid(0,0)(-1,-1)(22,12)
 \end{pspicture}
\end{quote}
\caption{Counter-dependent PRNG}
\label{fig:cntdpd}
\end{figure}
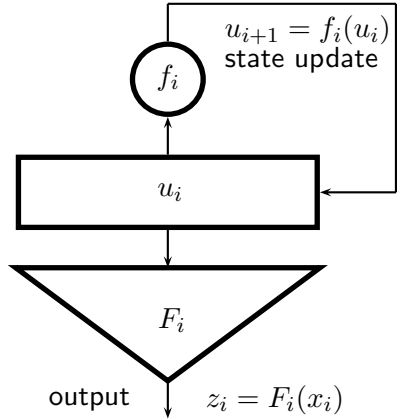

Automata of the form $\mathfrak
A$ could be used either as pseudorandom generators per se, or as components
of more complicated pseudorandom generators, 
%which are introduced in Section
%\ref{sec:Constr}
the so called \textit {counter-dependent generators} (see Figure \ref{fig:cntdpd}); the latter produce %pseudorandom
sequences 
$\{z_0,z_1,z_2,\ldots\}$ over $M$ according to the rule
\begin{equation}
\label{eq:cntdpd}
z_0=F_0(u_0),u_1=f_0(u_0);\ldots
z_{i}=F_i(u_i), u_{i+1}=f_i(u_i);\ldots
\end{equation}
That is, at the $(i+1)$\textsuperscript{th} step the automaton 
$\mathfrak A_i=\langle N,M,f_i,F_i,u_i\rangle $
is applied to the state $u_i\in N$, producing a new state $u_{i+1}=f_i(u_i)\in
N$, and
outputting a symbol $z_{i}=F_i(u_i)\in M$.

%Both types of PRNG are studied within the course; we start with ordinary
%PRNG for simplicity.

%\footnote{

Now to make our considerations more practical, we must impose certain restrictions
on these state update and output functions. As we want our generators to be
implemented in software and to demonstrate good performance, these functions can not be arbitrary, they must be finally written as more or less short
programs. That is, these functions must be represented as (not too complicated) compositions of basic 
instructions of a contemporary processor. Then,  what are these basic instructions?
\subsection{Basic instructions} A contemporary processor is word-oriented.
That is, it works with words of zeroes and ones of a certain fixed length
$n$ (usually $n=8,16,32,64$). Each binary word $z\in\mathbb B^n$ of
length $n$ could be considered as a base-2
expansion of a number $z\in\{0,1,\ldots,2^n-1\}$ and vise versa:
$$%\mathbb Z/2^n\ni 
z=\zeta_0+\zeta_1\cdot 2+\zeta_2\cdot 2^2+\cdots \longleftrightarrow
\zeta_0 \zeta_1 \zeta_2\ldots\in\mathbb B^n$$
We also can identify the set $\{0,1,\ldots,2^n-1\}$ with residues modulo
$2^n$; that is with the elements of the residue ring $\mathbb Z/2^n\mathbb
Z$ modulo $2^n$. 
Actually, \emph{arithmetic} (numerical) instructions of a processor are just {\sl operations
of the residue ring} $\Z/2^n\Z$: An $n$-it word processor performing a single instruction of addition (or multiplication)
of two $n$-bit numbers just deletes more significant digits of a sum (or
of a 
product) of these numbers thus merely reducing the result modulo $2^n$. Note
that to calculate  a sum of two integers (i.e., without reducing the result
modulo $2^n$) a `standard' processor uses not a single instruction but a program tt consists of basic instructions! 

Other sort of basic instructions of a processor are \emph {bitwise logical} operations:
$\XOR$, $\OR$, $\AND$, $\NOT$, which are clear from their definitions. It
worth notice only that the set $\mathbb B^n$ with respect to $\XOR$ could
be considered also as an
$n$-dimensional vector space over a field $\Z/2\Z=\mathbb B$. 

The third type of instructions could be called a \emph{machine} ones, since
they depend on a processor. But usually they include such standard instructions
as shifts (left and right) and circular rotations of an $n$-bit word.

Some more formal sample definitions: Let 
$$z=\delta_0(z)+\delta_1(z)\cdot 2+\delta_2(z)\cdot 2^2+\delta_3(z)\cdot 2^3+\cdots$$ 
be
% be represented in 
a base-2 expansion for $z\in\mathbb N_0=\{0,1,2,\ldots\}$.%; then %(sometimes %the $i$\textsuperscript{th}
%bit 
%$\delta_i(z)$ is denoted via $[z]_i$). %
Then, according to the respective
definitions, we have
\begin{itemize}
\item $y\XOR z=y\oplus z$ is a bitwise
addition modulo 2: $\delta_j(y\XOR z)\equiv\delta_j(y)+\delta_j(z)\pmod 2$;%$\bmod 2$
\item %$\delta_j(y\AND z)\equiv\delta_j(y)\cdot\delta_j(z)\pmod 2$, 
$y\AND z$ is a bitwise
multiplication modulo 2: $\delta_j(y\AND z)\equiv\delta_j(y)\cdot\delta_j(z)\pmod 2$;
\item $\lfloor \frac{z}{2}\rfloor$, the integral part of $\frac{z}{2}$,  is a shift towards less significant
bits;
\item $2\cdot z$ is a shift towards more significant
bits;
\item $y\AND z$ is  masking of $z$ 
with the mask $y$; %in particular, %reduction
%modulo $2^k$ is just 
\item $z\pmod{2^k}=z\AND(2^k-1)$ is a reduction of $z$
modulo $2^k$ %is just 
%\item $z\pmod{2^k}=z\AND(2^k-1)$
\end{itemize}

Let us make the first important observation:
\begin{quote} 
{\sl
Basic instructions of a processor, with the exception of rotations, are well
defined
on the whole set of positive integers.  
}
\end{quote}

Now we look at the basic instructions from a bit another point. 
\subsection{$T$-functions} From a school textbook algorithm of addition
of base-2 expansions of positive integers it immediately follows that each $i$-th bit of the sum does not depend
on higher order bits of summands, i.e., on $j$-th bits with $j>i$. The same
holds for products, bitwise logical operations, and shifts towards higher
order bits. This observation gives rise to the following definition:
\begin{defn}[T-function]
\label{def:T-fun}   
%Speaking formally, a {\it
An ($m$-variate) $T$-function is any mapping
%\onlySlide*{1}{%
%\begin{boxitpara}{box 0.9 setgray fill}
$$F\colon(\ldots, \alpha_2^{\downarrow},\alpha_1^{\downarrow},\alpha_0^{\downarrow})\mapsto 
(\ldots,\Phi_2(\alpha_0^{\downarrow},\alpha_1^{\downarrow},\alpha_2^{\downarrow}),
\Phi_1(\alpha_0^{\downarrow},\alpha_1^{\downarrow}),\Phi_0(\alpha_0^{\downarrow}))$$
%\end{boxitpara}
where $\alpha_i^{\downarrow}\in\mathbb B^m$ is a Boolean columnar $m$-dimensional vector; $\mathbb B=\{0,1\}$;
$\Phi_i\colon(\mathbb B^{m})^{(i+1)}\rightarrow
\mathbb B^n$ maps $(i+1)$ Boolean columnar $m$-dimensional vectors 
$\alpha_i^{\downarrow},\ldots,\alpha_0^{\downarrow}$
to $n$-dimensional
Boolean vector $\Phi_i(\alpha_0^{\downarrow},\ldots,\alpha_i^{\downarrow})$.
\end{defn}
For instance, a univariate $T$-function
$F\colon\mathbb B^n\rightarrow\mathbb B^n$ %first}; that is
is a mapping of $\mathbb B^n$ into itself such that 
$$(\ldots,\chi_2,\chi_1,\chi_0)\stackrel{F}{\mapsto} (\ldots;\psi_2(\chi_0,\chi_1,\chi_2);\psi_1(\chi_0,\chi_1);\psi_0(\chi_0)),$$
where $\chi_j\in\{0,1\}$, and each  $\psi_j(\chi_0,\ldots,\chi_j)$ is a Boolean function
%от переменн%is a Boolean
%function 
in Boolean variables 
$\chi_0,\ldots,\chi_j$. 

Thus, we state that
\begin{quote}
{\sl
Basic instructions of a processor, with the exception of rotations and shifts
towards low order bits, are $T$-functions.}
\end{quote}

Obviously, a composition of $T$-functions is a $T$-function; so while combining
basic instructions into a program, we very often can say that the resulting
mapping (that is, a program) is a $T$-function. So, it seems to be a good
idea to study the above mentioned automata under a restriction that both
their state update and output functions are $T$-functions, and try to design
a stream cipher on their base.

Few words about terminology: Despite the term `$T$-function' was suggested only in 2002 by A.~Klimov and A.~Shamir,
see
\cite{KlSh}, these mappings are well-known mathematical objects dating back
to 1960\textsuperscript{th} (however, under other names: Compatible mappings in algebra, determined
functions in automata theory, triangle boolean mappings in the theory of Boolean
functions, 
%ergodic and measure preserving 
functions that satisfy Lipschitz condition
with constant 1 in $p$-adic analysis; see e.g. \cite{LN}, \cite{Yb}, \cite{anashin1}). 
% Possible usefulness of these functions in cryptography
% was directly pointed out in 1993 by V.S. Anashin \cite{anashin1}. The name "T-function" was suggested by   A. Klimov and A.
% Shamir in 2002 .  
Throughout the course we use the term `$T$-function' as the most accepted
by cryptographic community; however, we will be interested in those properties
of $T$-functions that are explored in other areas of mathematics. The mentioned
$p$-adic analysis appears to be the most important one.
\subsection{Preparations to $p$-adic Calculus} We can calculate a sum of two
positive integers represented by their base-2 expansions with a `school textbook'
algorithm. Note that the summands are represented as finite strings of 0's
and 1's (or, better to say, as {\sl infinite} strings of 0's and 1's that contain only {\sl finite} number of 1's). Let us look what happens if we
apply this %school-textbook
algorithm to {\sl arbitrary infinite} strings of 0's and 1's. 

Consider an example:
\begin{align*}
&\mbox{}&{}&\ldots 1&{}& 1&{}&1&{}&1&{}&\\
&\mbox{+}&{}&{}\\
&\mbox{}&{}& \ldots 0&{}&0&{}&0&{}&1&{}&\\
%&{}&{}&\hbox to 2cm{\hrulefill}&\hbox to .1cm{\hrulefill}&\hbox to 2cm{\hrulefill}&{}&\\
\intertext{\hbox to 3cm{}\hbox to 6.6cm{\hrulefill}}
&\mbox{}&{}&\ldots 0&{}&0&{}&0&{}&0&{}&
\end{align*}

Obviously, the string $\ldots000$ is merely $0$, and the string $\ldots001$
is $1$. But then we {\sl must} conclude that $\ldots111=-1$; that is, the
infinite string $\ldots111$ is a base-2 expansion of a {\sl negative} integer
$-1$. With this in mind, we continue our investigations. Let's try multiplication
now: 
\begin{align*}
&\mbox{}&{}&\ldots 0&{}&1&{}& 0&{}&1&{}&0&{}&1\\
&\mbox{$\times$}&{}&{}\\
&\mbox{}&{}& \ldots 0&{}&0&{}&0&{}&0&{}&1&{}&1\\
%&{}&{}&\hbox to 2cm{\hrulefill}&\hbox to .1cm{\hrulefill}&\hbox to 2cm{\hrulefill}&{}&\\
\intertext{\hbox to 3cm{}\hbox to 8.4cm{\hrulefill}}
&\mbox{}&{}&\ldots 0&{}&1&{}& 0&{}&1&{}&0&{}&1\\
&\mbox{$+$}&{}&{}\\
&\mbox{}&{}& \ldots 1&{}&0&{}&1&{}&0&{}&1&{}&\\
%&{}&{}&\hbox to 2cm{\hrulefill}&\hbox to .1cm{\hrulefill}&\hbox to 2cm{\hrulefill}&{}&\\
\intertext{\hbox to 3cm{}\hbox to 8.4cm{\hrulefill}}
&\mbox{}&{}&\ldots 1&{}&1&{}&1&{}&1&{}&1&{}&1
\end{align*}

As we know that $\ldots0011=3$, and, as we have agreed,  $\ldots111=-1$,
then we are forced to conclude that $\ldots01010101=-\frac{1}{3}$. This sounds somewhat
odd
for us, but {\sl not so} for a computer! These calculations could be made
with an ordinary Windows built-in calculator, up to the best precision it
admits, 64 bits.\footnote{Don't forget to switch the calculator into {\tt
scientific mode} and choose {\tt bin}.}

Now denote $\mathbb Z_2$ the set of all infinite binary strings. We could
define addition and multiplication on $\mathbb Z_2$ with the said school-textbook
algorithms, thus turning $\mathbb Z_2$ into a ring. Obviously, any $T$-function
is well defined on $\mathbb Z_2$. Summing it up, we conclude that
\begin{quote} 
{\sl Basic processor instructions, with the only exception of rotations,
as well as $T$-functions, are well defined functions on the set $\mathbb
Z_2$ of all infinite binary sequences; these functions are evaluated in $\mathbb
Z_2$. 
}
\end{quote}

As a matter of fact, these functions turn out to be {\sl continuous} in some
well-defined sense. Moreover, very often they are {\sl differentiable} functions,
and we can use a special sort of Calculus to study their properties that
are crucial for cryptography with the techniques similar to that of classical
Calculus. That is what we are going to do within the course.   

% { These mappings are of interest for software-oriented ciphers, since both
% arithmetic and bitwise logical operations, which are basic instructions
% for most processors, are obviously $T$-functions.} 

What we are thinking about when saying `Calculus'? Well, of derivations, for instance.
And what notion do we use in the definition of a derivative? Evidently, a
notion of limit. But saying that `$a$ is a limit of the sequence $\{a_i\}_{i=0}^\infty$
of numbers as $i$ goes to infinity' we just mean that these $a_i$ are {\sl
approximations} of $a$, and we can achieve an {\sl arbitrarily} good precision
of these approximations
by taking {\sl sufficiently large} $i$. 

Now we are going to understand what does this `precision' means, or, better
to say, what a computer thinks of what `precision' means. A computer can
not work with arbitrarily long binary words. Actually, its basic instructions
work with words of certain length, a {\it bitlength}. Usual values of
bitlengths of contemporary processors are 8,16,32, 64. 

Now take some binary string, e.g., a string $\underbrace{1\ldots111}_{64\
\text{times}}$; that is, a number $2^{64}-1=18446744073709551615$. A 8-bit processor
%with bitlength 8 
can work only with 8-bit string, so it can store only 8
less significant bits of this string; that is, the number $2^8-1=255$. A 16-bit
processor stores 16 bit, that is, the number $2^{16}-1=65535$; a 32-bit processor
stores this string as $2^{32}-1=4294967295$, etc. It is reasonable to say
that 255 is an approximation with 8-bit precision of the number $2^{64}-1$, 65535 is an approximation with 16-bit precision, etc. 

Following this logic, we finally conclude that the sequence 
$$255, 65535, 4294967295,\ldots, 2^{2^n}-1,\ldots$$ 
tends to $-1=\ldots111$ as $k$ goes
to infinity, and the same does the sequence $2^n-1$. That is, $\lim\limits_{n\to\infty}^2(2^
n-1)=-1$, where $\lim\limits^2$ is something that behaves like an ordinary
limit, but with respect to the `$n$-bit precision'. Further, in case we want
this $\lim\limits^2$ behave similarly to an ordinary limit, we must conclude
that $\lim\limits_{n\to\infty}^2 2^n=0$, which is extremely odd!\footnote{Not
too odd, however. Intuitively, the sequence $\ldots0001$, $\ldots0010$, $\ldots0100$,
$\ldots$, which is the sequence of base-2 expansions of $1, 2, 4, 8,\ldots$,
tends to $\ldots0000=0$!}

To discover the underlying reality, we now must understand on what notion
is the notion of limit based. Recalling the classical definition, we see
that the notion of limit is stated in terms of `how close the two numbers
are'. That is, the notion of limit is based on the notion of distance!

The above examples demonstrate that for human beings and for computers, `distance'
means quite different things, or, better to say, is measured in different
ways. For us, human beings, a number $2^{32}=4294967296$ lies at a bigger
distance from $0$ than the number $2^8=256$; on the contrary, $2^{32}$ is
closer to $0$ than $2^8$, for a computer. What a peculiar distance a computer
uses?
    
\section{The notion of $p$-adic integer}

\subsection{The notion of distance} Actually, when we measure a distance
between two points, we associate a non-negative real number to the pair of
points.
Obviously, this number is 0 if and only if these points coincide, and
the distance measured from the first point to the second one is equal to the
distance measured in the opposite direction, from the second point towards
the
first. The distance obeys the `law of a triangle'; that is, the distance
from the first point $A$ to the second point $B$ is not greater than the sum of two distances,
from the first point $A$ to an arbitrary third point $C$, and from this third point $C$
to the point $B$. These observations are summarized in the following definition\footnote{Mathematicians
used to speak of metric rather than of distance, but distance is also OK}:
\begin{defn}[Metric]
\label{def:metr} 
Let $M$ be a non-empty set, and let $d\colon M\times M\>\R_{\ge
0}$ be a function valuated in  non-negative real numbers. The function $d$ is
called a \emph{metric} (and $M$ is called a \emph{metric space}) whenever
$d$ obeys the following laws:
\begin{enumerate}
\item For every pair $a,b\in M$, $d(a,b)=0$ if and only if $a=b$.
\item For every pair $a,b\in M$, $d(a,b)=d(b,a)$.
\item For every triple $a,b,c\in M$, $d(a,b)\le d(a,c)+d(c,b)$.
\end{enumerate}
\end{defn}

For example, the set $\mathbb R$ of all real numbers is a metric space with
metric $d(a,b)=|a-b|$, where $|\cdot|$ is absolute value. The latter notion
also could be defined for arbitrary commutative ring $R$.
\begin{defn}[Norm]
\label{def:norm} 
A function $\|\cdot\|$ defined on the  $R$ and valuated in
$\R_{\ge 0}$ is called a \emph{norm} whenever $\|\cdot\|$ satisfies the following
conditions:
\begin{enumerate}
\item For every $a\in R$, $\|a\|=0$ if and only if $a=0$.
\item For every pair $a,b\in R$, $\|a\cdot b\|=\|a\|\cdot\|b\|$.
\item For every pair $a,b\in R$, $\|a+b\|\le\|a\|+\|b\|$.
\end{enumerate}
\end{defn} 

It is easy to verify that assuming $d(a,b)=\|a-b\|$ we define metric $d$
on the ring  $R$. This metric $d$ is called a metric induced by the norm
$\|\cdot\|$. 

Note that once the norm (whence, metric) on the ring $R$ is defined, we immediately define
a notion of convergent sequence over $R$, a notion of limit, a notion of
continuous function defined on $R$ and valuated in $R$, a notion of derivative
of a function, etc. For instance, element $a\in R$ is a derivative of the
function $f\colon R\> R$ at the point $x\in R$ if and only if for all \emph{sufficiently
small} $h\in R$, $h\ne 0$, (that is, for $\|h\|<\delta$ for some real $\delta>0$)  
%for every $\epsilon >0$ there is some $\delta >0$ such that 
%$$\bigg\|\frac{f(x+h)-f(a)}{h} -a\bigg\|<\epsilon$$
$$f(x+h)=f(x)+a\cdot h+\lambda(h),$$
where $\frac {\|\lambda(h)\|}{\|h\|}$ goes to $0$ as $\|h\|$ goes to $0$.
%as soon as $0<\|h\|<\delta$. 
Thus, loosely speaking, every new norm %on a field 
leads to a new Calculus. %on the field.
\subsection{Norms on $\Z$} We know that absolute value $|\cdot |$ is a norm
on the ring $\Z$ of all integers. The question arises, is $|\cdot |$  {\sl
the only} norm on $\Z$? Surprisingly, {\sl not}!

Let $p$ be a prime number. Using this $p$, we define now a norm $\|\cdot\|_p$
on $\mathbb Z$. Obviously, since $\|-a\|=\|a\|$ for every $a\in R$ (this
is an exercise to deduce the identity from Definition \ref{def:norm}!), it
suffices to define $\|\cdot\|_p$ on the set $\N_0$ of all non-negative integers.
We assume $\|0\|_p=0$. Now, if $n>0$ is a natural number, it has a unique
representation as a product of powers of pairwise distinct primes. Denote $\ord_pn$ exponent of $p$ in this representation and put $\|n\|_p=p^{-\ord_pn}$.
It is an exercise to verify that the so defined function is a norm.

Indeed, (1) and (2) of Definition \ref{def:norm} obviously hold for the so
defined norm. Moreover, (3) holds in a {\sl stronger} form:
$$(3^\prime)\ \text{For every pair}\ a,b\in \Z, \|a+b\|_p\le\max\{\|a\|_p,\|b\|_p\}.$$
From here it obviously follows that the metric $d_p$ defined by the norm $\|\cdot\|_p$ also satisfies a
stronger relation than (3) of Definition \ref{def:metr}:
$$(3^\prime)\ \text{For every triple}\ a,b,c\in\Z,\ d_p(a,b)\le\max\{d_p(a,c),d_p(c,b)\}.$$

The latter relation is called a \emph{strong triangle inequality}, and a metric
that satisfies this inequality is called a \emph{non-Archimedean} metric,
or an \emph{ultrametric}. Accordingly, a metric space equipped with a non-Archimedean
metric is called a non-Archimedean metric space, or an ultrametric space.
\subsection{$p$-adic integers}
Clearly, for natural $n\in\N$  one can calculate $\ord_pn$ according
to the following rule: Represent  $n$ in its base-$p$ expansion, find the
least significant non-zero digit (let it be the $i$-th digit; enumeration starts
with zero); then $\ord_pn=i$. That is, 
$$n=\ldots a_{i+1}a_{i}\underbrace{0\ldots
0}_{i\
\text{zeros}}; a_i\ne 0 \Rightarrow \|n\|_p=\frac{1}{p^i}.$$

The latter definition could be expanded on the whole set $\Z_p$ of {\sl infinite}
strings of digits $0,1,\ldots,p-1$ in an obvious manner. Now it is not difficult
to prove that the set $\Z_p$ is a commutative ring with respect to addition
and multiplication defined by `school-textbook' algorithms, and, moreover,
the so defined function $\|\cdot\|_p$ is a norm on this ring!\footnote{Prove
this.} Elements of the ring $\Z_p$ are
called \emph{$p$-adic integers}. Actually, we think of the infinite string $\ldots a_ia_{i-1}\ldots
a_0$
over an alphabet $\{0,1,\ldots, p-1\}$ as of base-$p$ expansion
of a $p$-adic integer $a$:
\begin{equation}
\label{eq:base-p}
a=\ldots a_ia_{i-1}\ldots
a_0=\sum_{i=0}^\infty a_i\cdot p^i
\end{equation}

Note that for $a,b\in\Z_p$ $d_p(a,b)=\frac{1}{p^i}$ for some $i=0,1,2,\ldots,\infty$
(case $i=\infty$ just means that $d_p(a,b)=0$, whence, $a=b$). Moreover, $d_p(a,b)=\frac{1}{p^i}$
if and only if
\begin{gather*}
a=\ldots a_{i+1}a_{i}%\underbrace
{c_{i-1}\ldots
c_0}%_{i}
;\\
b=\ldots b_{i+1}b_{i}%\underbrace
{c_{i-1}\ldots
c_0}
%_{i}
,
\end{gather*}
and $a_{i}\ne b_{i}$. Using an obvious analogy with non-negative {\sl rational}
integers we write in this case that $a\equiv b\pmod {p^i}$. Thus, $d_p(a,b)=\frac{1}{p^i}$
where $i$ is the biggest non-negative rational integer such that $a\equiv b\pmod {p^i}$, and $a\not\equiv b\pmod {p^{i+1}}$. Throughout the course
we denote the $i$-th digit ($i=0,1,2,\ldots$) in a base-$p$ expansion of a $p$-adic integer
$a\in\Z_p$
via $\delta_i^p(a)$; that is, $\delta_i^p(a)=a_i$, cf. \eqref{eq:base-p}.
We omit the superscript (especially
in case $p=2$) when it does not lead to misunderstandings.

The ring $\Z_2$ of infinite binary strings mentioned above  corresponds
to the case $p=2$. Thus, $\Z_2$ is an ultrametric space with respect to the metric $d_2$
defined by the norm $\|\cdot\|_2$. And, indeed, with respect to this metric
$d_2$ the sequence $1,2,4,\ldots,
2^n,\ldots$ converges to $0$ as $n$ goes to infinity; whence\footnote{To
prove this we must prove a theorem on limit of sum of two convergent sequences
before. It is a good exercise to re-prove all classical theorems about limits
of compositions of sequences in general case, for arbitrary metric!}, the sequence
$1,3,7,\ldots,
2^n-1,\ldots$ indeed converges to $-1$.

%Now we state  a definition:  
%Thus,  a
Actually {\sl a processor works
with approximations of 2-adic integers with respect to 2-adic metric}:
When
one tries to load a number which base-2 expansion contains more than $n$
significant bits into a registry of an $n$-processor, the processor just
writes only $n$ low order bits of the number in a registry {\sl thus reducing
the number modulo} $2^n$. Thus, 
precision of the approximation is defined by the bitlength of the processor.

Since the ring (metric space) $\Z_2$ is of most importance for us, we %end the section
proceed with some examples that illustrate our main notions with respect to $\Z_2$.

Sequences that contain only finite number of 1's correspond  to {%\yellow
non-negative
rational integers} represented by their base-2 expansions:   $$\ldots 00011=3$$
Sequences that contain only finite number of 0's correspond  to {%\yellow
negative
rational integers}\footnote{Prove this}: $$\ldots 111100=-4$$
Sequences that are (eventually) periodic correspond to  { rational
numbers that could be represented by irreducible fractions with odd denominators}\footnote{Prove this}: $$\ldots1010101=-\frac{1}{3}$$
Non-periodic sequence correspond to no rational number. %: $$\ldots01111011101101$$

An example one how we measure distances in $\Z_2$:
\begin{equation*}
\left.
\begin{aligned}
\ldots1010{1}{\underbrace {0101}}&={{-\frac{1}{3}}}\\
\ldots0000{0}{\underbrace{0101}}&=5\\
\end{aligned}
\right\}
\Longrightarrow d_2\bigg(-\frac{1}{3},5\bigg)=\frac{1}{2^4}=\frac{1}{16}
\end{equation*}

%Пример: 
%{\yellow $d_2(5,-\frac{1}{3})=\frac{1}{8}$}, т.к. $\ldots000101=5$, $\ldots010101=-\frac{1}{3}$;
That is,
$-\frac{1}{3}\equiv 5\pmod{16}; -\frac{1}{3}\not\equiv 5\pmod{32}$.
\subsection{Odd world} Finally we conclude that our computers live in
the world other than we human beings. This virtual world is very odd. In
this subsection we only mention some facts about this virtual world to make it
more familiar to us. Proofs (and other peculiar facts) could be found in
the above mentioned books and monographs on $p$-adic analysis.

Our world, the world of real numbers $\R$ is Archimedean. That is, it satisfies
the Archimedean Axiom which read:
\begin{quote}
Given a segment $S$ of real line of length $s$, and another (smaller) segment
$L$ of length $\ell$, $\ell<s$, there exists a natural number $n$ such that
$n\cdot \ell>s$. (That is, if we append a short segment to itself sufficient
number of times, we can make the resulting segment arbitrarily long). 
\end{quote}

This axiom {\sl does not hold} in the $p$-adic world $\Z_p$: Appending a segment
to itself we could make the resulting segment {\sl shorter} than the original
one!  For instance,
let $p=2$ and let $L$ be some `segment of length $\frac{1}{2}$', say, $L=2$ then doubling the
segment (`appending' it to itself) we, obtain a `segment' $2\cdot L=4$, and
for which we have $\|4\|_2=\frac{1}{2}$. The `doubled segment' is twice as
{\sl short} as the original! 

Of course, origin of this fact is hidden in a {strong triangle inequality}
$(3^\prime)$ that governs the non-Archimedean world. This inequality implies
other odd-looking facts, e.g.,
\begin{itemize}
\item All triangles are  isosceles!
\item Every point inside a ball is a center of this ball!
\item The series $\sum_{i=0}^\infty z_i$ of $p$-adic integers are convergent
{\sl if and only if} $\lim\limits_{i\to\infty}^p z_i=0$ (where $\lim\limits_{i\to\infty}^p$)
is a limit with respect to the $p$-adic norm $\|\cdot\|_p$).
\end{itemize}
By the way, this implies that, say, $\ln(-3)=-\sum_{i=1}^\infty\frac{4^i}{i}$ %=0$
{is a 2-adic integer!}

If you are going to prove these statements (which  is a good exercise!) note that
every ball of radius $\frac{1}{p^k}$ in $\Z_p$ is of the form $a+p^k\cdot\Z_p=\{a+p^k\cdot
z\colon z\in\Z_p\}$. By the way, from here it follows that, in case $p=2$,
a boundary of
a (closed) ball is itself a ball of radius  $\frac{1}{p^{k+1}}$; e.g., a sphere of radius $\frac{1}{2}$
is a ball of radius $\frac{1}{4}$! Actually,
the whole metric space $\Z_p$ is a ball of radius 1 (and is a $p$-adic analog
of a real unit interval). For those who is familiar with functional analysis
we mention also that the space  $\mathbb Z_p$ is {\sl  complete} with respect to
the $p$-adic distance (metric) $d_p$, and {\sl compact}.
%First of all,   
%We conclude this section with some facts about
%the 

\section{Elements of applied $2$-adic analysis}
\label{sec:Ap_an}
The main goal of this section is to provide some experience in Calculus
on $\Z_2$. We are not going to do this too formally since there are a number
of excellent books and monographs on $p$-adic analysis, e.g. \cite{Shik,Mah,Kobl,Kat}.
We rather focus %(more or less informally) 
on those functions and techniques
that later in the course will be used in our cryptographic applications,
stream cipher design.
\subsection{$T$-functions revisited}
\label{sub:T-rev}
We start with 2-adic extensions of %formal definitions of 
what we called `basic instructions'. These are primarily arithmetic operations
(addition, subtraction, multiplication) and bitwise logical operations. These two set
of operations are not mutually independent, some of them could be expressed
via others. The following identities could be proved: For all
$u,v\in\mathbb Z_2$
\begin{equation}
\label{eq:id}
\begin{split}
&\NOT (u)=u \XOR (-1);\\
&\NOT (u)+u=-1;\\
&u \XOR v = u+v-2(u \AND v);\\ 
&u \OR v = u+v-(u \AND v);\\
&u \OR v=(u \XOR v)+(u \AND v).
\end{split}
\end{equation}

During the course we often write $\oplus$ instead of $\XOR$, also $\odot$,
$\&$ or $\wedge$
instead of $\AND$, and $\vee$ instead of $\OR$. These operations (with the only exception
of $\NOT$) are functions of two 2-adic variables. To work with these functions
we need to define 2-adic metric on a Cartesian square $\Z_2^2$. %This could
%be done  
Having already defined
metric on $\Z_2$ we define metric  on a Cartesian product $\Z^n_2=\underbrace{\Z_2\times\cdots\times\Z_2}_{n\
\text{times}}$ in a standard manner: For $\mathbf a=(a^{(1)},\ldots,a^{(n)}),
\mathbf b=(b^{(1)},\ldots,b^{(n)})\in\Z^n_2$ we put $\|\mathbf a\|_2=\max\{\|a^{(1)}\|_2,\ldots,\|a^{(n)}\|_2\}$
and, respectively, $d_2(\mathbf a, \mathbf b)=\max\{d_2(a^{(1)},b^{(1)}),\ldots,d_2(a^{(n)},b^{(n)})\}$.
We also write $\mathbf a\equiv\mathbf b\pmod{2^i}$ whenever $a^{(j)}\equiv
b^{(j)}\pmod{2^i}$ for all $j=1,2,\ldots, n$. 

Now it is a right time to consider $T$-functions 
as 2-adic mappings. Actually (see Definition \ref{def:T-fun}) we define $T$-function as a special mapping that puts into a correspondence  to  every sequence
of columnar $m$-dimensional Boolean vectors certain sequence of $n$-dimensional
columnar Boolean vectors. Now we can read these sequence not column after
column, but as a row after a row, starting with a top one. Each this row
is an infinite sequence of zeros and ones; that is, a 2-adic integer. Thus,
\begin{quote}
{\sl
we can consider a 
$T$-function $F$ from Definition \ref{def:T-fun} as a mapping from $\Z_2^m$ into $\Z_2^n$ such that  $F(\mathbf a)\equiv F(\mathbf b)\pmod{2^i}$ 
whenever $\mathbf a\equiv\mathbf b\pmod{2^i}$.
}
\end{quote}
From this observation immediately follows a very important theorem:
\begin{thm}
\label{thm:T-Lip}
$T$-functions are mappings from $\Z_2^m$ into $\Z_2^n$ that satisfy Lipschitz
condition with a constant 1:
$$\|F(\mathbf a)-F(\mathbf b)\|_2\le\|\mathbf a -\mathbf b\|_2$$
and vise versa, all mappings that satisfy this condition are $T$-functions. 
\end{thm}
\begin{cor}
\label{cor:T-cont}
All $T$-functions are continuous $2$-adic functions.\footnote{Any function
that satisfy Lipschitz condition with respect to a certain metric is continuous
with respect to this metric. Prove this!}
\end{cor}

These easy claims are a hint that 2-adic analysis could be useful in study
of
%properties of 
$T$-functions; of course, only of properties that are of `analytic nature',
which could
be properly stated in terms of analysis; that is, in terms of limits, convergence,
derivatives, etc. We have not stated still {\sl what} are these properties
of $T$-functions
that are crucial for cryptography. Yet, when we state these properties a
bit later, we
see that fortunately they are of this `analytic nature'.
%With this in mind, we are going to make more explorations of the ultrametric
%world.

By the way, the above observation reflects a very specific {\sl algebraic}
nature of $T$-functions. In general algebra, a {\it congruence} of an algebraic
system  is an equivalence 
relation which is preserved by all operations of this system; that is, if
replacing operands by equivalent elements the result of the operation is
equivalent to the original one. A function defined on (and valuated in) the
algebraic system is called {\it compatible} whenever this function preserves
all congruences of this algebraic system. The {\sl only} congruences of the
ring $\mathbb Z_p$ are congruences modulo $p^k$ for $k=1,2,\ldots$. Thus, {\sl $T$-functions
are merely compatible functions on the ring $\mathbb Z_2$}, so we start using
the term `compatible' along with (or instead of) the term `$T$-function'.

Actually, `$T$-function' just means `compatible on the ring $\mathbb Z_2$',
and many further results holds for functions that are compatible on $\Z_p$,
$p$ prime. A $p$-adic compatible function is the function that satisfies
$p$-adic Lipschitz condition with a constant 1, and vise versa.      
\subsection{More compatible functions} We already know that arithmetic operations
(addition, subtraction, and multiplication), as well as bitwise logical operations
($\XOR$, $\AND$, etc.) are $T$-functions (that is, compatible 2-adic functions).
Obviously, a {\sl composition of compatible functions is a compatible function}.
Whence, natural examples of compatible functions are {\sl polynomials} with
$p$-adic integer coefficients. That is, all polynomials with integer coefficients
{\sl are} $T$-functions! 

With some extra efforts one
could prove also that some other `natural' functions are also $T$-functions: 
\begin{equation}
\begin{split}
& {\text {\it exponentiation,}}\ \uparrow:\ (u,v)\mapsto u\uparrow v=(1+2\cdot
u)^v;
\ {\text{\rm in particular,}}
\\ 
& {\text {\it raising to negative powers}},\ u\uparrow(-r)=(1+2\cdot u)^{-r}, r\in\mathbb N;
\ {\text{\rm and}}
\\ 
& {\text {\it division,}}\ /: u/v=u\cdot (v\uparrow(-1))=\frac{u}{1+2\cdot
v}. 
\label{eq:opAr}
\end{split}
\end{equation}

That is, these functions are well defined on $\Z_2$, and satisfy 2-adic Lipschitz
condition with a constant 1. Use of compositions of these functions with the
above mentioned bitwise logical instruction results in very wild-lloking
functions, like this one:
$$%g(x)=
(1+x)\XOR 4\cdot\Biggl(1-2\cdot\frac{x\AND x^{2}+x^3\OR x^4}{3 - 4\cdot(5+6x^5)^{x^6\XOR x^7}}\Biggr)^{7-\frac{8x^8}{9+10x^9}}$$

Despite this function could be {\sl easily} evaluated on every digital computer
(since this function is continuous in a computer's 2-adic world), we do
not insist on using it (and similar) functions in applications:
Compositions of the above mentioned  functions may not be of big importance
for cryptography since their program implementations are usually slow, yet
they are of theoretical interest and often arise in studies. The $p$-adic
analogs of the above functions could be naturally defined (write $p$ instead of 2). 

It also worth notice here that $(1+p\cdot v)^{-1}=\sum_{i=0}^\infty(-1)^{i+1}p^iv^i$,
and the series in the right-hand part of this equality are convergent for
every $v\in\Z_p$.

We can describe univariate $T$-functions in some general way. It turns out
that each function $f\colon{\mathbb N}_{0}\rightarrow {\mathbb Z}_{p}$
(or, respectively, $f\colon{\mathbb N}_{0}\rightarrow {\mathbb Z}$) %from
admits one and only one representation in the form of   
so-called {\it Mahler interpolation series}
\begin{equation}
\label{eq:Mah}
f(x)=\sum^{\infty }_{i=0}a_{i}{\binom{x}{i}},
%\eqnum{\diamondsuit}
\end{equation}
where $
\binom{x}{i}
%\left(\matrix x\\ i\endmatrix\right)
%{{x}\choose{i}}
=\frac{x(x-1)\cdots  (x-i+1)}{i!}$ 
for $i=1,2,\ldots$, and
$
\binom{x}{0}
%\left(\matrix x\\ 0\endmatrix\right)
=1$;  $a_{i}\in {\mathbb Z}_{p}$ 
(respectively, $a_{i}\in {\mathbb Z}$), $i=0,1,2,\ldots $ .

If $f$ is uniformly continuous on ${\mathbb N}_{0}$ with respect to
$p$-adic distance, it can be uniquely expanded to a uniformly continuous
function on ${\mathbb Z}_{p}$. Hence the interpolation series for 
$f$ converges uniformly on 
${\mathbb Z}_{p}$. The following is true: The series
$
f(x)=\sum^{\infty }_{i=0}a_{i}
{\binom{x}{i}}
%\left(\matrix x\\ i\endmatrix\right)
, \quad$ 
($a_{i}\in {\mathbb Z}_{p}$, $i=0,1,2,\ldots \ $)
converges uniformly on ${\mathbb Z}_{p}$ iff 
$ \lim\limits^p_{i \to \infty }a_{i}=0,$
where $\lim\limits^p$ is a limit with respect to the $p$-adic distance; 
hence uniformly convergent series defines a uniformly continuous function
on ${\mathbb Z}_{p}$. 

The following theorem holds:
\begin{thm}
\label{thm:Mah-comp}
%{ 2.1 Theorem} { {\rom {(See 4.3 of \cite{11}; cf. \cite 5)}} 
The function
$f\colon{\mathbb Z}_{p}\rightarrow {\mathbb Z}_{p}$ represented by \eqref{eq:Mah}
is compatible if and only if
$$
a_{i}\equiv 0\pmod {p^{\lfloor\log_{p}i\rfloor}}
$$
%\noindent для 
for all $i=p,p+1,p+2,\ldots  $ . {\rm{ (Here and after
for a real $\alpha $ we denote
$\lfloor\alpha \rfloor$ an integral part of $\alpha $, i.e., the nearest
to $\alpha $ rational integer  not exceeding $\alpha $.)}}
\end{thm}
% \begin{note*} This theorem holds for arbitrary prime $p$, and not necessarily
% for $p=2$.
% \end{note*} 
\subsection{Derivatives modulo $p^k$}
\label{sub:der}
In this subsection we generalize  the main notion of
Calculus, a derivative. By  the definition, for $ \mathbf a=(a_{1},\ldots  ,a_{n})$ and
 $\mathbf b=(b_{1},\ldots  ,b_{n})$ of
${\mathbb Z}^{(n)}_{p}$ 
the congruence $\mathbf  a\equiv  \mathbf b\pmod{p^{s}}$ means that
$\|a_{i}-b_{i}\| _{p}\le p^{-s}$ (or, the same, that $a_{i}=b_{i}+c_{i}p^{s}$ 
for suitable $c_{i}\in {\mathbb Z}_{p}$, $i=1,2,\ldots  ,s$); that is $\|\mathbf
a-\mathbf b\|_p\le p^{-s}$. 
\begin{defn}[Derivations modulo $p^k$]
\label{def:der-mod}
A function 
$$F=(f_{1},\ldots  ,f_{m})\colon{\mathbb Z}^{n}_{p}\rightarrow {\mathbb Z}^{m}_{p}$$
is called {\it differentiable modulo $p^k$} at the point 
$ \mathbf u=(u_{1},\ldots  ,u_{n})\in {\mathbb Z}^{n}_{p}$ iff there exist a positive
integer rational
$N$ and an $n\times m$ matrix $F^{\prime}_{k}(\mathbf u)$ over ${\mathbb Z}_{p}$
(which is called {\it  the Jacobi matrix modulo} $p^{k}$ of the function $F$ at the
point
$\mathbf u$) such that for each positive rational integer  
$K\ge N$ and each $ \mathbf h=(h_{1},\ldots  ,h_{n})\in {\mathbb Z}^{n}_{p}$  
the inequality 
$\|\mathbf h\| _{p}\le     p^{-K}$ implies a congruence  
\begin{equation}
\label{eq:der-mod}
F( \mathbf u+\mathbf h)\equiv F(\mathbf u)+ \mathbf h\cdot F^{\prime}_{k}(\mathbf u)\pmod{p^{k+K}}.
\end{equation}
%\eqnum{\heartsuit } $$

In case $m=1$ the
Jacobi matrix modulo $p^k$ is called a {\it differential modulo $p^k$}. In
case $m=n$ a determinant of the Jacobi matrix modulo $p^k$ is called the {\it Jacobian
modulo $p^k$}. The entries of the Jacobi matrix modulo $p^k$
are called {\it partial derivatives modulo} $p^k$ of the function $F$ at
the point $\mathbf u$. 
A partial derivative (respectively, a differential) modulo $p^k$ we
sometimes  denote as 
$\frac{\partial_k f_i (\mathbf u)}{\partial_k x_j}$ (respectively, as
$d_{k}F(\mathbf u)=\sum^n_{i=1} \frac {\partial_k F(\mathbf u)}{\partial_k x_i}d_{k}x_{i}$).
\end{defn} 

It could be proved  that whenever $F$ is compatible, then, if $F$ is differentiable modulo
$p^k$ at
some point, the entries of the Jacobi matrix are {\sl necessarily}
$p$-adic integers (such functions are said to have {\it integer-valued} derivative).

Since the notion of function that is differentiable modulo $p^k$ is of high
importance in theory that follows, we discuss this notion in details. First
of all, we compare this notion to a classical notion of differentiable function.

Compare to differentiability, the {differentiability modulo $p^k$} is a weaker restriction. As a matter of fact, in a univariate case ($m=n=1$), definition \ref{def:der-mod} just yields
that   
$$
\frac{F( \mathbf u+\mathbf h)- F(\mathbf u)}{\mathbf h}\approx F^{\prime}_{k}(\mathbf u)$$
Note that this $\approx$ (`approximately') implies the following: 
%\textcolor{blue}{
\begin{align*}
{\approx}\ &\text{with {\sl arbitrarily high} precision}\Rightarrow \text{{differentiability}};\\
{\approx}\ &\text{with precision {\sl not worse than}}\ p^{-k}\Rightarrow{\text{differentiability}\bmod p^k}.
\end{align*}

It is obvious that whenever a function is differentiable (and its derivative
is a $p$-adic integer), it is differentiable
modulo $p^k$ for all $k=1,2,\ldots$, and in this case the derivative modulo
$p^k$ is just a {\sl reduction} of a derivative modulo $p^k$ (note that according
to definition \ref{def:der-mod} partial derivatives modulo $p^k$ are determined
up to a summand that is 0 modulo $p^k$).

For functions with integer-valued derivatives modulo $p^k$  
the `rules of derivation
modulo $p^k$' have the same (up to congruence modulo $p^k$ instead of equality)
form as for classical derivations.
For instance, if both functions 
$G\colon{\mathbb Z}^{s}_{p}\rightarrow {\mathbb Z}^{n}_{p}$ and
$F\colon{\mathbb Z}^{n}_{p}\rightarrow {\mathbb Z}^{m}_{p}$ 
are differentiable modulo 
$p^{k}$ at the points, respectively, $\mathbf v=(v_{1},\ldots  ,v_{s})$
and $\mathbf u=G(\mathbf v)$, and their partial derivatives modulo $p^{k}$ at
these points are $p$-adic integers, then a composition 
$F\circ G\colon{\mathbb Z}^{s}_{p}\rightarrow {\mathbb Z}^{m}_{p}$ 
of these functions is uniformly differentiable modulo $p^{k}$ at the point
$\mathbf v$, all its partial derivatives 
modulo $p^{k}$ at this point are $p$-adic integers, and 
$(F\circ G)^\prime_k (\mathbf v)\equiv G^\prime_k (\mathbf v) F^\prime_k (\mathbf u)\pmod
{p^k}$.

By the analogy with classical case we can give the following
\begin{defn}
\label{def:uniDer}
A function $F\colon{\mathbb Z}^{n}_{p}\rightarrow {\mathbb Z}^{m}_{p}$ is said
to be 
{\it uniformly differentiable modulo $p^k$ on $\mathbb Z_p^{(n)}$} iff there
exists $K\in\mathbb N$ such that \eqref{eq:der-mod} holds simultaneously for all 
$\mathbf u \in \mathbb Z_p^{n}$ as soon as
$\| h_{i}\| _{p}\le     p^{-K}$, $(i=1,2,\ldots  ,n)$. The
least such 
$K\in\mathbb N$
is denoted via $N_k(F)$. 
%The latter number plays an important role in
%further coniderations.
\end{defn}
It could be shown that   all  partial derivatives
 modulo $p^k$ of a uniformly differentiable modulo $p^k$ function $F$
 are periodic functions with period  
$p^{N_k(F)}$ (see \cite[Proposition  2.12]{anashin2}). 
This in particular implies that each partial derivative modulo
$p^k$ could be considered as a function defined on the residue ring $\mathbb Z/p^{N_k(F)}\Z$ modulo $p^{N_k(F)}$. 
Moreover, if a continuation $\tilde F$ of the function
$F=(f_{1},\ldots , f_{m})\colon{\mathbb N}^{n}_{0}\rightarrow {\mathbb N}^{m}_{0}$  
to the space $\mathbb Z_p^{n}$ is uniformly differentiable modulo $p^k$ on the
$\mathbb Z_p^{n}$, then one could continue both the function $F$  and all its
(partial) derivatives modulo $p^k$ to the space $\mathbb Z_p^{n}$
simultaneously. This implies that we could study if necessary (partial) 
derivatives modulo $p^k$
of the function $\tilde F$ instead of studying those of $F$ and vise versa.
For example, a partial derivative $\frac{\partial_k f_i (\mathbf u)}{\partial_k x_j}$
modulo $p^k$ vanishes modulo $p^k$ at no point of  $\mathbb Z_p^{n}$
(that is,
$\frac{\partial_k f_i (\mathbf u)}{\partial_k x_j}\not\equiv 0\pmod{p^k}$
for all $u\in \mathbb Z_p^{n}$, or, the same
$\big\|\frac{\partial_k f_i (\mathbf u)}{\partial_k x_j}\big\|_p> p^{-k}$
everywhere on $\mathbb Z_p^{n}$) if and only if 
$\frac{\partial_k f_i (\mathbf u)}{\partial_k x_j}\not\equiv 0\pmod{p^k}$
for all $u\in\{0,1,\ldots,p^{N_k(F)}-1\}$.

To calculate a derivative of, for instance, a $T$-function %state transition function, which
that is a composition of basic instructions %functions, see \ref{erg-comp},
one needs to know derivatives of these basic instructions (i.e., arithmetic,
bitwise logical, etc.)  %`elementary' functions, 
%such as \eqref{eq:opBinLog}
%and \eqref{eq:opAr}. 
Thus, we briefly introduce a $p$-adic analog of  
a `table of derivatives' of classical Calculus.
\begin{exmps} Derivatives of bitwise logical operations.
%\nopagebreak
\label{DerLog}
%\nopagebreak
\begin{enumerate}
%\nopagebreak
\item {\it the function $f(x)=x\AND c$ is uniformly differentiable on $\mathbb
Z_2$ for any $c\in
\mathbb Z$; $f^\prime(x)=0$ for $c\ge 0$, and $f^\prime(x)=1$ for $c<0$,} since
$f(x+2^ns)=f(x)$, and 
$f(x+2^ns)=f(x)+2^ns$ for $n\ge l(|c|)$, where $l(|c|)$ is the bit length
of absolute value of $c$
(mind that for $c\ge 0$ the $2$-adic representation
of $-c$ starts with $2^{l(c)}-c$ in less significant bits followed by 
$\ldots11$:
$-1=\ldots11$, $-3=\ldots11101$
%$11\ldots$:
%$-1=11\ldots$, $-3=10111\ldots$
, etc.).
\item {\it the function $f(x)=x\XOR c$ is uniformly differentiable on $\mathbb
Z_2$ for any $c\in
\mathbb Z$; $f^\prime(x)=1$ for $c\ge 0$, and $f^\prime(x)=-1$ for $c<0$.} This
immediately
follows from (1) since $u\XOR v=u+v-2(x\AND v)$ (see \eqref{eq:id}); thus
$(x\XOR c)^\prime=x^\prime+c^\prime-2(x\AND c)^\prime=1+2\cdot(0,\ \text{for}\
c\ge 0;\ \text{or}\
-1,\ \text{for}\ c<0)$.
\item in the same manner it could be shown that {\it functions $(x\bmod
2^n)=x\AND(2^n-1)$ {\rm (a reduction modulo $2^n$)}, $\NOT(x)$
and $(x\OR c)$ for $c\in \mathbb Z$ are uniformly differentiable on $\mathbb
Z_2$, and $(x\bmod 2^n)^\prime=0$, $(\NOT x)^\prime=-1$, 
$(x\OR c)^\prime=1$ for $c\ge 0$, 
$(x\OR c)^\prime=0$ for $c< 0$.}
\item {\it the function $f(x,y)=x\XOR y$ is not uniformly differentiable on 
$\mathbb Z_2^{2}$,
yet it is uniformly differentiable modulo $2$ on $\mathbb Z_2^{2}$};
from (2) it follows that its partial derivatives modulo 2 are 1 everywhere
on $\mathbb Z_2^{2}$.
\end{enumerate}
\end{exmps} 
% The examples of functions which are not uniformly differentiable on $\mathbb Z_p^{(n)}$,
% yet are  uniformly differentiable on $\mathbb Z_p^{(n)}$ modulo $p$, are 
% the function $f(x,y)=x\XOR y$ for $p=2$
% and its corresponding analogons for $p\ne 2$; all partial derivatives modulo
% $p$ of
% these functions are congruent to 1 modulo $p$ at all points (see \cite{me-1}). 
% Note by the way, that previously introduced  function 
% $\bmod{\,p^n}\colon \mathbb Z_p\rightarrow\mathbb Z/p^n$, the `reduction modulo $p^n$',
% is uniformly differentiable on  $\mathbb Z_p$ (its derivative is $0$ at all
% points);
% the function $f(x,y)=x\AND y$ is differentiable modulo  $2$ at no point
% of $\mathbb Z_2^{(2)}$, yet it is uniformly differentiable with respect to 
% $x$ for each  $y\in \mathbb Z$: its derivative is 0 for $y\ge 0$, and it is
% 1 in the opposite case.

%To clarify how it all works consider the following
Here how it works altogether:
\begin{exmp*}
%\label{exDer}
The function $f(x)=x+(x^2\OR 5)$ is uniformly differentiable
on $\mathbb Z_2$, 
%$N_1(f)=N_2(f)=\ldots =3$, 
and $f^\prime (x)=1+2x\cdot
(x\OR 5)^\prime=1+2x$.

The function $F(x,y)=(f(x,y),g(x,y))=
(x \oplus 2(x \wedge y ),(y +3 x^3 )\oplus x )$ 
is uniformly differentiable modulo $2$ as bivariate
function, and $N_1(F)=1$; namely
$$F(x+2^nt,y+2^ms)\equiv F(x,y)+(2^nt,2^ms)\cdot
\begin{pmatrix}
1&x+1\\
0&1
\end{pmatrix}
\pmod{2^{k+1}}$$ 
for all $m,n\ge 1$ (here $k=\min\{m,n\}$). The matrix
$\begin{pmatrix}
1&x+1\\
0&1
\end{pmatrix}
=F^\prime_1(x,y)$ is Jacobi matrix modulo 2 of $F$; here how we calculate
partial derivatives modulo $2$: for instance,  
$\frac{\partial_1 g(x,y)}{\partial_1 x}=\frac{\partial_1 (y +3 x^3)}{\partial_1 x}
\cdot \frac{\partial_1 (u\oplus x)}{\partial_1 u}\big|_{u=y +3 x^3}+
\frac{\partial_1 x}{\partial_1 x}\cdot 
\frac{\partial_1 (u\oplus x)}{\partial_1 x}\big|_{u=y +3 x^3}=9x^2\cdot 1+1\cdot
1\equiv x+1\pmod 2$.
Note that a partial derivative modulo 2 of the function 
$2(x \wedge y )$ is always $0$ modulo 2 because of the multiplier 2:
The function $x \wedge y$ is not differentiable modulo 2 as bivariate function,
yet $2(x \wedge y )$ is. So the Jacobian of the function $F$ is 
$\det F^\prime_1\equiv 1\pmod 2$.
\end{exmp*}
% 
%Here and after till the end of this section 
Now let  $F=(f_{1},\ldots , f_{m})\colon{\mathbb Z}^{n}_{p}\rightarrow {\mathbb Z}^{m}_{p}$  
and $f\colon{\mathbb Z}^{n}_{p}\rightarrow {\mathbb Z}_{p}$ be compatible
functions, 
which are uniformly differentiable on $\mathbb Z_p^{n}$  modulo $p$. This is a
relatively
weak restriction since all uniformly differentiable on $\mathbb Z_p^{n}$ functions,
as well as functions, which are uniformly differentiable on $\mathbb Z_p^{n}$
modulo $p^k$ for some $k\ge
1$, are uniformly differentiable on $\mathbb Z_p^{n}$ modulo $p$;
note that 
$\frac{\partial F}{\partial x_i}\equiv \frac{\partial_k F}{\partial_k x_i}\equiv
\frac{\partial_{k-1} F}{\partial_{k-1} x_i}\pmod{p^{k-1}}$.  
Moreover, as it was mentioned,
all values of all partial derivatives modulo $p^k$ (and thus, modulo $p$)
of $F$ and $f$ are $p$-adic integers everywhere on, 
%all points of 
respectively, $\mathbb Z_p^{n}$ and $\mathbb Z_p$, %(see \ref{intDer}),
so to calculate these values one can use the techniques considered above.
\section{Stream ciphers and 2-adic ergodic theory}
In this section we discuss what conditions state update and output functions
of a pseudorandom  generator should satisfy to guarantee some crucial cryptographic
properties of the produced sequence. It turns out that whenever these functions
are $T$-functions, the properties are tightly connected with the behaviour
of the functions with respect to a natural probabilistic measure on
the space $\Z_2$. We start with defining this measure.
\subsection{Notions of $p$-adic dynamics} When we measure a square of a figure  on a plane (or a volume of a body in a space),
we associate a real number to the figure (resp., to the body). These are
natural examples of {\it measures}. We are not going to recall basic notions
of measure theory here, referring to any book on this topic. We only mention
that we could define a measure $\mu$  on some set $\mathbb S$ by assigning non-negative real numbers
to some subsets that are called elementary. All other {\it measurable} subsets
are compositions of these elementary subsets with respect to countable unions,
intersections, and complements. Actually, if a measurable subset $S\subset
\mathbb S$   is a
disjoint union of elementary measurable subsets $E_j$, $S=\cup_{j=0}^\infty E_j$, then $\mu(S)=\sum_{j=0}^\infty
\mu(E_j)$, and the series in the right-hand part must be convergent. The set
$\mathbb S$ with so defined measure $\mu$ is called a {\it measurable space}.

The elementary subsets in $\Z_p$ are balls $B_{p^{-k}}(a)=a+p^k\Z_p$. To each such ball we assign a number $\mu_p(B_{p^{-k}}(a))=\frac{1}{p^k}$. It
could be verified that we indeed define a measure on the space $\Z_p$, and
this measure is a probabilistic measure, $\mu_p(\Z_p)=1$. This measure $\mu_p$
is called a (normalized) {\it Haar measure} on $\Z_p$.

We say that we have a {\it dynamical system} on a measurable space $\mathbb
S$, whenever
we consider a triple $(\mathbb S;\mu; f)$, where $\mathbb S$ is a measurable space with measure
$\mu$, and $f\colon \mathbb S\> \mathbb S$ is a {\it measurable function}; that is, an
$f$-preimage of every measurable subset is a measurable subset. Dynamical
system theory is a reach mathematical theory which is applied in different
parts of science and industry. As a matter of fact, in this course we will
discuss {\sl applications of 2-adic dynamical systems theory to stream cipher design}. 

A {\it trajectory} of a dynamical system is a sequence
$$x_0, x_1=f(x_0),\ldots, x_i=f(x_{i-1})=f^i(x_0),\ldots$$
of points of the space $\mathbb S$, $x_0$ is called an {\it initial} point of the
trajectory. If $F\colon \mathbb S\> \mathbb T$ is a measurable mapping to some other measurable space
$\mathbb T$ with a measure $\nu$ (that is, an $F$-preimage of any $\nu$-measurable subset
of $\mathbb T$ is a $\mu$-measurable subset of $X$), the sequence $F(x_0), F(x_1), F(x_2),\ldots$ is called an {\it observable}. Note that the trajectory formally
looks like the sequence of states of a pseudorandom generator, whereas the
observable resembles the output sequence, cf. subsection \ref{sub:PRNG}.
Further we will see that is not just an analogy.

The two important notions of dynamical systems theory follow:
A mapping $F\colon\mathbb S\rightarrow\mathbb
Y$ of a measurable space  $\mathbb S$ into a measurable space $\mathbb Y$ endowed with probabilistic measure  $\mu$ and $\nu$, respectively,
is said to be 
{{\it measure-preserving}} (or, sometimes, {\it equiprobable})
whenever $\mu(F^{-1}(S))=\nu(S)$ for each measurable subset   $S\subset\mathbb Y$. In case $\mathbb S=\mathbb Y$ and $\mu=\nu$,
a measure-preserving mapping  $F$ is said to be 
{{\it ergodic}} whenever for each measurable subset 
$S$ such that $F^{-1}(S)=S$ holds either  $\mu(S)=1$ or $\mu(S)=0$.
Loosely speaking, %\underline{
any invariant
set of the ergodic  
mapping is either nothing, or everything. 

The $p$-adic ergodic theory  studies ergodic (with respect to the Haar measure)
transformations of the space of $p$-adic numbers, conditions that provide
ergodicity, etc. It is a rapidly developing mathematical theory, with
various
applications, see e.g. \cite{Khren-Nils}. Actually, as we will see,
the course is a development of $p$-adic ergodic theory with special interest
to pseudorandom number generators (particulary, stream ciphers).\footnote{By
the way, methods developed within this approach could be applied to solve
some problems of $p$-adic ergodic theory, see \cite{me-spher}}
And now it is a right time to discuss how the above notions are related to properties of
pseudorandom generators. %that are built on $T$-functions. 
\subsection{What is a good PRNG} A PRNG  which could be considered any good
%for cryptographic purposes 
obviously must meet the following conditions:
%\begin{boxitpara}{box 0.9 setgray fill}
%\begin{enumerate}
\begin{itemize} 
\item The output sequence must be  pseudorandom (i.e., must pass certain
statistical tests). 
%undistinguishable
%from a truly random sequence up to some pre-defined statistial tests of
%$\mathcal T$).
\item For cryptographic applications, given a segment $z_j, z_{j+1},\ldots, z_{j+s-1}$ of the output sequence, finding the corresponding initial state
(which usually is a key) 
%$x_0$
must
be
infeasible in some properly defined sense.
\item The PRNG must be suitable for software (or hardware) implementation; the
performance must be
sufficiently fast.
\end{itemize}

In case the PRNG is an automaton described by Figure \ref{fig:PRNG} we could
re-state these conditions as follows:

First of all, we state 
\begin{quote}
\underline{Condition 1:} The state update function $f$ must provide pseudorandomness; in
particular, it must guarantee 
{\sl uniform distribution} and {\sl long period} of the state update sequence
$\{u_i\}$.
\end{quote}
It would be great if this sequence is secure; that is,
given $u_i$, it is infeasible  neither  to find (or to
predict) $u_{i+1}$, nor to find $u_0$. Unfortunately, this is not easy to
provide these properties: Generators that are `provably secure', that is,
supplied with proofs (which are based on some plausible, yet still unproven
conjectures) that their output sequences can not be predicted by polynomial-time
algorithms, are too slow for most practical applications. In real life %design
one has to undertake additional efforts to make the algorithm secure. Usually
this could be achieved with the use of the output function. Thus, we need
\begin{quote}
\underline{Condition 2:} The output function $F$ must not spoil  pseudorandomness
(at least, the output sequence $\{z_i\}$ must
be {\sl uniformly distributed} and  must have {\sl long period}).

Moreover, in cryptographic applications
the function $F$
must make the {PRNG} secure:
(in particular, given $z_i$, it must be {\sl difficult to find $u_i$
from the equation $z_i=F(u_i)$}).
\end{quote}

Finally, we can formulate
\begin{quote}
\underline{Condition 3:} To make the {PRNG} any suitable for software/hardware implementations,  
{\sl both $f$ and $G$ must be 
compositions of  %\hyperlink{instr}
basic processor instructions}.
\end{quote}

In section \ref{sec:Prelm} we already have discussed how one could satisfy condition
3: It is sufficient to choose both $f$ and $F$ from the class of $T$-functions.
Thus, we can assume that $f\colon\mathbb Z/2^n\Z\rightarrow\mathbb
Z/2^n\Z$ and $F\colon\mathbb Z/2^n\Z\rightarrow\mathbb
Z/2^m\Z$ (usually, $m\le n$).

Now, to satisfy condition 1, one could take  the state update function $f\colon\mathbb Z/2^n\Z\rightarrow\mathbb
Z/2^n\Z$ with a {\it single cycle property}; that is, $f$   permutes
elements of $\mathbb Z/2^n\Z$ cyclically. %These mappings are also called
%{\it transitive modulo} $2^n$. 

The state update sequence %$\{x_i\in\mathbb
%Z/2^n\}$ 
$$u_0,\ u_1=f(u_0), %x_2=f(x_1), 
\ldots,u_{i+1}=f(u_i)=f^{i+1}(u_0),\ldots$$
of $n$-bit words will have then the  {\sl longest possible period}
(of length $2^n$),
and {\it strict uniform distribution}; that is, {each $n$-bit word
will
occur
at the period exactly once}.

To satisfy the first part of {condition 2}, one could
take the output function  $F\colon\mathbb Z/2^n\Z\rightarrow\mathbb
Z/2^m\Z$ to be {\it balanced}:
That is, to {each} $m$-bit word the mapping $F$ maps {the
same number} of $n$-bit words (that's why $m\le n$). 
For %instance, \textcolor{red}{\sl latin squares} are balanced mappings with $k=\frac{n}{2}$;
%for 
$m=n$
balanced mappings are just {\sl invertible} (that is,
bijective, one-to-one) mappings. Obviously, if a balanced output function
is applied to a strictly uniformly distributed sequence of states, the output
sequence (of $m$-bit words) is also strictly uniformly distributed: It is {\sl periodic with a period
of length $2^n$, and each $m$-bit word occurs at the period exactly $2^{n-m}$
times}.

For $m\ll n$, balanced functions could serve us to
satisfy 
the second part of {condition 2}, since
the equation $y_i=G(x_i)$ has too many solutions then, $2^{n-m}$ (so it
is infeasible to an attacker to try them all).

Thus, we must know how to construct balanced (or single-cycle) functions out of basic
processor instructions. This is where the non-Archimedean analysis comes into play!
\subsection{A bridge} Now we make our studies more formal. Let $F\colon\Z_p^n\>\Z_p^m$
be a compatible function; that is, let $F$ satisfy the $p$-adic Lipschitz
condition with a constant 1 (see section \ref{sec:Ap_an}). In other words,
for every $k=1,2,\ldots$, and for every $\mathbf a,\mathbf b\in\Z_p^n$,  
$F(\mathbf a)\equiv F(\mathbf b)\pmod{p^k}$ whenever $\mathbf a\equiv\mathbf
b\pmod{p^k}$ (see subsection \ref{sub:der} for the definition of $\bmod p^k$).
This means that, given a compatible mapping $F\colon\Z_p^n\>\Z_p^m$, its
{\it reduction $F\bmod{p^k}$ modulo} $p^k$ is a {\sl well defined mapping} $$F\bmod{p^k}\colon(\Z/p^k\Z)^n\>(\Z/p^k\Z)^m$$ 
of respective Cartesian powers
of the residue ring $\Z/p^k\Z$. We call the mapping $F\bmod{p^k}$ the {\it
induced} mapping. The idea is quite clear: Reduction modulo $p^k$ just
deletes all most significant digits (starting with the $k$-th digit) both
of arguments
and of values of the function $F$.  

%We start with
%the following:
\begin{defn}
A compatible mapping $F\colon\mathbb Z_p\rightarrow\mathbb
Z_p$ is said to be  {{\it bijective} (resp., {\it transitive}) modulo $p^k$} iff the induced mapping  $x\mapsto
F(x)\pmod {p^k}$ is a (single-cycle) permutation of the elements of the ring %(resp., a permutation
$\mathbb Z/p^k\Z$.
\end{defn}  
{\it Balance modulo $p^k$} could be defined by an analogy. Now we can state
the central result of this section:

\begin{thm} [see \cite{anashin3}]%\textup{(Anashin)} %, 2002)}
\label{thm:erg-tran}
%\hypertarget{thm:erg-tran}{% 
For $m=n=1$, a compatible mapping  $F\colon\mathbb Z_p^n\rightarrow\mathbb
Z_p^m$ %\psboxit{box
%.9 setgray fill}{\spbox{
{preserves the normalized Haar measure} $\mu_p$ on $\mathbb Z_p$ %}} 
\textup{(}resp., %\psboxit{5 cartouche}{\spbox{
{is ergodic} with respect to $\mu_p$%}}
\textup{)}
if and only if it is %{\yellow  $%\Longleftrightarrow
%F$}
{
bijective %}} 
\textup{(}resp., %\psboxit{5 cartouche}{\spbox{
transitive%}}
\textup{)} modulo $p^k$} for all $k=1,2,3,\ldots$   %тогда и только тогда,
%когда оно
%\psboxit{box
%.9 setgray fill}{\spbox{
.
%относительно .

%\bigskip

For $n\ge m$,  the mapping $F$ {preserves the measure} $\mu_p$ if and only if it induces a {balanced} mapping of $(\mathbb Z/p^k\Z)^n$ onto $(\mathbb Z/p^k\Z)^m$, for all 
$k=1,2,3,\ldots$.

\end{thm}

This theorem acts like a bridge between $p$-adic ergodic theory and stream
cipher design: We consider  the corresponding PRNG as approximation with
respect to 2-adic metric of some ergodic dynamical system on 2-adic integers.
In a {pseudorandom generator}, we can take
compatible ergodic functions for state update functions; also we can
take compatible
measure-preserving functions for output functions. The reduction modulo $2^n$ a computer performs automatically.
In particular, for $p=2$ from theorem \ref{thm:erg-tran} we obtain:

\begin{itemize}
% \psboxit{box
% .9 setgray fill}{\spbox{\textcolor{blue}{$\surd $} \textcolor{red}{
\item {measure preservation $=$ invertibility} modulo ${2^k}$ for
all
$k\in\mathbb N$;
 
\item in dimensions $>1$, i.e., for $F\colon\mathbb Z_2^n\rightarrow\mathbb
Z_2^m$,

{measure preservation $=$ balance} modulo ${2^k}$ for all 
$k\in\mathbb N$;  
\item 
{ergodicity $=$ single cycle property} modulo  ${2^k}$ for all
$k\in\mathbb N$.
\end{itemize}
In other words, a compatible function $F\colon\mathbb Z_2\rightarrow\mathbb Z_2$ is measure-preserving (respectively, ergodic)
if and only if the corresponding $T$-function $F\pmod {2^n}$ on $n$-bit words (which
is merely an approximation of $F$
with precision $\frac{1}{2^n}$) is invertible or, respectively, has a single cycle property!

Now the problem is how to describe these measure-preserving (in particular,
ergodic) mappings in the class of all compatible mappings. We start to develop some
theory to answer the following questions: What compositions of  {basic  instructions}
are measure-preserving? are ergodic? Given a composition of basic instructions, is it measure-preserving? is it ergodic?
%, and we start to study
%these techniques.
\section{Tools}
\label{sec:Tools}
The main goal of this section is to describe some tools with the use of which we could answer
the above stated questions. However, we start with some historical observations.
\subsection{A phenomenon} Study of pseudorandom generators has a
long history. You can read about this issue in, for instance, an excellent
book of Donald Knuth \cite{knuth}. Here we discuss briefly a short passage
of this long story, aiming to make some important observations.

One could notice that behavior of a mapping modulo $p^N$, where $N$ is {\Large
big}, is totally
determined by the behavior of this mapping modulo $p^n$, where $n$ is {{\tiny
 small}}. One of the first generators that demonstrate this behaviour is
 
%\begin{quote}
\underline{Linear Congruential Generator} (Hull and
Dobell, 1962):

%\bigskip

{\it The mapping 
$$x\mapsto a\cdot x+b\pmod{p^N},$$ 
where $a,b\in\mathbb Z$, {$N\ge
2$}, is a 
permutation with a {single cycle property} if and only if  $x\mapsto a\cdot x+b\pmod{p^n}$ is a permutation
with a {single cycle property for $n=1$} in case $p$ odd, or for { $n=2$}, otherwise.}
%\end{quote}

The following important example is

%\begin{quote}
\underline{Bijectivity Criterion for Polynomials with Integer Coefficients} (proved
and re-proved by a number of authors; known since 1960\textsuperscript{th}):
%A folklore; easily follows from Hensel's Lemma)%(mid 60\textsuperscript{th})

%\bigskip

{\it
The mapping 
$$x\mapsto f(x)\pmod{p^N},$$ 
where  {$N\ge
2$} and $f$ is a polynomial with rational integer coefficients, is 
{bijective} % with a single cycle 
if and only if $x\mapsto f(x)\pmod{p^n}$ is {bijective}
%with a single cycle 
for {$n=2$.}}
%\end{quote}

Yet another one example:

\underline{Quadratic Generator} (Coveyou, 1969):
%Transitivity Criterion for Polynomials over
%Integer Rationals} (Larin) %; %much later 
%also Zieve and DesJardine)%, mid 80\textsuperscript{th})

%\bigskip

{\it
The mapping 
$$x\mapsto f(x)\pmod{p^N},$$ 
where {$N\ge
3$} and $f$ is a quadratic polynomial with rational integer coefficients, is a 
{permutation  with a single cycle property} 
iff $x\mapsto f(x)\pmod{p^n}$ is a {permutation
with a single cycle property}
for {$n=3$} 
in case $p\in\{2,3\}$ , or for {$n=2$}, otherwise.}

%\bigskip

It worth notice here that in 1980\textsuperscript{th} M.~V.~Larin proved that the {\sl  word `quadratic' in the
statement could
be omitted}! The result was spread as a manuscript that time, a journal publication
\cite{larin}
appeared much later.

\subsection{Explanation: $p$-adic derivations} Looking at the examples of
the preceding subsection, we naturally start suspecting that some very strong reason for such behaviour must exist!
The following theorem, which
was published in 1993 \cite{anashin1, anashin2}, gives an explanation:
\begin{thm} %\textup{(Anashin)} %, 1993)}
\label{thm:erg_Der}
%\textcolor{yellow}
{
Let a compatible function $F\colon{\mathbb Z}_{p}\rightarrow {\mathbb Z}_{p}$ 
be uniformly differentiable modulo {$p^{2}$}. 
%and let $F$
%have integer-valued derivative modulo $p^{2}$. 
Then $F$ is
%asymptotically 
ergodic if and only if it is transitive modulo $p^{N_{2}(F)+1}$ for
odd prime $p$ or, respectively, modulo $2^{N_{2}(F)+2}$ for $p=2$.
}
%\textup{(Recall
%Definition \ref{def:der}.)} 
\end{thm}
This theorem works for a much wider class of functions that the ones mentioned
in the above examples. Actually, this class includes functions that are compositions
of not exceptionally arithmetic operations, but of logical operations as
well. To illustrate the techniques, consider the following example.
\begin{exmp}
\label{ergKlSh}
In their paper \cite{KlSh} of 2002 Klimov and Shamir write  that 
\begin{quote}
...neither the invertibility nor 
the cycle structure of
$x +(x^2\vee 5)$ could be determined by his ({\slshape i.e., mine --- V.A.}) techniques.
\end{quote}
See however how it could be immediately done with the use of Theorem
\ref{thm:erg_Der}:
The function $f(x)=x+(x^2\vee 5)$ is uniformly differentiable
on $\mathbb Z_2$, thus, it is uniformly differentiable modulo 4 
(see \ref{DerLog} and an example thereafter), and $N_2(f)=3$. Indeed, $(x+h)\OR 5=(x\OR 5)+h$ whenever $h\equiv 0\pmod
8$ (the latter congruence is obvious since the base-2 expansion of 5 is ...000101).

Now to
prove that $f$ is ergodic, in view of \ref{thm:erg_Der} it suffices 
to demonstrate that $f$ induces a permutation
with a single cycle on $\mathbb Z/32$. Direct calculations show that the
string
$$0,f(0)\bmod 32, f^2(0)\bmod 32=f(f(0))\bmod 32, \ldots, f^{31}(0)\bmod
32$$ is a permutation of the string $0,1,2,\ldots,31$, thus ending the proof.
\end{exmp}

In connection with Theorem \ref{thm:erg_Der}, the following natural question arises: What about ergodicity in higher
dimensions? Unfortunately, for uniformly differentiable modulo $p$ function
the answer is negative. The following result could be considered as  a non-existence theorem for compatible smooth ergodic
mappings in higher dimensions.
\begin{thm}[see \cite{anashin1,anashin2}]%\textup{(Anashin)} %, 1993)}
\label{thm:noTmvar}
{Let the function $F=(f_{1},\ldots  ,f_{n})\colon{\mathbb Z}^{{n}}_{p}\rightarrow {\mathbb Z}^{{n}}_{p}$ 
be compatible,
%asymptotically 
ergodic, 
and uniformly differentiable
modulo $p$ on $\mathbb Z_p$. %, and let it have integer-valued
%derivatives modulo $p$. 
Then {$n=1$}.} %\textup{(Non-differentiable $\bmod p$ ones do exist
%for $n>1$)}
\end{thm}
\begin{note*}
{Non-differentiable $\bmod p$} ones do exist
for $n>1$
\end{note*} 

The following theorem, which uses derivations modulo $p$ instead of $p^2$,  could be applied to construct balanced mappings to serve
as output
functions of {PRNG}.
\begin{thm}[see \cite{anashin3}]
\label{thm:MHL}
%\textup{(``Multidinensional Hensel's Lemma")} %, 2002)}
{ Let $F%=(f_{1},\ldots  ,f_{m})
\colon{\mathbb Z}^{n}_{p}\rightarrow {\mathbb Z}^{m}_{p}$ 
be a compatible function that is uniformly differentiable modulo {
$p$}. %function. %and let all its partial derivatives
%modulo $p$ be integer-valued on $\Bbb Z_p$. 
Then $F$ preserves measure %is %asymptotally
%equiprobable
whenever 
it is balanced modulo $p^{k}$ for some
$k\ge N_{1}(F)$ and the rank of its Jacobi matrix $F_1^\prime (u)$ modulo
$p$ is exactly $m$ at all points  $\mathbf u=(u_{1},\ldots  ,u_{n})\in ({\mathbb Z}/p^{k})^{n}$.}
\end{thm}
\begin{proof}
For $\xi \in ({\mathbb Z}/p^{s})^{m}$ denote  $$F^{-1}_{s}(\xi )=\{\gamma \in ({\mathbb Z}/p^{s})^{n}\colon F(\gamma )\equiv \xi \pmod{p^{s}}\}.$$ 
Let
$s\ge k\ge N_{1}(F)$. Since $F$ is compatible, and hence
$F$ is a sum of a compatible function and a periodic function with period
 $p^{N_1(F)}$ (see 2.10
of \cite{anashin2}), we conclude that if $\eta \in F^{-1}_{s+1}(\xi )$,  then  $\bar{\eta }\in F^{-1}_{s}(\bar{\xi })$. Here and further we denote
via
%Here, in accordance with our agreement in the introduction,  
$\bar{\alpha }=(\bar{\alpha }_{1},\ldots  ,\bar{\alpha }_{m})\in ({\mathbb Z}/p^{s})^{m}$ 
the residue modulo $p^s$,  $\alpha\bmod{p^s}=(\alpha_1\bmod{p^s},\ldots,\alpha_m\bmod{p^s})$, where $\alpha =(\alpha _{1},\ldots  ,\alpha _{m})\in ({\mathbb Z}/p^{s+1})^{m}$. 

Put $\lambda =\bar{\eta }+p^{s}\sigma\in(\mathbb Z/p^{s+1})^{n}$, where $\sigma \in ({\mathbb Z}/p)^{n}$. 
In view of the uniform differentiability of 
the function $F$ modulo $p$ (see \ref{def:der-mod}), we have
\begin{equation}
\label{eq:Tayl-md}
F(\lambda )\equiv F(\eta )+p^{s}\sigma F_1^{\prime}(\bar{\eta })\pmod{p^{s+1}}.
\end{equation}
\noindent  Since  $F(\bar{\eta })\equiv \bar{\xi }+p^{k}\beta \pmod{p^{s+1}}$
and $ \xi =\bar{\xi }+p^{s}\gamma $ 
for suitable $\beta ,\gamma \in ({\mathbb Z}/p)^{(m)}$, in view of \eqref{eq:Tayl-md}
we conclude that $\lambda \in F^{-1}_{s+1}(\xi )$ if and only if $\bar{\lambda }\in F^{-1}_{s}(\xi )$ 
(i.e., $\bar{\eta }\in F^{-1}_{s}(\xi ))$ and $\alpha $
satisfies the following system of linear equations over a finite field ${\mathbb Z}/p$:
\begin{equation}
\label{eq:lin-syst}
\beta +\alpha F_1^{\prime}(\bar{\eta })=\gamma .
\end{equation}
\noindent Thus, if columns of the matrix $F_1^\prime (\bar{\eta })$ are
linearly independent over ${\mathbb Z}/p$, then linear system \eqref{eq:lin-syst} has exactly
$p^{n-m}$ pairwise distinct solutions for arbitrary $\beta ,\gamma \in ({\mathbb Z}/p)^{(m)}$. 
From here it follows that
\begin{equation}
\label{eq:coim-num}
\vert F^{-1}_{s+1}(\xi )\vert =\vert F^{-1}_{s}(\xi )\vert p^{n-m}.
\end{equation}
\noindent Hence, if $F$ is equiprobable modulo $p^{k}$ (i.e., if $\vert F^{-1}_{s}(\bar{\xi })\vert $ does
not depend on $\bar{\xi })$
and if rank of the matrix $F_1^\prime (\bar{\eta })$ is $m$, then \eqref{eq:coim-num} implies
that $F$ 
is balanced modulo $p^{s+1}$.
\end{proof}
\begin{cor} Under assumptions of theorem \ref{thm:MHL}: 
\begin{itemize}
\item If $m=1$,  then
$F$ is measure-preserving whenever $F$ is balanced
modulo $p^{k}$ for some $k\ge N_{1}(F)$, and the differential $d_{1}F$  modulo
 $p$  of the function $F$
vanishes at no point of $({\mathbb Z}/p^{k}|Z)^{n}$.

\item Let $f(x_{1},\ldots  ,x_{n})$ be a polynomial in variables
 $x_{1},\ldots  ,x_{n}$, and let all coefficients of $f$ 
are $p$-adic integers. The polynomial 
$f$ preserves measure whenever it is balanced modulo
$p$ and all its partial derivatives vanishes simultaneously modulo $p$ at no point
of $({\mathbb Z}/p\Z)^{n}$ {\rm (i.e., are simultaneously congruent to $0$ modulo $p$ 
nowhere) on $({\mathbb Z}/p\Z)^{n}$.}
\end{itemize}
\end{cor}
For $m=n$ the above stated sufficient conditions of measure preservation becomes also necessary ones.

\begin{thm}
\label{thm:MHL-bj}  
A compatible and uniformly differentiable modulo $p$ function
$$F=(f_{1},\ldots  ,f_{m})\colon{\Z}^{n}_{p}\rightarrow {\mathbb Z}^{n}_{p}$$
%with integer-valued derivatives modulo $p$ asymptotically 
preserves
measure if and only if it is bijective modulo 
$p^{N_{1}(F)}$ and its Jacobian modulo $p$ vanishes at no point of 
$({\Z}/p^{N_1(F)}\Z)^{n}$ {\rm (Equivalent condition: If and only if $F$ is bijective
modulo $p^{N_1{(F)+1}}$)}.
\end{thm}
\begin{proof}
If $F$ is bijective modulo  $p^{N_{1}(F)}$, and if its Jacobian modulo
$p$ vanishes nowhere on $({\mathbb Z}/p^{N_1(F)})^{n}$, then in view of Theorem \ref{thm:MHL}  $F$ %is asymptotically equiprobable,
%hence, asymptotically 
preserves measure.
%, since $m=n$.
\par
Vise versa, let $F$ preserve measure, i.e., let $F$ be
bijective modulo $p^{k}$ for all $k\ge N$, where $N$ is some positive rational
integer.
Now take $k\ge \max\{N,N_{1}(F)\}$, then the definition of uniform differentiability
modulo $p$ implies that
\begin{equation}
\label{eq:Tayl}
F(u+p^{k}\alpha )\equiv F(u)+p^{k}\alpha F_1^{\prime}(u)\pmod{p^{k+1}}
\end{equation}
\noindent for all $u,\alpha \in {\mathbb Z}_{p}$. Here $F_1^\prime (u)$ is
an $n\times n$ matrix over a field ${\mathbb Z}/p$. If $\det  F_1^\prime (u)\equiv 0\pmod{p}$
for some $u\in {\mathbb Z}^{n}_{p}$ (or, the same, for 
some $u\in \{0,1,\ldots  ,p^{N_{1}(F)}-1\}^{n}$ in view of the periodicity
of partial derivatives modulo $p$), 
then there exists $\alpha \in \{0,1,\ldots  ,p-1\}^{n}, \alpha \not\equiv (0,\ldots  ,0)\pmod{p}$, 
such that
$\alpha F_1^\prime(u)\equiv  (0,\ldots,0)\pmod{p}$. But then \eqref{eq:Tayl} implies that 
$F(u+p^{k}\alpha )\equiv F(u)\pmod{p^{k+1}}$. The latter contradicts the
bijectivity modulo $p^{k+1}$ of the function
$F$, since for $u\in \{0,1,\ldots  ,p^{N_{1}(F)}-1\}^{n}$ we have
$u,u+p^{k}\alpha \in \{0,1,\ldots  ,p^{k+1}-1\}^{n}$ and $u+p^{k}\alpha \neq u$.
\par
Now we prove the criterion in the equivalent form. Let $F$ be bijective 
modulo $p^{N_{1}(F)}$. Then assuming $k=N_{1}(F)$ in the above argument,
we conclude that $\det F_1^\prime(u)\not\equiv 0\pmod{p}$ for all $u\in {\mathbb Z}^{n}_{p}$. 
According to Theorem \ref{thm:MHL}, this implies that $F$ preserves measure.
\par
Let $F$ preserve measure, and let $F$ be not 
bijective modulo $p^{k}$ for some $k\ge N_{1}(F)$. We prove that in this
case $F$ is not bijective modulo $p^{k+1}$. 

Choose  $u,v\in \{0,1,\ldots  ,p^{k}-1\}^{n}$ 
such that $u\neq v$ и
$F(u)\equiv F(v)\pmod{p^{k}}$. Then either $F(u)\equiv F(v)\pmod{p^{k+1}}$ (i.e., $F$ 
is not bijective modulo $p^{k+1})$, or $F(u)\not\equiv F(v)\pmod{p^{k+1}}$. 
Yet in the latter case we have
$F(u)\equiv F(v)+p^{k}\alpha \pmod{p^{k+1}}$ for some $\alpha \in \{0,1,\ldots  ,p-1\}^{n}$, $\alpha \not\equiv (0,\ldots  ,0)\pmod{p}$.
Consider $u_{1}=u+p^{k}\beta $, where $\beta \in \{0,1,\ldots  ,p-1\}^{n}$
with $\beta \not\equiv (0,\ldots  ,0)\pmod{p}$ and
$\beta F_1^\prime(u)+\alpha \equiv (0,\ldots  ,0)\pmod{p}$. Such $\beta $ exists,
since $F$ preserves measure and, consequently,  
$\det F_1^\prime(u)\not\equiv 0\pmod{p}$, as
this have been proven already.
Now the definition of uniform differentiability modulo $p$ implies that 
\begin{equation}
\label{eq:Tayl-2}
F(u+p^{k}\beta )\equiv F(u)+p^{k}\beta F_1^{\prime}(u)\equiv F(v)+p^{k}\alpha +p^{k}\beta F_1^{\prime}(u)\equiv F(v)\pmod{p^{k+1}},
\end{equation}
\noindent where $u+p^{k}\beta \in \{0,1,\ldots  ,p^{k+1}-1\}^{(n)}$ and $u+p^{k}\alpha \neq v$ (since $u\neq v$). 
Thus \eqref{eq:Tayl-2} in combination with our assumption imply that $F$ is not bijective
modulo 
$p^{k+1}$. Applying this argument sufficient number of times, we conclude
that $F$ is not bijective 
modulo $p^{s}$ for all $s\ge k$. But at the same time $F$ %asymptotically
preserves measure. A contradiction.
\end{proof}
Comparing theorems \ref{thm:MHL} and \ref{thm:MHL-bj} one may ask whether
sufficient conditions of theorem \ref{thm:MHL} are also necessary. The answer
is negative: In \cite{anashin3} it is proved that the function $f(x,y)=2x+y^{3}$ on ${\mathbb Z}_{2}$ provides a counter-example.
\begin{opq*} Characterize all compatible measure-preserving mappings $$F=(f_{1},\ldots  ,f_{m})\colon{\Z}^{n}_{p}\rightarrow {\mathbb Z}^{n}_{p}$$ with $m<n$. The
 answer is not known even under restriction that all $f_i$ are polynomials
 over $\mathbb Z_p$.
\end{opq*}

The technique presented in this subsection is rather effective: Actually, all the examples of preceding subsection could be deduced from
the results of this subsection. Moreover, all results of \cite{KlSh} also
could be proved by these techniques. We re-prove these results to illustrate
our techniques: 
\begin{exmps} 
\label{KlSh-ex}
%(see \cite{KlSh})
The following is true:
\begin{enumerate}
\item {\it A mapping 
$$(x,y ) \mapsto F(x,y)=(x \oplus 2(x \wedge y ),(y +3 x^3 )\oplus x )\bmod{2^r}$$
of $\mathbb (Z/2^r)^{2}$ onto $\mathbb (Z/2^r)^{2}$ 
is bijective for all $r=1,2,\ldots$}

Indeed, the function $F$ is bijective modulo $2^{N_1(F)}=2$ (direct verification)
and  $\det(F_1^\prime (\mathbf u))\equiv 1\pmod 2$ for all $\mathbf u\in(\mathbb
Z/2)^{2}$ (see \ref{DerLog} and example thereafter).
\item {\it The following mappings of $\mathbb Z/2^r$ onto $\mathbb Z/2^r$ 
are bijective for all $r=1,2,\ldots$}:
$$ 
\begin{array}{lcr} 
x\mapsto& (x +2x^2)&\bmod{2^r},\\ x\mapsto& (x +(x^2\vee 1))&\bmod{2^r},\\
x\mapsto& (x \oplus (x^2\vee 1))&\bmod{2^r}.
\end{array}
$$

Indeed, all three mappings are uniformly differentiable
modulo 2, and $N_1=1$ for all of them. So it suffices to prove that
all three mappings are bijective modulo 2, i.e. as mappings of the residue
ring $\mathbb Z/2$ modulo 2 onto itself (this could be checked by direct calculations), 
and that
their derivatives modulo 2 vanish at no point of $\mathbb Z/2$. The latter
also holds, since  the derivatives are, respectively,
%$$1+2x=1\pmod 2,\ 1+2x\cdot 0=1\pmod 2,\ 1+2x\cdot 0\pmod 2$$
$$
\begin{array}{lcr}
1+4x&\equiv& 1\pmod 2,\\ 1+2x\cdot 1&\equiv& 1\pmod 2,\\ 1+2x\cdot 1&\equiv&
1\pmod 2,
\end{array}
$$
%since $(x^2\vee 1)^\prime=2x\cdot 0=0$ , and $(x\oplus C)^\prime_1=1$,
since $(x^2\vee 1)^\prime=2x\cdot 1\equiv 1\pmod 2$, and $(x\oplus C)^\prime_1\equiv
1\pmod 2$,
(see \ref{DerLog}).
%where $(x\oplus C)^\prime_1$ is derivative
%modulo 2 of the function $x\oplus C$
\item {\it The following closely related variants of the previous mappings
of 
$\mathbb Z/2^r$ onto $\mathbb Z/2^r$ 
are {\sl not} bijective for all $r=1,2,\ldots$}:
$$
\begin{array}{lcr}
x\mapsto& (x +x^2)&\bmod{2^r},\\  x \mapsto& (x +(x^2\wedge 1))&\bmod{2^r},\\  x\mapsto& (x +(x^3\vee 1))&\bmod{2^r},
\end{array}
$$ 
since they are compatible but
not bijective modulo 2.
\item (see \cite{Riv}, also \cite[Theorem 1]{KlSh}) {\it Let $P (x )=a_0 +a_1 x + \cdots+a_d x^d$ be a polynomial with integral
coefficients. Then $P (x )$ is a permutation polynomial } (i.e., is bijective)
{\it modulo $2^ n$,
$n>1$ if and
only if $a_1$ is odd, $(a_2 +a_4 + \cdots)$ is even, and $(a_3 +a_5 +\cdots)$
is even.}

In view of \ref{thm:MHL-bj} we must verify whether the two conditions
hold: first, whether $P$ is bijective modulo 2, and second,
whether
$P^\prime(z)\equiv 1\pmod 2$ for $z\in\{0,1\}$.
The first condition implies that $P(0)=a_0$ and $P(1)=a_0+a_1+a_2+\cdots a_d$
must be distinct modulo 2; hence $a_1+a_2+\cdots a_d\equiv 1\pmod 2$. 
The second condition implies that
$P^\prime(0)=a_1\equiv 1\pmod2,\ P^\prime(1)\equiv a_1+a_3+a_5+\cdots\equiv 1\pmod 2$.
Now combining all this together we get $a_2+a_3+\cdots a_d\equiv 0\pmod 2$ and 
$a_3+a_5+\cdots\equiv 0\pmod 2$, hence $a_2 +a_4 + \cdots\equiv 0\pmod 2$.
\item As a bonus, we can use exactly the same proof to
get exactly the same characterization of bijective modulo $2^r$ $(r=1,2,\ldots)$
mappings of the form $x\mapsto P (x )=
a_0\oplus  a_1x\oplus \cdots\oplus  a_dx^d\bmod 2^r$ since $u\oplus v$ is uniformly
differentiable modulo 2 as bivariate function, and its derivative modulo
2 is exactly the same as the derivative of $u+v$, and besides, $u\oplus v\equiv
u+v\pmod 2$. 
\end{enumerate}
\end{exmps}
Note that in general theorems \ref{thm:MHL} and \ref{thm:MHL-bj} could be applied to a class of
functions that is narrower than the class of all compatible functions.
However, it turns out that for $p=2$ this is not the case. Namely, the
following proposition holds:%, which in fact is just a restatement of a 
%corresponding assertion of \ref{ergBool}.
\begin{prop}
\label{mpDer}
{\rm(\cite[Corollary 4.6]{anashin2}, \cite[Corollary 4.4]{anashin1})}
If a compatible function $g\colon\mathbb Z_2\rightarrow\mathbb Z_2$ preserves
measure then it is  uniformly differentiable modulo $2$, and %has integer 
its  derivative
modulo $2$ %(which 
is always $1$ modulo $2$.
\end{prop}

The above results are good to verify whether a given function preserves measure
or is ergodic. However, we need more tools to construct
measure-preserving, (respectively, ergodic) mappings in explicit form. 
\subsection{Mahler's series} We already have mentioned that uniformly continuous
functions defined on (and valuated in) $\Z_p$ could be uniquely represented
as Mahler's interpolation series \eqref{eq:Mah}.  So, it is natural to express
conditions of measure-preservation or ergodicity in terms of coefficients
of these series.
\begin{thm}[\cite{anashin2, anashin1, anashin3}]
\label{thm:ergBin} 
For $p=2$ a function $f\colon{\mathbb Z}_{p}\rightarrow {\mathbb Z}_{p}$  
is compatible and measure-preserving if and only if it could be represented as 
\begin{equation*}
f(x)=c_0+x+\sum^{\infty }_{i=1}c_{i}\,p^{\lfloor \log_p i \rfloor +1}\binom{x}{i}
\qquad (x\in\mathbb Z_p);
\end{equation*}
The function 
$f$
is compatible and ergodic if and only if it could be represented as 
\begin{equation*}
f(x)=1+x+\sum^{\infty}_{i=1}c_{i}p^{\lfloor \log_{p}(i+1)\rfloor+1}\binom{x}{i}
\qquad (x\in\mathbb Z_p),
\end{equation*}
where $c_0,c_1, c_2 \ldots \in {\mathbb Z}_p$. For $p\ne2$ these conditions
remain sufficient, and {\sl not} necessary.
\end{thm}
Thus, in view of theorem \ref{thm:ergBin} one can choose a state transition
function to be a polynomial with rational (not necessarily integer)
key-dependent coefficients setting $c_i=0$ for all but finite number of $i$.
Note that to determine whether a given polynomial $f$ with rational (and not
necessarily integer) coefficients is integer valued (that is, maps $\mathbb
Z_p$ into itself), compatible and ergodic, it is sufficient to determine
whether it
induces a cycle on $O(\deg f)$ integral points. To be more exact, the following
proposition holds.
\begin{prop}[\cite{anashin3}]
\label{prop:Qpol} 
A polynomial $f(x)$ with rational, and not necessarily integer coefficients,
is integer valued, compatible, and ergodic
{\rm (}resp., measure preserving{\rm)} if and only if 
$$z\mapsto f(z)\bmod p^{\lfloor
\log_p (\deg f)\rfloor +3},$$ 
where $z$ 
runs through $0,1,\ldots,p^{\lfloor
\log_p (\deg f)\rfloor +3}-1$,  is compatible and transitive 
{\rm (}resp., bijective{\rm)} mapping
%задает  совместимую и транзитивную функцию на кольце вычетов 
of the residue ring $\Z/p^{\lfloor
\log_p (\deg f)\rfloor +3}$ onto itself. 
\end{prop}
Theorem \ref{thm:ergBin} enables one to use exponentiation in design of  
generators that are transitive modulo $2^n$ for all $n=1,2,3,\ldots$. 
%(on exponential generators see e.g. \cite{LinRec}). 

\begin{exmp} 
\label{expGen}
For any odd $a=1+2m$ a function $f(x)=ax+a^x$ defines 
a transitive modulo $2^n$ 
generator $x_{i+1}=f(x_i)\bmod 2^n$.

Indeed, in view of \ref{thm:ergBin} the function $f$ defines a compatible and ergodic
mapping of $\mathbb Z_2$ onto $\mathbb Z_2$
since $f(x)=(1+2m)x+(1+2m)^x=x+2mx+\sum_{i=0}^\infty m^i 2^i\binom{x}{i}=
1+x+4m\binom{x}{1}+
\sum_{i=2}^\infty m^i 2^i\binom{x}{i}$ and $i\ge\lfloor\log_2(i+1)\rfloor+1$
for all $i=2,3,4,\ldots$. 

Such a generator could be of practical value since it uses not more than
$n+1$ multiplications modulo $2^n$ of $n$-bit numbers; of course, one should
use calls to the table 
$a^{2^j}\bmod{2^n}$, $j=1,2,3,\ldots,n-1$. The latter table must be precomputed,
 corresponding calculations involve $n-1$ multiplications modulo $2^n$. Obviously,
one can use $m$ as a long-term key, with the initial state $x_0$ being
a short-term
key, i.e., one changes $m$ from time to time, but uses  new $x_0$ for each
new message. Obviously, without a properly
chosen output function such a generator is not secure. The choice of output
function in more details is discussed further. %in the paper. 

\end{exmp}
\begin{note*}
A similar argument shows that for every prime
$p$ and  every $a\equiv 1\pmod
p$ the function $f(x)=ax+a^x$ defines a compatible and ergodic mapping
of $\mathbb Z_p$ onto itself.
\end{note*}

For polynomials with  (rational or $p$-adic) integer coefficients 
theorem \ref{thm:ergBin} may be restated in the following form.
\begin{prop}[\cite{anashin1,anashin2}]
\label{ergPol}
%{\rm (See  \cite[Corollary 4.11]{}, \cite[Corollary 4.7]{me-conf})}
Represent a polynomial $f(x)\in\mathbb Z_2[x]$  in a basis of descending 
factorial powers
$$
x^{\underline 0}=1,\ x^{\underline 1}=x,\ x^{\underline 2}=x(x-1),\ldots,\
x^{\underline i}=x(x-1)\cdots(x-i+1),\ldots,$$
i.e., let
$$f(x)=\sum^{d}_{i=0}c_i\cdot x^{\underline i}$$
for $c_0,c_1,\dots,c_d\in\mathbb Z_2$. Then the polynomial $f$ induces
an ergodic {\rm (and, obviously, a compatible)} mapping of $\mathbb Z_2$ onto
itself iff its coefficients $c_0,c_1,c_2, c_3$ satisfy the following congruences:   
$$c_0\equiv 1\ (\bmod\, 2),\quad c_1\equiv 1\ (\bmod\, 4),\quad c_2\equiv 0\
(\bmod\, 2),\quad c_3\equiv 0\ (\bmod\, 4).$$
The polynomial $f$ induces a measure preserving mapping iff
$$c_1\equiv 1\ (\bmod\, 2),\quad c_2\equiv 0\ (\bmod\, 2),\quad c_3\equiv 0\ (\bmod\, 2).$$
\end{prop}
Thus, to provide ergodicity of the polynomial mapping $f$  
it is necessary and sufficient to hold fixed $6$ bits only, while the other bits of 
coefficients of $f$ may vary (e.g., may be key-dependent). This guarantees transitivity
of the state transition function $z\mapsto f(z)\bmod 2^n$ for each $n$, and hence,
uniform distribution of the output sequence.

Proposition \ref{ergPol} implies that the polynomial $f(x)\in\mathbb
Z[x]$ is ergodic (resp., measure preserving) iff it is transitive modulo 8
(resp., iff it is bijective modulo 4). A corresponding assertion
holds in general case, for arbitrary prime $p$.
\begin{thm}[\cite{larin, anashin3}]
\label{ergPolGen}
%{\rm (See \cite{Lar}, \cite{me-2})} 
A polynomial $f(x)\in\mathbb Z_p[x]$
induces an ergodic mapping of $\mathbb Z_p$ onto itself iff it is transitive
modulo $p^2$ for $p\ne 2,3$, or modulo $p^3$, for $p=2,3$. The polynomial
$f(x)\in\mathbb Z_p[x]$ induces a measure preserving mapping of 
$\mathbb Z_p$ onto itself iff it is bijective
modulo $p^2$.
\end{thm}

\begin{exmp} 
The mapping $x\mapsto f(x)\equiv x+2x^2\pmod{2^{32}}$ (which is used in
RC6, see \cite{RC6}) is bijective, since it is bijective modulo 4: 
$f(0)\equiv 0\pmod4$, $f(1)\equiv 3\pmod4$, $f(2)\equiv 2\pmod4$, 
$f(3)\equiv 1\pmod4$. Thus, the mapping 
$x\mapsto f(x)\equiv x+2x^2\pmod{2^{n}}$ is bijective for all $n=1,2,\ldots$. 
\end{exmp}
Hence, with the use of the theorem \ref{ergPolGen} it is possible to
obtain transitive modulo $q>1$ mappings for arbitrary natural $q$: one can
just take $f(z)=(1+z+\hat qg(z))\bmod q$, where $g(x)\in\mathbb Z[x]$ is
an arbitrary polynomial, and $\hat q$ is a product of $p^{s_p}$ for all
prime factors $p$ of $q$, where $s_2=s_3=3$, and $s_p=2$ for $p\ne 2,3$. Again,
the polynomial $g(x)$ may be chosen, roughly speaking, `more or less at random',
i.e., it may be key-dependent, but the output sequence will be uniformly
distributed for any choice of $g(x)$. This assertion may be generalized
either.
\begin{prop}[\cite{anashin3}]
\label{ergAn} 
%{\rm (\cite[Lemma 4.4 and Proposition 4.5; resp., Lemma
%4.11 and Proposition 4.12 in the preprint]{me-2})} 
Let
$p$ be a prime, and let
$g(x)$ be an arbitrary composition of arithmetic operations  and mappings listed in \eqref{eq:opAr}.
Then the mapping $z\mapsto 1+z+p^2g(z)$\ $(z\in\mathbb Z_p)$ is ergodic. 
\end{prop}

In fact, both propositions \ref{ergPol}, \ref{ergAn} and theorem \ref{ergPolGen} 
are 
special cases of the following general
\begin{thm}[\cite{anashin3}]
\label{ergAnGen}
%{\rm (\cite[Theorem 4.2, or 4.9 in the preprint]{me-2})}
Let $\mathcal B_p$ be a class of all functions defined by series of
a form $f(x)=\sum^{\infty}_{i=0}c_i\cdot x^{\underline i}$, where 
$c_0,c_1,\dots$ are $p$-adic integers, and
%Represent a polynomial $f(x)\in\mathbb Z_2[x]$  in a descending factorial basis
$x^{\underline i}$ $(i=0,1,2,\ldots)$  
%$(x)_1=x$, $(x)_2=x(x-1)$, $\ldots$, $(x)_i=x(x-1)\cdots(x-i+1),\ldots$
are descending factorial powers {\rm(see \ref{ergPol})}.
%i.e., let
Then
the function $f\in \mathcal B_p$ preserves measure iff it is bijective
modulo $p^2$; $f$ is ergodic iff it is transitive modulo $p^2$
{\rm(}for $p\ne 2,3${\rm)}, or modulo $p^3$ {\rm(}for $p\in\{2,3\}${\rm)}.

\end{thm}
\begin{note*} As it was shown in \cite{anashin3}, the class $\mathcal B_p$
contains all polynomial functions over $\mathbb Z_p$,
as well as analytic (e.g., rational, entire) functions that are convergent everywhere
on $\mathbb Z_p$. \footnote{More information about this class could be found
in \cite{me-spher}} As a matter of fact, every mapping that is  a composition 
of arithmetic operators (addition, subtraction,  multiplication, and operators
listed in \eqref{eq:opAr})  belong to $\mathcal B_p$; thus, every
such mapping modulo $p^n$ could be induced by a polynomial with rational
integer coefficients (see the end of Section 4 in \cite{anashin3}). For instance,
the mapping $x\mapsto (3x+3^x) \bmod 2^n$ (which is transitive modulo $2^n$,
see \ref{expGen}) could be induced by a polynomial $1+x+4\binom{x}{1}+
\sum_{i=2}^{n-1} 2^i\binom{x}{i}=1+5x+\sum_{i=2}^{n-1} \frac{2^i}{i!} \cdot
x^{\underline i}$ --- just note that $c_i=\frac{2^i}{i!}$ are $2$-adic integers
since the exponent of maximal power of $2$ that is a factor of $i!$ 
is exactly $i-\wt_2i$,
where $\wt_2 i$ is a number of $1$'s in the base-2 expansion of $i$
(see e.g. \cite[Chapter 1, Section 2, Exercise 12]{Kobl}); thus 
$\|c_i\|_2=2^{-\wt_2 i}\le 1$, i.e. $c_i\in \mathbb Z_2$ and so $c_i\bmod
{2^n}\in \mathbb Z$.  
\end{note*}

Theorem \ref{ergAnGen} implies that, for instance, the state transition
function $f(z)=(1+z+\zeta(q)^2(1+\zeta(q)u(z))^{v(z)})\bmod q$ is transitive
modulo $q$ for each natural $q>1$ and arbitrary polynomials $u(x),v(x)\in\mathbb
Z[x]$,  where $\zeta(q)$ is a product of all prime factors of $q$. So the
one can choose as a state transition function not only polynomial functions, but
also rational functions, as well as analytic ones. It should be mentioned,
however, that this is merely a form the function is represented (which
could be suitable for some cases and unsuitable for the others), yet, for a
given $q$, all the
functions of this type may also be represented as polynomials over $\mathbb
Z$ (see \cite[Proposition 4.4; resp., Proposition 4.10 in the preprint]{anashin3}). 
For instance, certain generators
of inversive kind (i.e., those using taking the inverse modulo $2^n$) could
be considered in such manner.
\begin{exmp}
\label{Invers}
For $f(x)=-\frac{1}{2x+1}-x$ a generator $x_{i+1}=f(x_i)\bmod{2^n}$ is transitive.
Indeed, the function $f(x)=(-1+2x-4x^2+8x^3-\cdots)-x=-1+x-4x^2+8(\cdots)$ 
is analytic
and defined everywhere on $\mathbb Z_2$; thus $f\in\mathcal B_p$. Now the
conclusion follows in view of \ref{ergAnGen} since by direct calculations
it could be easily verified that the function $f(x)\equiv -1+x-4x^2\pmod
8$ is transitive modulo 8. Note that modulo $2^n$ the mapping $x\mapsto
f(x)\bmod 2^n$ could be induced by a polynomial $-1+x-4x^2+8x^3+\cdots+(-1)^n
2^{n-1}x^{n-1}$.
\end{exmp}
\subsection{Explicit expressions}It turns out that there is an easy way to construct a measure preserving or ergodic
mapping out of an arbitrary compatible mapping, i.e., out of an arbitrary
composition of both arithmetic (including \eqref{eq:opAr}) 
and bitwise logical %logical \eqref{eq:opLog}
operators. 
%Namely, the following
%proposition holds.
\begin{thm}[\cite{anashin3}]
\label{Delta} %\cite[Lemma 2.1 and Theorem 2.5]{me-2}. 
Let $\Delta$ be a difference operator, i.e., $\Delta g(x)=g(x+1)-g(x)$
by the definition. Let, further, $p$ be a prime, let $c$ be a coprime with
$p$, $\gcd(c,p)=1$, and let $g\colon\mathbb Z_p\rightarrow
\mathbb Z_p$ be a compatible mapping. Then the mapping $z\mapsto c+z+p\Delta
g(z)\ (z\in\mathbb Z_p)$ is ergodic, and the mapping $z\mapsto d+cx+pg(x)$, 
%where $d,c\in\mathbb Z_p$, $g$ is compatible, and $c$ is coprime with $p$, 
%defines a 
preserves measure for arbitrary $d$.
%mapping $f\colon\mathbb Z_p\rightarrow\mathbb
%Z_p$ 

Moreover, if $p=2$, then the converse also holds: Each compatible and ergodic
\textup {(}respectively each compatible
and measure preserving \textup {)}
mapping $z\mapsto f(z)\ (z\in\mathbb Z_2)$ could be represented as
$f(x)=1+x+2\Delta g(x)$  \textup {(}respectively as
$f(x)=d+x+2g(x)$\textup {)} for suitable $d\in\mathbb Z_2$ and compatible 
$g\colon\mathbb
Z_2\rightarrow \mathbb Z_2$.
\end{thm}
% \begin{proof} 
% The first assertion of the proposition is just Lemma
% 2.1 of \cite{me-2}.
% To prove the second assertion 
% recall that 
% each compatible mapping $g\colon\mathbb
% Z_p\rightarrow\mathbb Z_p$
% could be represented as
% $$g(x)=c_0+\sum_{i=1}^{\infty}c_ip^{\lfloor\log_pi\rfloor}\binom{x}{i} \quad (x\in\mathbb
% Z_p)$$
% for suitable $c_0, c_1,\ldots,\in\mathbb Z_p$ (see e.g. 4.3 of \cite{me-1}).
% Now, as $\Delta\binom{x}{i}=\binom{x}{i-1}$ for $i=1,2,\ldots$, we finish
% the proof in view of \ref{ergBin}.
% \end{proof}
\begin{note*} The case $p=2$ is the only case the converse of the first
assertion of theorem \ref{Delta} holds.
\begin{proof}
To start with, by induction on  $l$ we show that 
$g$ is bijective modulo $p^{l}$ for all  $l=1,2,3,\ldots  $ . The assumption
is obviously true for $l=1$.
\par
Assume it is true for $l=1,2,\ldots  ,k-1$. Prove that it holds for
$l=k$ either.
Let $g(a)\equiv g(b)\pmod{p^{k}}$ for some $p$-adic integers $a,b$. 
Then $a\equiv b\pmod{p^{k-1}}$ by the induction hypothesis. Hence  
$pv(a)\equiv pv(b)\pmod{p^{k}}$ since $v$ is compatible. 
Further, the congruence $g(a)\equiv g(b)\pmod {p^{k}}$ implies that 
$ca+pv(a)\equiv cb+pv(b)\pmod{p^{k}}$, and consequently, $ca\equiv cb\pmod{p^{k}}$. 
Since $c\not\equiv 0\pmod p$,
the latter congruence implies that $a\equiv b\pmod{p^{k}}$, proving the
first assertion of the lemma.
\par
To prove the rest part of the first assertion we note that the 
just proven claim implies that $h$ preserves measure.
To prove the transitivity of $h$ modulo $p^{k}$ for all $k=1,2,3,\ldots  $ 
we apply induction on $k$ once again.
\par
It is obvious that $h$ is transitive modulo $p$. Assume that $h$ is transitive
modulo $p^{k-1}$.
Then, since $h$ induces a permutation on the residue ring ${\mathbb Z}/p^{k}\Z$
and since $h$ is a compatible function, we conclude that the length of each
cycle of this permutation must be a multiple of $p^{k-1}$. Thus, to prove
this permutation is single cycle it suffices to prove that the function 
$$h^{p^{k-1}}(x)=\underbrace {h(h\ldots  (h}_{p^{k-1}\;\text{ раз}}(x))\ldots)$$ 
induces a single cycle permutation on the ideal $p^{k-1}\Z$, generated by
the element $p^{k-1}$
of the ring ${\mathbb Z}/p^{k}\Z$.  In other words, it is sufficient to demonstrate
that the function ${\frac {1} {p^{k-1}}}h^{p^{k-1}}(p^{k-1}x)$ 
is transitive modulo $p$.
\par
Applying obvious direct calculations, we successively obtain that 
\begin{gather*}
h^{1}(x)=c+x+pv(x+1)-pv(x),\\
\ldots \qquad \ldots \qquad \ldots \\
%\begin{multline}
h^{j}(x)=h(h^{j-1}(x))=cj+h^{j-1}(x)+pv(h^{j-1}(x)+1)-pv(h^{j-1}(x))=\\
 cj+x+p\sum^{j-1}_{i=0}v(h^{i}(x)+1)-p\sum^{j-1}_{i=0}v(h^{i}(x)),
%\end{multline}
\end{gather*}
\par
\noindent and henceforth. We recall that $h^{0}(x)=x$ by the definition. So,
\begin{equation}
\label{eq:iter}
h^{p^{k-1}}(x)=cp^{k-1}+x+p\sum^{p^{k-1}-1}_{i=0}v(h^{i}(x)+1)-p\sum^{p^{k-1}-1}_{i=0}v(h^{i}(x)).
\end{equation}
\par
Since $h$ is transitive modulo $p^{k-1}$ and compatible, we get now that 
$$
\sum^{p^{k-1}-1}_{i=0}v(h^{i}(x)+1)\equiv \sum^{p^{k-1}-1}_{i=0}v(h^{i}(x))\equiv \sum^{p^{k-1}-1}_{z=0}v(z)\pmod{p^{k-1}},
$$
\noindent and \eqref{eq:iter} implies then 
$h^{p^{k-1}}(x)\equiv cp^{k-1}+x\pmod{p^{k}}$. But  $c\not\equiv 0\pmod p$, 
so we conclude that the function $cp^{k-1}+x$ induces on the ideal $p^{k-1}\Z$ a
single cycle permutation, thus proving the first assertion of the theorem.

To prove the second assertion, note that as $g$ is compatible, its Mahler's
interpolation series are of the form of Theorem \ref{thm:Mah-comp}; noe note
that $\Delta\binom{x}{i}=\binom{x}{i-1}$ and apply Theorem \ref{thm:ergBin}.
\end{proof} 
% Moreover, from \ref{ergBin}
% it follows immediately that a mapping
% $f\colon\mathbb Z_2\rightarrow\mathbb Z_2$ is compatible and measure preserving
% iff it could be represented as $f(x)=d+x+2g(x)$, where $d\in\mathbb Z_2$
% and $g$ is compatible. A formula $f(x)=d+c\cdot
% x+pg(x)$, where $d,c\in\mathbb Z_p$, $g$ is compatible, and $c$ is coprime with $p$, 
% defines a measure preserving mapping $f\colon\mathbb Z_p\rightarrow\mathbb
% Z_p$ for arbitrary prime $p$ (see a note after \ref{ergBin}), yet only
% for $p=2$  this
% formula describes {\it all} compatible measure preserving mappings. 
\end{note*}
\begin{exmp}
\label{KlSh-2} 
Theorem \ref{Delta} immediately implies
Theorem 2 of \cite{KlSh}: For any composition $f$ of primitive functions, 
the mapping $x\mapsto x +2f(x )\pmod {2^n}$ is invertible --- just note
that
% The assertion
% follows immediately from \ref{Delta} since each 
a composition of primitive
functions is compatible (see \cite{KlSh} for the definition of primitive
functions).\qed
\end{exmp}
Theorem \ref{Delta} is maybe one of the most important tools in design of pseudorandom
generators such that  both their state transition functions and output functions are
key-dependent.
The corresponding schemes are rather flexible: In fact, one may use nearly
arbitrary composition of arithmetic and logical operators to  produce a
strictly uniformly distributed sequence:
% We emphasize, in the  example just mentioned the transitivity modulo $m=2^k$
% {\it does not depend} neither on $k$ nor on actual form of the composition
% $g$ --- 
Both for 
$g(x)=x\XOR(2x+1)$ and for 
$$g(x)=\Biggl(1+2\frac{x\AND x^{2}+x^3\OR x^4}{3 + 4(5+6x^5)^{x^6\XOR x^7}}\Biggr)^{7+\frac{8x^8}{9+10x^9}}$$   
a sequence $\{x_i\}$ defined by recurrence relation
$x_{i+1}=(1+x_i+2(g(x_i+1)-g(x_i)))\bmod {2^n}$ is strictly uniformly distributed
in $\mathbb Z/2^n\Z$  for each $n=1,2,3\ldots$, i.e., the sequence $\{x_i\}$ 
%Actually, this sequence
is 
purely periodic with {\slshape period length exactly} $2^n$, and  {\slshape each} element
of 
$\{0,1,\ldots,2^n-1\}$ occurs at the period {\slshape exactly once}. We will
demonstrate further that a designer could vary the function $g$ in a very
wide scope without worsening prescribed values of 
some important indicators of security. In fact,  choosing the proper arithmetic
and bitwise logical operators 
%\eqref{eq:opBinLog} and \eqref{eq:opAr} 
the designer is restricted
only by desirable performance, since any compatible ergodic mapping could
be produced in this way:
\begin{cor}
\label{erg-comp} 
%Under assumptions of the proposition \ref{Delta}, 
Let $p=2$,
and let $f$
be a compatible
and ergodic mapping of $\mathbb Z_2$ onto itself. Then for each $n=1,2,\ldots$
the state transition function $f\bmod 2^n$ could be represented as a finite
composition of arithmetic and bitwise logical operators. %\eqref{eq:opBinLog} and \eqref{eq:opAr}.
\end{cor}
\begin{proof}
In view of proposition \ref{Delta} it is sufficient to prove that for
arbitrary compatible $g$ the function $\bar g=g\bmod 2^n$ could be represented
as a finite composition of operators mentioned in the statement. %\eqref{eq:opBinLog} and \eqref{eq:opAr}.
In view of Definition \ref{def:T-fun},%\ref{Bool}, 
one could
represent $\bar g$ as 
$$\bar g(x)=\gamma_0(\chi_0)+2\gamma_1(\chi_0,\chi_1)+\cdots
+2^{n-1}\gamma_{n-1}(\chi_0,\ldots,\chi_{n-1}),$$
where $\gamma_i=\delta_i(\bar g)$, $\chi_i=\delta_i(x)$, $i=0,1,\ldots,n-1$.
Since each $\gamma_i(\chi_0,\ldots,\chi_i)$ is a Boolean function in Boolean
variables $\chi_0,\ldots,\chi_i$, it
%could be expressed as a Boolean polynomial.
%i.e., as a square-free polynomial over $\mathbb Z/2$ in variables $\chi_0,\ldots,\chi_i$. 
%Thus, each 
%$\gamma_i(\chi_0,\ldots,\chi_i)$
could be expressed via finite number of $\XOR$s and $\AND$s of these variables
$\chi_0,\ldots,\chi_i$. Yet each variable $\chi_j$ could be expressed as
$\chi_j=\delta_j(x)=x\AND(2^j)$, and the conclusion follows. 
\end{proof}
\subsection{Using Boolean representations}
As we just have seen, in case $p=2$ we have two equivalent descriptions of the class of all
compatible ergodic mappings, namely, theorems \ref{thm:ergBin} and 
\ref{Delta}. They enable one to express {\slshape any} compatible and transitive
modulo $2^n$ state transition function either as a polynomial of special kind
over a field
$\mathbb Q$ of rational numbers, or as a special composition of arithmetic
and bitwise logical operations. %, \eqref{eq:opAr} and \eqref{eq:opBinLog}.
Both these representations are suitable for programming, since they involve
only standard machine instructions. However, we
need one more representation, in a Boolean form, which we have already used in the definition
of $T$-function (see Definition \ref{def:T-fun}). Despite this
representation is not very convenient for programming,
it could be used %further for better understanding of certain important properties
%of the considered generators, as well for proving 
to prove the ergodicity of some
simple
mappings, see e.g. \ref{KlSh-3} below.
%\begin{note*} 
The following theorem is just a restatement in our terms of a known (at least 30 years
old) result from the theory 
of Boolean functions, the so-called bijectivity/transitivity criterion  for triangle
Boolean mappings.
However, the latter result is a  mathematical folklore, and thus it is somewhat
difficult to
attribute it. %Here we give a re-statement (and a proof) of this result using
%our terminology. %, yet a reader could find a proof in, e.g.,
%\cite[Lemma 4.8]{me-1}.
%\end{note*}

Recall that the
{\it algebraic normal form}, ANF, of the Boolean function $\psi_j(\chi_0,\ldots,\chi_j)$
is the representation of this function via $\oplus$ (addition modulo 2, that
is,
logical `exclusive or') and $\cdot$ (multiplication modulo 2, that is, logical `and', or
conjunction). In other words, the ANF of the Boolean function $\psi$ is its representation
in the form {
$$\psi(\chi_0,\ldots,\chi_j)= \beta\oplus\beta_0\chi_0\oplus\beta_1\chi_1\oplus\ldots
\oplus\beta_{0,1}\chi_0\chi_1\oplus\ldots,$$
where $\beta,\beta_0,\ldots\in\{0,1\}$.} The ANF is sometimes called a {\it
Boolean polynomial}.

%\bigskip

Recall that the {\it weight} of the Boolean function $\psi_j$ in $(j+1)$
variables is the number of $(j+1)$-bit words that {\it satisfy} $\psi_j$;
that is, weight is the cardinality of the truth set of $\psi_j$.

\begin{thm}
\label{ergBool} 
A mapping $T\colon\mathbb Z_2\rightarrow\mathbb Z_2$ is
compatible and measure preserving iff for each $i=0,1,\ldots$ the ANF of
the Boolean function 
$\tau^T_i=\delta_i(T)$ 
in Boolean variables $\chi_0,\ldots,\chi_{i}$ is %could be represented as a
%Boolean
%polynomial\footnote{that is, {\it algebraic normal form}} of the form
$$\tau^T_i(\chi_0,\ldots,\chi_i)=\chi_i\oplus\varphi^T_i(\chi_0,\ldots,\chi_{i-1}),$$ 
where $\varphi^T_i$
is an ANF. The mapping $T$ is compatible  and ergodic iff,
additionally, the Boolean function
$\varphi^T_i$ is of odd weight, that is,
takes value $1$ exactly at the odd number of points 
$(\varepsilon_0,\dots,\varepsilon_{i-1})$, where
$\varepsilon_j\in\{0,1\}$ for $j=0,1,\ldots,i-1$. The latter takes place if and only
if $\varphi^T_0=1$, and the degree of the ANF $\varphi^T_i$ for
$i\ge 1$ is exactly
$i$, that is, $\varphi^T_i$ contains a monomial
$\chi_0\cdots\chi_{i-1}$.
\end{thm}
\begin{proof} Represent the value of the function  $T$ at the 2-adic integer
point $x=\chi_0+\chi_1\cdot 2+\chi_2\cdot 2^2+\cdots$ as a $2$-adic integer:
$$T(\chi_0+\chi_1\cdot 2+\chi_2\cdot 2^2+\cdots)=\sum_{i=0}^\infty \delta_i(x)\cdot
2^i.$$
The function $T$ is compatible (that is, a $T$-function) if and only if $\delta_i(x)$
does not depend on $\chi_{i+1},\chi_{i+2},\ldots$ for every $i=0,1,2,\ldots$,
see Definition \ref{def:T-fun}. Thus, each $\delta_i(x)$ is a Boolean function
$\tau_i^T$
in Boolean variables $\chi_0,\chi_1,\ldots,\chi_i$. Re-write the  ANF of the
function $\tau_i^T$ in the following form:
$$\tau_i^T(\chi_0,\ldots,\chi_i)=\chi_i\cdot\psi^T_i(\chi_0,\ldots,\chi_{i-1})\oplus\varphi^T_i(\chi_0,\ldots,\chi_{i-1}),$$
where both $\psi^T_i(\chi_0,\ldots,\chi_{i-1})$ and $\varphi^T_i(\chi_0,\ldots,\chi_{i-1})$
are Boolean functions in Boolean variables $\chi_0,\ldots,\chi_{i-1}$. 

Obviously, whenever all $\psi^T_i(\chi_0,\ldots,\chi_{i-1})$ are identically
1, the function is measure-preserving since it is bijective modulo $2^{k+1}$
for each $k=0,1,2,\ldots$: To find a  co-image of the mapping $T\bmod
2^k$ one must solve a system of Boolean equations
$$
\begin{cases}
\chi_0+\varphi^T_0&=\alpha_0,\\
\chi_1+\varphi^T_1(\chi_1)&=\alpha_1,\\
\ldots\ldots\ldots\ldots\ldots\ldots\ldots&\ldots\ldots\\
\chi_k+\varphi_k^T(\chi_0,\ldots,\chi_{k-1})&=\alpha_k,
\end{cases}
$$
which has a unique solution given any  $\alpha_0,\ldots,\alpha_k\in\{0,1\}$.

Conversely, in let $i$ be the smallest number such that $\psi_i(\chi_0,\ldots,\chi_{i-1})=0$
for a certain set $\chi_0,\ldots,\chi_{i-1}$ of zeros and ones. Then 
$$T(\chi_0+\chi_1\cdot 2+\cdots\chi_2\cdot 2^{i-1}+0\cdot 2^i)\equiv T(\chi_0+\chi_1\cdot 2+\cdots\chi_2\cdot 2^{i-1}+1\cdot 2^i)\pmod{2^{i+1}}.$$
Thus, $T$ can not be measure-preserving in view of Theorem \ref{thm:erg-tran}.

Further, to prove the ergodicity part of the statement we note that $T$ is
transitive modulo 2 if and only if $\tau_0^T(\chi_0)=\chi_0\oplus 1$. In
case $T$ is transitive modulo $2^k$,
$$
\delta_i(T^{2^k})(x)=
\begin{cases}
\chi_i, &\text{if $i<k$;}\\
\chi_k\oplus \sigma, &\text{if $i=k$},
\end{cases}
$$
where $\sigma$ is a sum modulo 2 of all values of the Boolean function $\varphi^T_k$
at all points of $\mathbb B^k$; that is, $\sigma$ is the weight modulo 2
of the function $\varphi^T_k$. Clearly, to provide transitivity of the function
$T$ modulo $2^{k+1}$, (cf. Theorem \ref{thm:erg-tran} must be $\sigma=1$.
That is, weight of the function $\varphi_k^T$ must be odd.

The rest of the statement of the theorem is a well-known result in the theory
of Boolean functions; the proof is left to a reader.    
\end{proof}
\begin{note*}The bit-slice techniques
of Klimov and Shamir, which they introduced in 2002 in \cite{KlSh} is just a re-statement of the above stated folklore theorem \ref{ergBool}.
\end{note*}
This is how Theorem \ref{ergBool} works:

% Also, it worth noticing here that mappings $T$
% such that
% $\tau^T_i(\chi_0,\ldots,\chi_i)=\chi_i+\varphi^T_i(\chi_0,\ldots,\chi_{i-1})$
% for all
% $i=0,1,2\ldots$ are known in dynamical systems theory as skew shifts on (infinite
% dimensional) torus $\mathbb Z/2\times \mathbb Z/2\times\cdots$.

\begin{exmp}
\label{KlSh-3} 
With the use of \ref{ergBool} it is possible to give another proof of the
main result of \cite{KlSh}, namely, of Theorem 3:
{\it The mapping $f (x)=x +(x^2\vee C )$ over $n$-bit words is invertible
if and only if the least significant bit of $C$ is 1. For $n\ge 3$ it is a permutation
with a single cycle if and only if both the least significant bit and the third least
significant bit of $C$ are $1$.}

{\it Proof of theorem 3 of \cite{KlSh}.} 
% Note that $f$ is compatible, so bijectivity (resp.,
% transitivity) of $f$ modulo $2^n$ implies bijectivity (resp.,
% transitivity) of $f$ modulo $2^{n-1}$. 
Recall that for $x\in\mathbb Z_2$ and $i=0,1,2,\ldots$ 
%consider it base-$2$ expansion
%$x=x_0+x_1\cdot 2+x_2\cdot 4+x_i\cdot 2^i+\ldots$, i.e., 
we denote $\chi_i=\delta_i(x)\in\{0,1\}$; also we denote $c_i=\delta_i(C)$. 
We will calculate $\delta_i(x+(x^2\vee C))$ as an ANF %Boolean
%polynomial 
in Boolean variables $\chi_0,\chi_1,\ldots$ and we start with the following easy claims:
\begin{itemize}
\item $\delta_0(x^2)=\chi_0$,\ $\delta_1(x^2)=0$,\ $\delta_2(x^2)=\chi_0\chi_1\oplus\chi_1$,
\item $\delta_n(x^2)=\chi_{n-1}\chi_0\oplus\psi_{n}(\chi_0,\ldots,\chi_{n-2})$ for all
$n\ge 3$, where $\psi_{n}$ is a Boolean function in $n-1$
Boolean
variables $\chi_0,\ldots,\chi_{n-2}$.
\end{itemize}

The first of these claims could be easily verified by direct calculations. To prove
the second one represent $x=\bar x_{n-1}+2^{n-1}s_{n-1}$ (where we recall
$\bar x_{n-1}=x\bmod
2^{n-1}$) and
%a reduction
%of $x$ modulo $2^{n-1}$. 
%(i.e., the base-2 expansion of $\bar x_{n-1}$ is the
%first $n-1$ less significant bits of $x$). 
calculate $x^2=(\bar x_{n-1}+2^{n-1}s_{n-1})^2=
\bar x_{n-1}^2+2^{n}s_{n-1}\bar x_{n-1}+2^{2n-2}s_{n-1}^2=\bar x_{n-1}^2+2^n\chi_{n-1}\chi_0
\pmod{2^{n+1}}$ for $n\ge 3$ and note that $\bar x_{n-1}^2$ depends only on 
$\chi_0,\ldots,\chi_{n-2}$.

This gives
\begin{enumerate}
\item $\delta_0(x^2\vee C)=\chi_0\oplus c_0\oplus\chi_0c_0$
\item $\delta_1(x^2\vee C)=c_1$
\item $\delta_2(x^2\vee C)=\chi_0\chi_1\oplus\chi_1\oplus c_2\oplus c_2\chi_1\oplus c_2\chi_0\chi_1$
\item $\delta_n(x^2\vee C)=\chi_{n-1}\chi_0\oplus\psi_{n}\oplus c_n\oplus c_n\chi_{n-1}\chi_0\oplus
c_n\psi_{n}$
for $n\ge 3$
\end{enumerate} 
From here it follows  
that if $n\ge 3$, then $\delta_n(x^2\vee C)=
\lambda_n(\chi_0,\ldots,\chi_{n-1})$, 
%where 
%$\lambda_n=\lambda_n(\chi_0,\ldots,\chi_{n-1})$ is a Boolean polynomial in Boolean
%variables $\chi_0,\ldots,\chi_{n-1}$, 
and $\deg \lambda_n\le
n-1$, since $\psi_{n}$ depends only on, may be, $\chi_0,\ldots,\chi_{n-2}$.

Now successively calculate $\gamma_n=\delta_n(x+(x^2\vee C))$ for $n=0,1,2,\ldots$.
We have $\delta_0(x+(x^2\vee C))=c_0\oplus\chi_0c_0$ so necessarily $c_0=1$
since otherwise $f$ is not bijective modulo 2. Proceeding further with
$c_0=1$ we obtain $\delta_1(x+(x^2\vee C))=c_1\oplus\chi_0\oplus\chi_1$, since
$\chi_1$ is a carry. Then $\delta_2(x+(x^2\vee C))=(c_1\chi_0\oplus c_1\chi_1\oplus\chi_0\chi_1)\oplus
(\chi_0\chi_1\oplus\chi_1\oplus c_2\oplus c_2\chi_1\oplus c_2\chi_0\chi_1)\oplus\chi_2=
c_1\chi_0\oplus c_1\chi_1\oplus \chi_1\oplus c_2\oplus c_2\chi_1\oplus c_2\chi_0\chi_1\oplus
\chi_2$,
here $c_1\chi_0\oplus c_1\chi_1\oplus \chi_0\chi_1$ is a carry. From here in view of \ref{ergBool}
we immediately have $c_2=1$ since otherwise $f$ is not transitive
modulo 8. 
%(but bijectivity modulo 8 impose no restrictions on $c_2$ since conditions
%of the above mentioned criterion $\vartriangle$ are satisfied). 
%Now denoting 
%$\delta_n(x+(x^2\vee C))=\gamma_n$ 
Now for $n\ge 3$ 
one has $\gamma_n=\alpha_{n}+\lambda_n
\oplus\chi_n$, where $\alpha_n$ is a carry, and $\alpha_{n+1}=\alpha_n\lambda_n
\oplus\alpha_n\chi_n\oplus\lambda_n\chi_n$. But if $c_2=1$ then $\deg\alpha_3=\deg
(\mu\nu\oplus\chi_2\mu\oplus\chi_2\nu)=3$, where $\mu=c_1\chi_0\oplus c_1\chi_1\oplus
\chi_0\chi_1$, 
$\nu=(\chi_0\chi_1\oplus\chi_1\oplus c_2\oplus c_2\chi_1\oplus c_2\chi_0\chi_1)=
0$. This implies inductively in view of (4) above that 
$\deg\alpha_{n+1}=n+1$ and that $\gamma_{n+1}=\chi_{n+1}\oplus\xi_{n+1}(\chi_0,\ldots,\chi_{n})$,
$\deg\xi_{n+1}=n+1$. So the conditions of \ref{ergBool} are satisfied, thus
finishing the proof of theorem 3 of \cite{KlSh}.\qed
\end{exmp}

There are some more applications of Theorem \ref{ergBool}.
\begin{prop}
\label{compBool}
Let 
$F\colon\mathbb Z_2^{n+1}\rightarrow\mathbb Z_2$ be a compatible mapping
such that for all $z_1,\ldots,z_n\in\mathbb
Z_2$ the mapping $F(x,z_1,\ldots,z_n)\colon\mathbb Z_2\rightarrow \mathbb
Z_2$ is  measure preserving. Then $F(f(x),2g_1(x),\ldots,2g_n(x))$ preserves
measure for all compatible $g_1,\ldots,g_n\colon\mathbb Z_2\rightarrow \mathbb
Z_2$ and all compatible and measure
preserving $f\colon\mathbb Z_2\rightarrow \mathbb
Z_2$. Moreover, if 
%$f,g$ are compatible
$f$ is ergodic then $f(x+4g(x))$, $f(x\oplus (4g(x)))$, $f(x)+4g(x)$, and
$f(x)\oplus  (4g(x))$ are ergodic for any compatible $g\colon\mathbb Z_2\rightarrow \mathbb
Z_2$ \textup{(here $\oplus$ stands for $\XOR$)}.
\end{prop}
\begin{proof} Try to prove this yourself!
\end{proof} 
\begin{exmp}
\label{XOR}
With the use of \ref{compBool} it is possible to construct very fast generators
$x_{i+1}=f(x_i)\bmod 2^n$ that are transitive modulo $2^n$. 
For instance, take
%let the state transition function
%of the generator be 
$$f(x)=(\ldots((((x+c_0)\oplus d_0)+c_1)\oplus d_1)+\cdots +c_m)\oplus
d_m,$$
where $c_0\equiv 1\pmod 2$, and the rest of $c_i,d_i$ are 0 modulo 4.
By the way, this generator, looking somewhat `linear', is as a rule rather
`nonlinear': the corresponding polynomial over $\mathbb Q$ is of high degree.
The general case of these functions $f$ (for arbitrary $c_i, d_i$) 
was studied by the author's student Ludmila Kotomina: She proved that such
a function
is ergodic iff it is transitive modulo 4.
\end{exmp}

Yet another application of Theorem \ref{ergBool} are multivariate single
cycle $T$-functions. We already know that there are no such functions among
uniformly differentiable modulo 2 functions, see Theorem \ref{thm:noTmvar}.
However, the non-differentiable modulo 2 multivariate ergodic functions
on $\Z_2$ exist.

In 2004 Klimov
and Shamir introduced a multivariate $T$-function $H$ with a single cycle property.
%\onlySlide*{1}{%
The %\hypertarget{KlShMult}
{$m$-variate mapping} 
$$H\colon (\overrightarrow x_{0},\overrightarrow x_1,\ldots,\overrightarrow x_{m-1})
\mapsto(h_0,h_1,
\ldots,h_{m-1})$$
over $n$-bit words $\overrightarrow x_{0},\overrightarrow x_1,\ldots,\overrightarrow x_{m-1}$,
%which is 
defined by
\begin{multline*}
h_s%(x^0,\ldots,x^{m-1})
=\overrightarrow x_s\oplus((h(\overrightarrow x_0\wedge\cdots\wedge \overrightarrow
x_{m-1})\oplus \\
(\overrightarrow x_0\wedge\cdots\wedge \overrightarrow x_{m-1}))\wedge \overrightarrow
x_0\wedge\cdots\wedge \overrightarrow x_{s-1},
\end{multline*}
$s=0,1,\ldots,m-1$, has a single cycle property whenever $h$ is a univariate
$T$-function with a single cycle property. Here $\wedge$ stands for $\AND$,
bitwise logical `and' (a conjunction). 
We assume that a bitwise conjunction over
an empty set of indices is a string of all 1's.

Actually, this is just a %\hyperlink{trick}
{trick}: The $m$-variate
mapping %\hyperlink{KlShMult}
{$H$} on $n$-bit words  is 
%\textcolor{yellow}
{a multivariate representation of a univariate $T$-function over $mn$-bit
words.} Indeed, given %$x=(\chi_0,\chi_1,\chi_2, \ldots)$ and
a %\hypertarget{trick}
{univariate $T$-function $F$},
{
$$x=(\ldots,\chi_2,\chi_1,\chi_0)\stackrel{F}{\mapsto} (\ldots;\psi_2(\chi_0,\chi_1,\chi_2);\psi_1(\chi_0,\chi_1);\psi_0(\chi_0)),$$ }
arrange this mapping in columns of height $m$, this way:
{
\begin{align*}
&\ldots\chi_{2m}&{}&\chi_{m}&{}&\chi_{0}&{}&\stackrel{f_{0}}{\mapsto}&{}&\ldots\psi_{2m}(x)&{}&\psi_{m}(x)&{}&\psi_{0}(x)\\
&\ldots\chi_{2m+1}&{}&\chi_{m+1}&{}&\chi_{1}&{}&\stackrel{f_1}{\mapsto}&{}&\ldots\psi_{2m+1}(x)&{}&\psi_{m+1}(x)&{}&\psi_{1}(x)\\
&\ldots&{}&\ldots&{}&\ldots&{}&\ldots\\
&\ldots\chi_{3m-1}&{}&\chi_{2m-1}&{}&\chi_{m-1}&{}&\stackrel{f_{m-1}}{\mapsto}&{}&\ldots\psi_{3m-1}(x)&{}&\psi_{2m-1}(x)&{}&\psi_{m-1}(x)
\end{align*}
}
Now just assume the left-hand rows are new variables: 
{ $$\overrightarrow x_{j}=(\ldots,\chi_{2m+j},\chi_{m+j},\chi_{j}), \qquad(j=0,1,\ldots,m-1).$$} 
Obviously,
%that %\textcolor{blue}
{the $m$-variate mapping $\mathbf F=(f_0,f_1,\ldots,f_{m-1})$
has a single cycle property iff a univariate mapping $F$ has a single %cycle
property}.

Consider the simplest example: $F(x)=1+x$. We have 
%the
%$j$\textsuperscript{th} bit $\delta_j(H(X))$ of a canonical $2$-adic representation 
% 
%of $H(X)$
%\footnote{i.e., the $j$\textsuperscript{th} bit of infinite binary string
%that represents
%$H(X)$}
% \footnote
% {We recall that $\delta_j(u)$ is the $j$\textsuperscript{th} bit
% in base-$2$ expansion of $u\in\Z_2$, $j=0,1,2,\ldots$, and the space $\Z_2$ 
% of all $2$-adic
% integers could be thought of as a set of all infinite sequences
% of $0$'s and $1$'s.} 
%could be expressed as 
{
$$\delta_j(F(x))\equiv\delta_j(x)+\prod _{s=0}^{j-1}\delta_s(x)\pmod 2$$
}
(we assume the product over the empty set is $1$); then the $m$-variate
representation $\mathbf F=(f_0,f_1,\ldots,f_{m-1})$ of this mapping is
%the conjugate $m$-variate
%mapping is given by
{
\begin{multline*}
f_k(\overrightarrow x_0,\ldots,\overrightarrow x_{m-1})= \overrightarrow
x_k\oplus
\bigg(\bigg(\bigwedge_{s=0}^{k-1} \overrightarrow x_s\bigg)
\wedge
\bigg(\bigwedge_{r=0}^{m-1}
((\overrightarrow x_r+1)\oplus \overrightarrow x_r)\bigg)\bigg)=\\
\overrightarrow x_k\oplus
\bigg(\bigg(\bigwedge_{s=0}^{k-1} \overrightarrow x_s\bigg)
\wedge
\bigg(\bigg(
\bigg(\bigwedge_{r=0}^{m-1}\overrightarrow x_r\bigg)+1\bigg)\oplus 
\bigg(\bigwedge_{r=0}^{m-1}\overrightarrow x_r\bigg)\bigg)\bigg).
\end{multline*}
}

With the use of this trick and with Theorem \ref{ergBool} the following multivariate
ergodic $T$-functions could be constructed:
\begin{prop}[\cite{anashin4a}]
\label{cor:WP:mult}
%\textup{(Anashin, 2004)}
Let $t,j\in\{0,1,\ldots, m-1\}$, 
%$j=0,1,\ldots, m-1$, 
let all 
$f^{(t)}_j$ (resp., $g^{(t)}_j$) be univariate ergodic (resp, measure-preserving)
%modulo $2^n$
%$T$-functions. 
compatible mappings from $\Z_2$ onto $\Z_2$.
%\:\Z_2\>\Z_2$ be ergodic 
%$(s\in\{0,1,\ldots, m-1\}$, $j=0,1,\ldots, m-1)$ be {\textup
%(}univariate{\textup)}
%ergodic functions, let
%and let all
%$g^{(j)}_t\:\Z_2\>\Z_2$ 
%$(s\in\{0,1,\ldots, j-1\}$ , $j=1,2,\ldots, m-1)$ be 
%{\textup(}univariate{\textup)}
%measure-preserving functions.
%preserve measure.
%, and let $\star\in\{+,\oplus\}$. 
Then the mapping
{
$\mathbf F(\mathbf x)=(f_{0}(\mathbf x),\ldots,f_{m-1}(\mathbf x))$ %where %пр-ва $\Z_2^{m}$,
%\:(\Z_2)^{(m)}\>(\Z_2)^{(m)},$$
%of $(\Z_2)^{(m)}$ onto $(\Z_2)^{(m)}$ such that
%which is
%заданное следующими равенствами, 
\begin{gather*}
f_{0}(\mathbf x)=
%x_0\mapsto 
\overrightarrow x_{0}\boxplus 
\bigg(\bigwedge_{r=0}^{m-1}
(f^{(r)}_0(\overrightarrow x_{r})\oplus \overrightarrow x_{r})\bigg)
%((f_0^0(x_0)\oplus x_0)\wedge\cdots\wedge
%(f_0^{m-1}(x_{m-1})\oplus x_{m-1}))
;\\
f_{1}(\mathbf x)=
%x_1\mapsto 
\overrightarrow x_{1}\boxplus \bigg(g^{(0)}_1(\overrightarrow x_{0})\wedge
\bigg(\bigwedge_{r=0}^{m-1}
(f^{(r)}_1(\overrightarrow x_{r})\oplus \overrightarrow x_{r})\bigg)\bigg)
%(f_1^0(x_0)\oplus x_0)\wedge\cdots\wedge
%(f_1^{m-1}(x_{m-1})\oplus x_{m-1}))
;\\
\ldots \ldots \ldots\ldots\ldots\ldots\ldots\ldots\ldots
\ldots\ldots\ldots\ldots\ldots\ldots\ldots\ldots\ldots\ldots\ldots\ldots\ldots\\
f_{m-1}(\mathbf x)%=\hspace{8cm}%\\
%x_{m-1}\mapsto
=\overrightarrow x_{m-1}\boxplus \bigg(\bigg(\bigwedge_{t=0}^{m-2} g^{(t)}_{m-1}(\overrightarrow
x_{t})\bigg)
\wedge
%\cdots\wedge
%g_{m-1}^{m-1}(x_{m-1})
\bigg(\bigwedge_{r=0}^{m-1}
(f^{(r)}_{m-1}(\overrightarrow x_{r})\oplus \overrightarrow x_{r})\bigg)\bigg),
%\wedge\cdots\wedge
%(f_{m-1}^{m-1}(x_{m-1})\oplus x_{m-1})),
\end{gather*}
where $\mathbf x=(\overrightarrow x_{0},\ldots,\overrightarrow x_{m-1})$, %\in\Z_2^{s}$, 
$\boxplus \in\{+,\oplus\}$,
} 
%is an $m$-variate $T$-function with 
%has a single
%cycle property.%\textup{(thus, \textcolor{yellow}{$\mathbf F$ produces %a cycle of length $2^{mn}$ on
%$n$-bit words}).}
is a compatible and ergodic mapping of $\mathbb Z_2^m$ onto $\mathbb Z_2^m$.
\end{prop}

\section{Wreath products of PRNGs}
In the preceding section we have developed some tools that enable us to construct
algorithms based on standard instructions of an $n$-bit word processor that
produce strictly uniformly distributed sequences of period length $2^n$.

To judge whether these sequences could be of use for stream encryption we
must study their properties that are crucial for stream ciphers. One of these
properties is long period. But is the period of are sequences long enough?
Not yet! In case $n=32$, which is a standard for most contemporary processors, we obtain a period of length $2^{32}$, which is too
small to satisfy contemporary safety conditions: At least some $2^{80}$ is
needed. Thus, we must make the period longer {\sl leaving the sequence uniformly
distributed}. In this section we consider corresponding techniques.
\subsection{What is wreath product} We start with a formal definition:
\begin{defn}
Given a mapping $U\colon Z\rightarrow Z$, and a set of mappings
$\mathcal V=\{(V_z\colon X\rightarrow X)\colon z\in Z\}$, a {{\it  wreath product}} (or, a {{\it skew product}} 
or, a {{\it skew 
shift}})  is a mapping 
$$U\rightthreetimes \mathcal V\colon(z,x)\mapsto (U(z),V_z(x))$$ 
of the Cartesian product $Z\times X$ into itself.
\end{defn}   
In other words, the wreath product is a bivariate mapping where the first
coordinate is a function of the variable $z$ {\sl only}, and the second coordinate
is a bivariate function of $z$ and $x$.

Most probably, you are already familiar with examples of wreath products; recall Feistel network:
The mapping it is based on is 
$(z,x)\mapsto (z,z\oplus f(x))$, where $z,x\in \mathbb B^n$, $f\colon \mathbb
B^n\rightarrow\mathbb B^n$, which is obviously a wreath product of $U(z)=z$ with $\mathcal V=\{V_z(x)=z\oplus f(x)\colon z\in \mathbb B^n\}$.
\begin{quote}
{\sl Obviously, the wreath product $U\rightthreetimes \mathcal
V$ is bijective whenever both $U$ and all $V_z$ are
bijective.}
\end{quote}

Some terminology notes: In automata theory (and in algebra) they used to
speak of wreath products, whereas in dynamical systems (and in ergodic theory) theory they prefer
the term skew product, or skew shift. Recall that ordinary %\hyperlink{PRNG}
{PRNG} corresponds to an {\it autonomous} dynamical system. %In the 
%In counter-dependent PRNG 
%clock
%state update  functions $f_i$ (and/or clock output functions $G_i$) could reside
%in memory, or they could be produced on-the-fly.
%\fromSlide{2}{% 

%\bigskip

This is  a {\it non-autonomous} dynamical system, which is a counterpart of a counter-dependent PRNG \ref{eq:cntdpd} in dynamics: %
%}
%
%is, 
%\onlySlide*{2}{%
%\bigskip
A non-autonomous dynamical system is a dynamical system driven by another dynamical system, and skew products are used to combine two dynamical systems
into a new one. 

Note that a $T$-function is a composition of wreath products: Let $F$ be
a $T$-function,
%\begin{boxitpara}{box 0.9 setgray fill}
$$(\chi_0,\chi_1,\chi_2, \ldots)\stackrel{F}{\mapsto} (\psi_0(\chi_0);\psi_1(\chi_0,\chi_1);\psi_2(\chi_0,\chi_1,\chi_2);\ldots),$$
%\end{boxitpara}
then
%\begin{boxitpara}{box 0.9 setgray fill}
$$
\begin{array}{rcl}
\chi_0&\mapsto& \psi_0(\chi_0)\\
(\chi_0,\chi_1)&\mapsto& (\psi_0(\chi_0),\psi_1(\chi_0,\chi_1))\\
((\chi_0,\chi_1),\chi_2)&\mapsto& ((\psi_0(\chi_0),\psi_1(\chi_0,\chi_1)),\psi_2(\chi_0,\chi_1,\chi_2))\\
\ldots\ldots\ldots\ldots &\ldots& \ldots\ldots\ldots\ldots\ldots\ldots\ldots\ldots\ldots\ldots\ldots\ldots\\
\end{array}
$$

Now we re-state the above definition for the case of wreath products of automata:
\begin{defn}
\label{def:WP}
Let $\mathfrak A_j=\langle N,M,f_j,F_j\rangle$  be a family of 
automata with the same state set $N$ and the same
output alphabet $M$ indexed by elements of
a non-empty (possibly, countably infinite) set $J$ 
(members of the family need not be necessarily pairwise distinct). 
Let $T\colon J\rightarrow J$ be an arbitrary mapping. A {\it wreath product}
% $\mathfrak A_j\Wr_{j\in J}T$
 of the family $\{\mathfrak A_j\}$ of automata
with respect to the mapping $T$ is an automaton with the state set $N\times J$, state
transition function $\breve f(j,z)=(f_j(z),T(j))$ and output function 
$\breve F(j,z)=F_j(z)$.  %\footnote{cf. {\it skew shift} in ergodic theory;
%It is bijective whenever all $f_j$ and $T$ are bijective; 
%cf. round function in the Feistel network. We are using a term from group %theory.}. %; it is denoted as $\breve f=f_j\Wr_{j\in J}T$.
We call $f_j$ (resp., $F_j$) {\it clock} state update (resp.,
output) functions.
\end{defn}
Obviously, the state transition function $\breve f(j,z)=(f_j(z),T(j))$
is a {wreath product of a family of mappings $\{f_j\colon j\in
J\}$ with respect to the mapping $T$}

It worth notice here  that if $J=\mathbb N_0$ and $F_i$ does not depend on $i$, this construction 
gives us a number of examples of counter-dependent generators 
in the sense of \cite[Definition 2.4]{ShTs}, where
the notion of a counter-dependent generator was 
originally introduced. %in \cite{ShTs}. 
However, we use this notion in a broader  sense
in comparison with that of \cite{ShTs}: In our counter-dependent
generators
not only the state
transition function, but also the output function depends on $i$. Moreover,
in \cite{ShTs} only a special case of counter-dependent generators
is studied; namely, counter-assisted generators and their cascaded and two-step
modifications. A state transition function of a counter-assisted generator  is
of the form $f_i(x)=i\star
h(x)$, where $\star$ is a binary quasigroup operation (in particular, group
operation, e.g., $+$ or
$\XOR$), and $h(x)$ does not depend on $i$. 
An output function of a counter-assisted generator does not depend
on $i$ either.
\subsection{Constructions}
\label{subsec:Constr}
In this subsection we introduce a method %several constructions that enable one to
%assemble 
to construct counter dependent pseudorandom 
generators 
%and stream ciphers 
out of %`building blocks' based on 
ergodic and measure-preserving mappings. The
method guarantees that
output sequences of these generators are always strictly uniformly
distributed. Actually, all these constructions are wreath
products of automata in the sense of \ref{def:WP}; the following results give us conditions these automata
should satisfy to produce a uniformly distributed output sequence. Our main technical tool is the following theorem, which actually could be considered
as a generalization of Theorem \ref{ergBool}:
%Other probabilistic and cryptographic properties of these generators are discussed
%in further sections.
\begin{thm}[\cite{anashin4}]
\label{thm:WP}
Let $\mathcal G=g_0,\ldots,g_{m-1}$ be a finite sequence of 
compatible measure preserving
mappings of $\mathbb Z_2$ onto itself such that
\begin{enumerate}
\item the sequence $\{(g_{i\bmod m}(0))\bmod 2\colon i=0,1,2,\ldots\}$ is 
purely periodic, its shortest period is of length $m$;
\item $\sum_{i=0}^{m-1}g_i(0)\equiv 1\pmod 2$;
\item $\sum_{j=0}^{m-1}\sum_{z=0}^{2^k-1}g_j(z)\equiv 2^{k}\pmod {2^{k+1}}$
for all $k=1,2,\ldots$ .
\end{enumerate}
Then the recurrence sequence $\mathcal Z$ defined by the relation $x_{i+1}=g_{i\bmod
m}(x_i)$ is strictly uniformly distributed modulo $2^n$ for all $n=1,2,\ldots:$
That is, modulo each $2^n$ the sequence $\mathcal Z$  is purely periodic, its
shortest period is 
of
length 
$2^nm$, and each element of $\mathbb Z/2^n\Z$ occurs at the period
exactly $m$ times. 
\end{thm}
\begin{note*}
In view of \ref{ergBool}
% a compatible mapping $g_i\colon\mathbb Z_2\rightarrow\mathbb Z_2$ preserves
% measure iff 
% $$\delta_k(g_i(x))\equiv \chi_k+\varphi_k^i(\chi_0,\ldots,\chi_{k-1})\pmod
% 2,$$
% where $\chi_s=\delta_s(x)$ $(s=0,1,2,\ldots)$, 
condition  (3)
of theorem \ref{thm:WP} could be replaced by the equivalent condition
%$$\sum_{j=0}^{m-1}\wt\varphi_k^j\equiv 1\pmod 2 \qquad (k=1,2,\ldots),$$}
%where $\wt\varphi_k^j$ is a weight of the Boolean polynomial $\varphi_k^j$
% in the variables $\chi_0,\ldots,\chi_{k-1}$. In turn, since for every Boolean
% polynomial $\varphi$ in variables $\chi_0,\ldots,\chi_{k-1}$ holds $\wt\varphi\equiv
% \Coef_{0,\ldots,k-1}(\varphi)\pmod 2$, where $\Coef_{0,\ldots,k-1}(\varphi)$
% stands for a coefficient of the monomial $\chi_0\cdots\chi_{k-1}$ in the
% Boolean polynomial $\varphi$, the {\it latter condition could be also replaced
% by
$$\sum_{j=0}^{m-1}\Coef_{0,\ldots,k-1}(\varphi_k^j)\equiv 1\pmod 2 \qquad (k=1,2,\ldots),$$
where $\Coef_{0,\ldots,k-1}(\varphi)$
is a coefficient of the monomial $\chi_0\cdots\chi_{k-1}$ in ANF $\varphi$.
% or by
% $$\sum_{j=0}^{m-1}\bigg\lfloor\frac{\deg\varphi_k^j}{k}\bigg\rfloor\equiv 1\pmod {2} \qquad (k=1,2,\ldots).$$} 
\end{note*}
It turns out that the sequence $\mathcal Z$ of \ref{thm:WP} is just the sequence $\mathcal
Y$ of the following
%To prove theorem \ref{thm:WP} we need the following lemma, which is of
%interest also:
\begin{lem}[\cite{anashin4}]
\label{le:WP-odd} 
Let $c_0,\ldots,c_{m-1}$ be a finite sequence of $2$-adic integers, and
let $g_0,\ldots,g_{m-1}$ be a finite sequence of compatible
mappings of $\mathbb Z_2$ onto itself such that 
%\begin{itemize}
\renewcommand{\theenumi}{\roman{enumi}}
\begin{enumerate}
\item $g_j(x)\equiv x+c_j\pmod 2$ for $j=0,1,\ldots,m-1$, 
%\item the sequence $\{d_j=(c_j+1)\pmod 2\}$ satisfy conditions of proposition \ref{WP-odd},
\item $\sum_{j=0}^{m-1}c_j\equiv
1\pmod 2$, 
\item the sequence 
$\{c_{i\bmod m}\bmod 2\colon i=0,1,2,\ldots\}$ is purely periodic, its shortest period is
of length $m$,
\item $\delta_k(g_j(z))\equiv \zeta_k+\varphi_k^j(\zeta_0,\ldots,\zeta_{k-1})\pmod
2$, $k=1,2,\ldots$,
where $\zeta_r=\delta_r(z)$, $r=0,1,2,\ldots$, 
\item for each $k=1,2,\ldots$ an odd number of ANFs %Boolean polynomials 
$\varphi_k^j$
%(\zeta_0,\ldots,\zeta_{k-1})$ 
in Boolean variables $\zeta_0,\ldots,\zeta_{k-1}$ are of odd weight.
\end{enumerate}
\renewcommand{\theenumi}{\arabic{enumi}}
%\end{itemize}
Then the recurrence sequence $\mathcal Y=\{x_i\in\mathbb Z_2\}$ defined by the relation 
$x_{i+1}=g_{i\bmod m}(x_i)$ is strictly uniformly distributed: 
It is purely periodic modulo $2^k$ for all $k=1,2,\ldots$; its  shortest period is
of length $2^km$; each element of $\mathbb Z/2^k\Z$ occurs at
the period exactly $m$ times. 
Moreover,
\begin{enumerate}
%\item the exact period length of the sequence $\mathcal Y_k$ is $2^km$
%(see definition \ref{def:strict}),
\item  %{\rm (not necessarily exact, see definition \ref{def:strict})} 
the sequence 
$\mathcal D_s=\{\delta_s(x_i)\colon i=0,1,2,\ldots\}$
% $(s=0,1,\ldots, k-1)$ 
is purely periodic; it has a period of length $2^{s+1}m$,
\item $\delta_s(x_{i+2^{s}m})\equiv\delta_s(x_{i})+1\pmod
2$ for all $s=0,1,\ldots, k-1$, $i=0,1,2,\ldots$, 
\item for each $t=1,2,\ldots,k$ and each $r=0,1,2,\ldots$ the sequence 
$$x_r\bmod 2^t,x_{r+m}\bmod 2^t,x_{r+2m}\bmod 2^t,\ldots$$
is purely periodic, its shortest period is of length $2^t$, each element
of $\mathbb Z/2^t\Z$ occurs at the period exactly once.
\end{enumerate} 
\end{lem}
% \begin{note*}
% In view of \ref{ergBool} the conditions of the lemma imply that all the
% mappings $g_j$ preserve measure.
% \end{note*}
% 
\begin{note}
\label{note:Bool} 
Assuming $m=1$ in \ref{thm:WP} one obtains ergodicity criterion
\ref{ergBool}.
%so this theorem could be considered as a generalization
%of that criterion.
\end{note}

\begin{cor}[\cite{anashin4}]
\label{cor:WP}
Let  a finite sequence  of mappings $\{g_0,\ldots,g_{m-1}\}$ of $\mathbb
Z_2$ into itself satisfy conditions
of theorem \ref{thm:WP}, and let $\{F_0,\ldots,F_{m-1}\}$ be an arbitrary
finite sequence of balanced {\rm (}and not necessarily compatible{\rm
)} mappings of $\mathbb Z/2^n\Z$ $(n\ge 1)$ onto $\mathbb
Z/2^k\Z$, $1\le k\le n$. Then the sequence
$\mathcal F=\{F_{i\bmod m}(x_i)\colon i=0,1,2\ldots\}$, where $x_{i+1}=g_{i\bmod m}(x_i)\bmod
2^n$, is strictly uniformly distributed over $\mathbb Z/2^k\Z$: It is purely
periodic with a period of length $2^nm$, and each element of $\mathbb Z/2^k\Z$
occurs at the period  exactly $2^{n-k}m$ times.
\end{cor}

Theorem \ref{thm:WP} and lemma \ref{le:WP-odd} together with corollary \ref{cor:WP} 
enables one to construct a counter-dependent generator out of the following components:
\begin{itemize}
\item A sequence $c_0,\ldots,c_{m-1}$ of integers, which we call a \textit{control
sequence}.
\item A sequence $h_0,\ldots,h_{m-1}$ of compatible mappings, which is
used to form a sequence of clock state update functions $g_i$ %(see e.g.
%examples \ref{WP-even}).
\item A sequence $H_0,\ldots,H_{m-1}$ of compatible mappings to produce
clock output functions $F_i$ %(see e.g. proposition \ref{prop:reverse}).
\end{itemize}
Note that ergodic functions that are needed %to meet conditions of \ref{prop:reverse}
%or \ref{WP-even} (3) 
could be produced out of compatible ones with the use of \ref{Delta}
or \ref{compBool}.
A control sequence could be produced by an external generator (which
in turn could be a generator of the kind considered in this course), or it
could be just a queue the state update and output functions are called
from a look-up table.
The functions $h_i$ and/or $H_i$  could be either precomputed to arrange
that look-up table, 
or they could be produced on-the-fly in a form that is determined
by a control sequence. This form may also look  `crazy', e.g., %; for instance, one
%may take
\begin{equation}
\label{eq:crazy}
h_i(x)=(\cdots((u_0(\delta_0(c_i))\bigcirc _{\delta_1(c_i),
\delta_2(c_i)}u_1(\delta_3(c_i)))\bigcirc _{\delta_4(c_i),\delta_5(c_i)}u_2(\delta_6(c_i)))\cdots,
\end{equation}
where $u_j(0)=x$, the variable, and $u_j(1)$ is a constant (which is determined
by $c_i$, or is read from a precomputed look-up table, etc.), while (say) $\bigcirc _{0,0}=+$,
an integer addition, $\bigcirc _{1,0}=\cdot$, an integer multiplication,
$\bigcirc _{0,1}=\XOR$, $\bigcirc _{1,1}=\AND$.  This is absolutely no matter
what these $h_i$ and $H_i$ look like or how they are obtained, 
{\slshape the above stated results give a general method to combine all the data together
to produce a uniformly distributed output sequence of a maximum period
length}. %Illustrative examples follow.
%From theorem \ref{thm:WP} one could deduce the following 

\begin{figure}
\begin{quote}\psset{unit=0.4cm}
 \begin{pspicture}(-1,5)(24,15)
%\pscustom[fillstyle=slopes,
%slopecolors=0 1 1 .9  17 .5 1 .5  23 0 0.5 0.5  3]{
%\psccurve(0,2.5)(12,3.5)(20,4)(23,2)(17,2.5)}
%\psaxes(0,0)(0,12)(25,0)
%\pscircle[linewidth=1pt](12,14){.5}
\psframe[linewidth=1pt](11.5,13.5)(12.5,14.5)
\pscircle[linewidth=2pt](12,10){1}
\pscircle[linewidth=2pt](4,10){1}
\pscircle[linewidth=2pt](9.9,14){1}
\psline{->}(8.4,14)(8.4,10)
\psline{<->}(11,10)(5,10)
\psline{->}(4,11)(4,13)
%\psline[linestyle=dotted](0,14)(-0.5,14)
  %\psline[linestyle=dotted](-0.5,14)(-0.5,3.5)
  %\psline[linestyle=dotted]{->}(-0.5,3.5)(10,3.5)
%\psline[linestyle=dotted](0,14)(-0.5,14)%Were dotted
  %\psline[linestyle=dotted](-0.5,14)(-0.5,1.5)%Were dotted
  %\psline[linestyle=dotted]{->}(-0.5,1.5)(11.5,1.5)%Were dotted
  \psline{->}(12,11)(12,13.5)
  \psline{->}(24,7)(16,7)
  \psline{->}(8,14)(9,14)
  \psline{->}(10.8,14)(11.5,14)
  \psline(24,14)(24,7)
  \psline(12.5,14)(24,14)
  \psline{->}(12,8)(12,9)
  %\psline{->}(12,2)(12,1)
  \psline{<-}(12,5)(12,6)
  \psframe[linewidth=2pt](0,13)(8,15)
  \psframe[linewidth=2pt](8,6)(16,8)
 % \pspolygon[linestyle=dotted,linewidth=2pt](8,5)(16,5)(12,2)
  \uput{0}[180](12.4,7){$x_i$}
  \uput{0}[90](12.1,9.6){$h_{y_i}$}
  \uput{0}[90](9.9,13.7){$W$}
  \uput{0}[90](4,9.7){$U$}
  \uput{0}[90](12,13.7){+}
  \uput{1}[90](12,3){$\mathcal Z$}
  \uput{1}[0](12.7,12.5){%\scriptsize 
  $x_{i+1}%=V(z_i,x_i)
  =c_i+h_{y_i}(x_i)$}
  \uput{1}[0](12.7,11.5){%\scriptsize 
  $c_i%x_{i+1}%=V(z_i,x_i)
  =W(y_i)%+h_{z_i}(x_i)
  $}
  %\uput{1}[0](-1.4,0.3){\scriptsize{$L$ --- \hypertarget{parsec}{T-функция} периода $2^s$; %(c)=2\cdot c\oplus \overline u\cdot\delta_{n-1}(c)$
 %$v_i$ --- произвольные Т-функции;
  %$\overline u$ вектор к-тов примитивно полинома $u$
  %}}
  %\uput{1}[0](-1.4,15.7){\scriptsize{\blue Text %Принципиальная криптосхема шифратора Парсек.
   %\hyperlink{abc}{{\bf Ср. с ABC.}} 
  %}}
  %}
  %\uput{1}[0](-1.3,-0.8)%(12.5,10)
  %{{\scriptsize$h_i(x_i)=x_i+4\cdot v_i(x_i)$}}
  \uput{1}[0](-3,11.5){%\scriptsize
  $y_{i+1}=U(y_i)$}
  %\uput{1}[0](8.5,14.7){$c_i$}
  \uput{1}[0](2.6,14){$y_i$}
  %\uput{1}[0](15.6,7.6){\scriptsize{\sf переход}}
 % \fromSlide{3}{\uput{1}[0](16.6,12.6){\red driven DS}}
 % \fromSlide{2}{\uput{1}[0](3.2,11.6){\red driving DS}}
 % \fromSlide{2}{\uput{1}[0](-2,7.6){\red driving DS}}
  %\uput{1}[0](12,1.5){\scriptsize{$y_i=G_i(x_i)$}}
  %\uput{1}[180](12,1.5){\scriptsize{\sf выход}}
%\psgrid(0,0)(-1,-1)(3,2)
 \end{pspicture}
\end{quote}
\caption{Wreath product basic circuit for Examples \ref{WP-even}.}%, (2)--(4).} 
\label{fig:wr}
\end{figure}
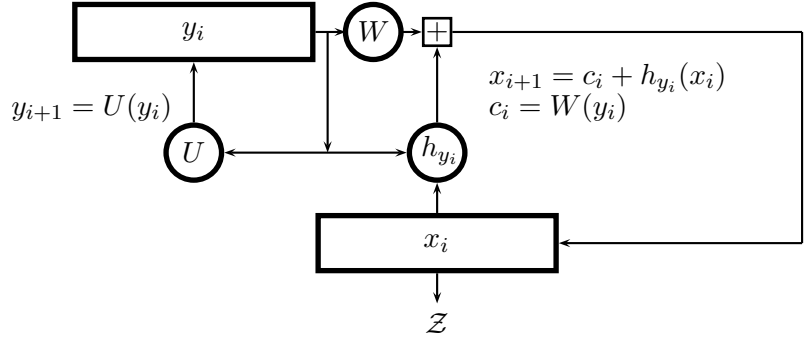

\begin{exmps}[\cite{anashin4}]
\label{WP-even}A basic circuit illustrating these example wreath products is given at Figure
\ref{fig:wr}.
%These are obtained with the use of \ref{le:WP-odd}, \ref{ergBool}, \ref{compBool},
%and \eqref{eq:sumBool}.

\begin{enumerate}
% \item A control sequence could be produced by the generator $${\mathfrak A}=\langle \mathbb
% Z/2^s, \mathbb Z/2^s,f,F,u_0\rangle$$ (see Section \ref{Prelm}) with ergodic
% state update function $f$ and measure-preserving
% output function $F$. Then length of the shortest period of the control sequence is $m=2^s$,
% see \ref{prop:Auto}. Take $m$ arbitrary ergodic functions $h_0,\ldots,h_{m-1}$ and arbitrary
% odd $k\in\{0,1,\ldots,m-1\}$, and 
% put
% $\breve g_0(x)=x\oplus(x+1)\oplus h_0(x),\ldots,\breve g_{k-1}=x\oplus(x+1)\oplus h_{k-1}(x)$, 
% $\breve g_k=h_k,\ldots,\breve g_{m-1}=h_{m-1}$, $g_i=\breve g_{c_i\bmod
% m}$ for $i=0,1,2,\ldots$. In other words, in this case the control sequence
% just define the queue the functions $\breve g_j$ are called,
% thus producing the output sequence 
% $$x_0,x_1=\breve g_{c_0}(x_0)\bmod 2^n, x_2=\breve g_{c_1}(x_1)\bmod2^n,
% \ldots$$
% Obviously, in this example a control sequence could be an arbitrary permutation
% of $0,1,\ldots,2^s-1$, and not necessarily an output of the generator $\mathfrak
% A$.
% %In view of \ref{Delta} and \ref{ergBool}
%it follows then that the conditions of lemma \ref{le:WP-odd} are satisfied.
\item
%$T\colon\mathbb Z/2^m\rightarrow\mathbb Z/2^m$, 
%$m\ge 1$, 
%be an arbitrary permutation
%with a single cycle, 
Let 
$c_0,\ldots,c_{m-1}$ be an arbitrary 
%finite 
sequence
of length $m=2^s$ %(i.e.,  $c_0,\ldots,c_{m-1}$ are not necessarily pairwise
%distinct)
, and  
%$2$-adic 
%integers, 
%such that $\sum_{j=0}^{2^m-1}c_j\equiv 1\pmod 2$,
let $\hat h_0,\ldots,\hat h_{m-1}$ be arbitrary compatible %and ergodic
mappings.
%of $\mathbb Z_2$ onto itself. 
%{\rm ($f_j$
%are not necessarily pairwise distinct)}. 
For $0\le j\le m-1$ put $h_j(x) =1+x+4\cdot \hat h_j(x)$ and let $g_j(x)=c_j+h_j(x)$. %{\rm (}
% If, additionally, $h_j$
% preserves measure, 
% {one may also put $g_j(x)=(c_j+x)\oplus(2\cdot h_j(x))$.}
% %or $g_j(x)=c_j+x+2\cdot h_j(x)$
% %{\rm)}. 
These mappings
$g_j$ satisfy
conditions of theorem \ref{thm:WP}
%$j=0,1,2,\ldots,2^m-1$. 
% Then the wreath product
% $H_j\Wr_{j=0}^{2^m-1}T$ defines a bijective mapping 
% $W\colon\mathbb Z_2\twoheadrightarrow \mathbb Z_2$ 
% %according to the folowing rule: 
% $$W(x)=T(x\bmod{2^m})+2^m\cdot H_{x\bmod{2^m}}
% \bigg(\Big\lfloor\frac{x}{2^m}\Big\rfloor\bigg);$$
% this mapping is asypmtotically compatible and asymptotically ergodic
% {\rm (i.e., $a\equiv b\pmod{2^k}\Rightarrow W(a)\equiv W(b)\pmod{2^k}$ and
% $W$ is transitive modulo $2^k$ for all sufficiently large $k$; in fact,
% for all $k>m$, see \cite{me-conf, me-1, me-2} for definitions)} if and only if
% $\sum_{j=0}^{2^m-1}c_j\equiv 1\pmod 2$.
% % That is, $a\equiv b\pmod{2^k}\Rightarrow W(a)\equiv W(b)\pmod{2^k}$ and
% % $W$ is transitive modulo $2^k$ for all sufficiently large $k$ {\rm(}in fact,
% % for all $k>m${\rm)}. 
%Then the
%In other words, 
%every 
%recurrence sequence $\mathcal U_n=\{x_i\}$ defined by the relation 
%$$x_{i+1}=H_{i\bmod{2^m}}(x_i)\bmod{2^n}$$
%{\rm (}$n=1,2,3,\ldots${\rm )} 
%is  strictly uniformly distributed sequence
%over $\mathbb Z/2^n$  of period length exactly $2^{n+m}$
%{\rm (i.e., with every element of $\mathbb Z/2^n$
%occurring at the period exactly $2^m$ times)} 
if and only if
$\sum_{j=0}^{2^m-1}c_j\equiv 1\pmod 2$.
%\newpage 
\item
For $m>1$ odd let $\{h_0,\ldots,h_{m-1}\}$ be a finite sequence of compatible 
and ergodic mappings; 
%of $\mathbb Z_2$ onto itself, and let
%$T\colon\mathbb Z/m\rightarrow\mathbb Z/m$, $m$ odd, 
%be an arbitrary permutation
%with a single cycle, 
%let further 
let $c_0,\ldots,c_{m-1}$  be a finite sequence of 
%$2$-adic 
integers such that 
\begin{itemize}
\item $\sum_{j=0}^{m-1}c_j\equiv
0\pmod 2$, and 
\item the sequence 
$\{c_{i\bmod m}\bmod 2\colon i=0,1,2,\ldots\}$ is purely periodic 
with the shortest period of length $m$.
\end{itemize}
%let $\{c_0,\ldots,c_{2^m-1}\}$ be a finite sequence
%of $2$-adic integers, 
%such that $\sum_{j=0}^{2^m-1}c_j\equiv 1\pmod 2$,. 
%{\rm ($f_j$
%are not necessarily pairwise distinct)}. 
Put $g_j(x)=c_j\oplus h_j(x)$ {\rm (}respectively, $g_j(x)=c_j+h_j(x)${\rm)}.
Then $g_j$ satisfy conditions of \ref{thm:WP}.
%\item The conditions of (2) are satisfied whenever 
\item The conditions of (2) are satisfied in case $m=2^s-1$ and $c_0,\ldots,c_{m-1}$
is the output sequence %(or the sequence of states) 
of a maximum period
linear feedback
shift register over $\mathbb Z/2\Z$ with $s$ cells.
%\item
%\label{ex:KlSh-2}
\end{enumerate}
\end{exmps}

\section{Properties of output sequences}
\label{Out}
In this section we study a structure and statistical properties of output sequences of wreath products
of automata, that is, sequences described by
Theorem \ref{thm:WP}. %as well as their more deep statistical properties.
%\begin{note*} In view of Note
Note that in view of \ref{note:Bool}, all the results of this section
%\ref{Out} 
remain true for compatible mappings $T\colon\Z_2\>\Z_2$ (i.e.,
for T-functions) either.
%\end{note*}
\subsection{Distribution of $k$-tuples} %In this subsection we study a
%distribution of overlapping binary $k$-tuples in output sequences of automata
%introduced above. 
The output sequence $\mathcal Z$ of any wreath product
of automata that satisfy 
\ref{thm:WP} 
%with
%output alphabet $\{0,1,2,\ldots, 2^n-1\}=\Z/2^n$ 
is strictly uniformly
distributed as a sequence over $\Z/2^n\Z$ for all $n$. That is, each sequence
$\mathcal Z_n$ of residues modulo $2^n$ of terms of the sequence $\mathcal Z$
is purely periodic,
and each element of $\Z/2^n\Z$
occurs at the period the same number of times. However, when this sequence
$\mathcal Z_n$
is used as a key-stream, that is, 
%the same sequence 
as a binary sequence $\mathcal Z^\prime_n$ obtained by a concatenation of successive $n$-bit words
of $\mathcal Z$, it
is important to know how
$n$-tuples are distributed in this binary sequence. %{\slshape 
Yet strict uniform distribution of an
arbitrary sequence $\mathcal T$
as a sequence over $\Z/2^n\Z$ does not necessarily imply uniform distribution
of $n$-tuples, if this sequence is considered as a binary sequence
$\mathcal T^\prime$.
%!} 

For instance, let $\mathcal T=0132013201321\ldots$. This sequence is strictly uniformly
distributed over $\Z/4\Z$; the length of its shortest period is $4$. Its binary representation
% \footnote{recall that according to
% our conventions in Section \ref{Prelm} we write senior bits right, and not left;
% i.e., $2=01$, $1=10$, etc.}
is $\mathcal T^\prime_2=000111100001111000011110\ldots$ Considering $\mathcal T$ as
a sequence over $\Z/4\Z$, each number of $\{0,1,2,3\}$ occurs in the
sequence with the same frequency $\frac{1}{4}$. Yet if we consider $\mathcal
T$ in its binary form $\mathcal T^\prime_2$, then $00$ (as well as $11$) occurs in this sequence with
frequency $\frac{3}{8}$, whereas $01$ (as well as $10$) occurs with frequency $\frac{1}{8}$.
%Thus, the sequence $\mathcal T$ is uniformly distributed over $\Z/4$, and
%it is not uniformly distributed over $\Z/2$.

In this subsection we show that such an effect does not take
place for output sequences of automata described in 
%\ref{WP-even}, \ref{WP-even-trunc}, \ref{WP-odd}, 
\ref{thm:WP}, \ref{le:WP-odd},
and \ref{WP-even}: 
{\slshape Considering any of these sequences in a
binary form, a distribution
of $k$-tuples is uniform, for all $k\le n$}. Now we state this property  formally.

% It turnes out that this binary sequence is
% %periodicThe sequence $\mathcal Z$ of theorem \ref{thm:WP} is 
% {\slshape strictly $k$-distributed
% for all $k=1,2,\ldots,n$}: That is,
% %The latter  being considered as a
% %binary sequence of cocnatented binary $n$-tuples, this sequence is purely periodic
% %of period length $2^nmn$, 
% each $k$-tuple occurs at the period of the sequence
% %at the cycle, which corresponds to the period of the sequence $\mathcal Z^\prime_n$,
% with the same frequency $\frac{1}{2^k}$. 
% 
% Now we state this  property more formally.
Consider a (binary) {\it $n$-cycle} 
$C=(\varepsilon_0\varepsilon_1\dots \varepsilon_{n-1})$, i.e., 
%over a set $\{0,1\}$}; 
an oriented
graph on vertices $\{a_0,a_1,\ldots, a_{n-1}\}$ and edges 
$$\{(a_0,a_1),(a_1,a_2),\ldots, (a_{n-2},a_{n-1}),(a_{n-1},a_0)\},$$ 
where
each vertex $a_j$ is labelled with $\varepsilon_j\in\{0,1\}$, $j=0,1,\dots,n-1$.
(Note that then $(\varepsilon_0\varepsilon_1\dots \varepsilon_{n-1})=
(\varepsilon_{n-1}\varepsilon_0\dots \varepsilon_{n-2})=\ldots$, etc.).
Clearly, each purely periodic sequence $\mathcal S$ over $\Z/2\Z$ with period 
$\alpha_0\ldots\alpha_{n-1}$
of length $n$
could be related to a binary $n$-cycle $C(\mathcal S)=(\alpha_0\ldots\alpha_{n-1})$.
Conversely, to each binary $n$-cycle $(\alpha_0\ldots\alpha_{n-1})$ we could
relate $n$ purely periodic binary sequences with periods of length $n$: Those
are $n$ shifted versions of the sequence
$$\alpha_0\ldots\alpha_{n-1}\alpha_0\ldots\alpha_{n-1}\ldots.$$
% that is
% \begin{align*}
% &\alpha_1\ldots\alpha_{n-1}\alpha_0\alpha_1\ldots\alpha_{n-1}\alpha_0\ldots,\\
% &\alpha_2\ldots\alpha_{n-1}\alpha_0\alpha_1\alpha_2\ldots\alpha_{n-1}\alpha_0\alpha_1\ldots,\\
% &\ldots\qquad\ldots\qquad\ldots\\
% &\alpha_{n-1}\alpha_0\alpha_1\alpha_2\ldots\alpha_{n-2}\alpha_{n-1}\alpha_0\alpha_1\alpha_2\ldots\alpha_{n-2}\ldots
% \end{align*}

Further, {\it a $k$-chain in a binary $n$-cycle}  
$C$ is a
binary string $\beta_0\dots\beta_{k-1}$, $k<n$, that satisfies the following
condition: There exists $j\in\{0,1,\ldots,n-1\}$ such that $\beta_i=\varepsilon_{(i+j)\bmod
n}$ for $i=0,1,\ldots, k-1$. Thus, a $k$-chain
is just a string of length
$k$ of labels that corresponds to a chain of length $k$ in a graph $C$.
We call a binary $n$-cycle $C$ {\it $k$-full}, if each $k$-chain
occurs in the graph $C$ the same number $r>0$ of times.

Clearly, if $C$ is $k$-full, then $n=2^kr$. For instance, a well-known
De Bruijn sequence is an $n$-full $2^n$-cycle. 
%(to be more exact, a periodic
%binary sequence that corresponds to this $2^n$-cycle), 
%see e.g. \cite{MrH}
%for further references. 
Clearly enough that a $k$-full $n$-cycle is $(k-1)$-full:
Each $(k-1)$-chain occurs in $C$ exactly $2r$ times, etc. Thus, if an $n$-cycle
$C(\mathcal S)$ is $k$-full, then each $m$-tuple (where $1\le m\le k$) occurs in
the sequence $\mathcal S$ with the same probability (limit frequency) $\frac{1}{2^m}$.
That is, the sequence $\mathcal S$ is {\it $k$-distributed}, see
\cite[Section 3.5, Definition D]{knuth}.
\begin{defn} A purely periodic binary sequence $\mathcal S$ with the shortest
period of length
$N$ is said to be {\it
strictly $k$-distributed} iff the corresponding $N$-cycle $C(\mathcal S)$
is $k$-full.
\end{defn}

Thus, if a sequence $\mathcal S$ is strictly $k$-distributed, then it is
strictly $s$-distributed, for all positive $s\le k$.

\begin{thm}[\cite{anashin4}]
\label{thm:distr}
For the sequence $\mathcal Z$ of theorem \ref{thm:WP} each binary sequence
$\mathcal Z^\prime_n$ is
%periodicThe sequence $\mathcal Z$ of theorem \ref{thm:WP} is 
strictly $k$-distributed
for all $k=1,2,\ldots,n$. 
% {\rm Here $\mathcal Z^\prime_n$ is a
% binary representation of  the sequence $\mathcal Z_n=\{z_i\bmod 2^n\}$ (hence $\mathcal Z^\prime_n$ is
% a purely periodic binary sequence of period length exactly $2^n nm$)}.
\end{thm} 

\begin{note}
\label{note:distr} 
Theorem \ref{thm:distr} remains true 
%for {\it wreath
%products of truncated automata}, i.e., 
for the sequence $\mathcal F$ of corollary
\ref{cor:WP}, where $F_j(x)=\big\lfloor\frac{x}{2^{n-k}}\big\rfloor\bmod 2^k$, 
$j=0,1,\ldots,m-1$, a truncation of $(n-k)$ less significant bits. Namely, {\it
a binary representation $\mathcal F^\prime_n$ of the sequence $\mathcal F$
is a purely periodic strictly $k$-distributed binary sequence with a period
of length $2^nm k$.}
\end{note}
Theorem \ref{thm:distr} treats an output sequence of a counter-dependent
automaton as an infinite (though, a periodic) binary sequence. However, in cryptography
only a part of a period is used during encryption. So it is natural
to ask how `random' is a finite segment (namely, the period) of this infinite sequence.
According to \cite[Section 3.5, Definition Q1]{knuth} %there is considered the following 
%``indicator of randomness"
%of a finite sequence over a finite alphabet $A$ (we formulate the corresponding
%definition for $A=\{0,1\}$): A 
a finite binary sequence 
$\varepsilon_0\varepsilon_1\dots \varepsilon_{N-1}$
of length $N$ is said to be random, iff
\begin{equation}
\label{eq:Q1}
\bigg|\frac{\nu(\beta_0\ldots\beta_{k-1})}{N}-\frac{1}{2^k}\bigg|\le\frac{1}{\sqrt
N}
\end{equation}
for all $0<k\le\log_2N$, where $\nu(\beta_0\ldots\beta_{k-1})$ is the number
of occurrences of a binary word $\beta_0\ldots\beta_{k-1}$ in a binary word
$\varepsilon_0\varepsilon_1\dots \varepsilon_{N-1}$. If a finite sequence
is random in the sense of this Definition Q1 of \cite{knuth}, we shall say
that this sequence %has {\it a property} Q1, or 
{\it satisfies} Q1. We shall also
say that an {\it infinite periodic sequence satisfy} Q1 iff its shortest
period satisfies Q1.
Note that, contrasting to the case of strict $k$-distribution, which implies
strict $(k-1)$-distribution, 
%$n$-cycle, where $k$-fullness implies
%$(k-1)$-fullness,
it is not enough to demonstrate only
that \eqref{eq:Q1}
holds for $k=\lfloor\log_2N\rfloor$ to prove a finite sequence of length $N$ 
satisfies Q1:
For instance, the sequence $1111111100000111$ satisfies \eqref{eq:Q1} for
$k=\lfloor\log_2N\rfloor=4$ and does not satisfy \eqref{eq:Q1} for $k=3$.

%For the output sequences of counter-dependent generators theorem \ref{thm:distr}
%implies the following
\begin{cor} [\cite{anashin4}]
\label{cor:distr}
The sequence $\mathcal
Z^\prime_n$ of theorem \ref{thm:distr} %. Then the finite sequence $\mathcal
%P$ 
satisfies {\rm Q1} if $m\le\frac{2^n}{n}$. Moreover, in this case 
under the conditions of \ref{note:distr} the output binary sequence still
satisfies {\rm Q1}
if one truncates 
$0\le k\le\frac{n}{2}-\log_2\frac{n}{2}$ 
lower order bits {\rm (that is, if one uses clock output functions $F_j$
%$\lfloor\frac{\cdot}{2^k}\rfloor\bmod
%2^{n-k}$
of \ref{note:distr})}.
%could be truncated
%without affecting property Q1 of 
%to guarantee that .
\end{cor}
We note here that according to \ref{cor:distr} a control sequence
of a counter-dependent automaton (see \ref{thm:WP}, \ref{le:WP-odd}, \ref{cor:WP},
and the text and examples thereafter) may not satisfy Q1 at all, yet nevertheless
a corresponding output sequence necessarily satisfies Q1. Thus, {\slshape
with the use of
wreath product
techniques one could stretch `non-randomly looking' sequences to `randomly
looking' ones}.

\subsection{Structure}
A recurrence sequence could be `very uniformly distributed', yet nevertheless could
have some mathematical structure that might be used by
an attacker to break the cipher. For instance, a clock sequence
$x_i=i$ is uniformly distributed in $\mathbb Z_2$. %; moreover,
%its counterpart in the field $\mathbb R$ of real numbers, the so-called %Van der Corput
%sequence $u_i={i}\cdot{2^{-\lfloor\log_2i\rfloor-1}}$,
%is uniformly distributed in a unit interval $[0,1]$ and 
%has the least (of the known)
%(among known)
%discrepancy. %, see \cite{KN}. 
We are going to study what structure could
have sequences outputted by our counter-dependent generators. 

Theorem \ref{thm:WP} immediately implies that the
{\it $j$\textsuperscript{th} coordinate sequence $\delta_j(\mathcal Z)=\{\delta_j(x_i)\colon i=0,1,2,\ldots\}$
$(j=0,1,2,\ldots)$ of the sequence $\mathcal Z$}, i.e., a sequence
formed by all $j$\textsuperscript{th} bits of terms
of the sequence $\mathcal Z$,  %of theorem \ref{thm:WP}
has a period not longer than $m\cdot 2^{j+1}$. Moreover, the following 
could be easily proved: %rom the proof
%of
%lemma \ref{le:WP-odd} (see Appendix) follows 
\begin{prop}[\cite{anashin4}]
\label{note:halfper-odd}
%During the proof of theorem \ref{thm:WP}
%proposition \ref{WP-odd} and lemma \ref{le:WP-odd}
%we have demonstrated that {\it 
{\rm (1)} The $j$\textsuperscript{th} coordinate sequence
% $j$\textsuperscript{th} coordinate sequence $\delta_j(\mathcal Z)=\{\delta_j(x_i)\colon i=0,1,2,\ldots\}$
% $(j=0,1,2,\ldots)$
$\delta_j(\mathcal Z)$
is a purely periodic binary sequence with a period of length $2^{j+1}m$, and {\rm
(2)} the
second half of the period is a bitwise negation of the first half: $\delta_j(x_{i+2^jm})\equiv
\delta_j(x_i)+1\pmod 2$, $i=0,1,2,\ldots$  %(see the proofs of claims (1)--(2) of 
%lemma \ref{le:WP-odd} in the Appendix).
\end{prop}
\begin{note*}
The $j$\textsuperscript{th} coordinate sequence of a sequence generated by
a single-cycle $T$-function %$\mathcal S_j$ 
is purely periodic, and
$2^{j+1}$ is the length of the shortest period of this sequence.
The second half of the period is a bitwise negation of the first half, i.e.,
%\begin{equation*} 
$\zeta_{i+2^j}\equiv \zeta_{i}+1\pmod 2$
%\end{equation*}
for each $i=0,1,2,\ldots$. 
\end{note*}
Proposition \ref{note:halfper-odd} means
that
%loosely speaking,  
the $j$\textsuperscript{th} coordinate sequence of the sequence of states of
a counter-dependent generator is completely determined by
the first half of its period; so, intuitively, it is as
`complex' as the first half of its period. Thus we ought to understand what
sequences of length $2^jm$ occur as the first half of the
period of the $j$\textsuperscript{th} coordinate sequence. 

For $j=0$ (and
$m>1$) the
answer immediately follows from \ref{thm:WP} and \ref{le:WP-odd} --- any binary
sequence $c_0,\ldots,c_{m-1}$ such that $\sum_{j=0}^{m-1}c_j\equiv
1\pmod 2$ does.  It turns out that for $j>0$ %We are going
%to demonstrate that 
{\slshape any binary sequence could be produced as
the first half of the period of the $j$\textsuperscript{th} coordinate sequence independently
of other coordinate sequences}.
%, that is, what
%values takes the rational integer $\gamma$ of \ref{2-comp}.

More formally, 
to each sequence $\mathcal Z$ described by theorem \ref{thm:WP}
we associate a sequence $\Gamma(\mathcal Z)=\{\gamma_1,\gamma_2,\ldots\}$
of non-negative rational integers $\gamma_j\in\N_0=\{0,1,2,\ldots\}$ such that $0\le\gamma_j\le
2^{2^{j}m}-1$ %iff \eqref{eq:num:coord} holds
and the base-$2$ expansion of $\gamma_j$
agrees with the first half
of the period of the $j$\textsuperscript{th} coordinate sequence $\delta_j(\mathcal
Z)$ for all $j=1,2,\ldots$; that is
$$\gamma_j=\delta_j(x_0)+2\cdot\delta_j(x_1)+
4\cdot\delta_j(x_2)+\dots+2^{2^jm-1}\cdot\delta_j(x_{2^jm-1}),$$
where $x_0$ is an initial state; $x_{i+1}=g_{i\bmod m}(x_i)$, $i=0,1,2,\ldots$.
Now
we take an arbitrary sequence %$\Gamma$
$\Gamma(\mathcal Z)=\{\gamma_1,\gamma_2,\ldots\}$
of non-negative rational integers $\gamma_j$ such that $0\le\gamma_j\le
2^{2^{j}m}-1$ %of this type 
and wonder
whether this sequence could be so associated to some sequence $\mathcal Z$
described by theorem \ref{thm:WP}.

% %consider a case $m=1$ first, and 
% let $\gamma_j(f,z)\in\N_0$ be such a number that its base-$2$ expansion 
% agrees with the first half
% of the period of the $j$\textsuperscript{th} coordinate sequence 
% %produced
% %by the
% of the automaton $\mathfrak A^\prime_j=\langle \Z_2,\Z_2,f,I,z\rangle $, 
% with ergodic state transition function $f$, identity output function $I$, 
% and initial state $z$.
% That is:
% $$\gamma_j(f,z)=\delta_j(f^{(0)}(z))+2\delta_j(f^{(1)}(z))+
% 4\delta_j(f^{(2)}(z))+\dots+2^{2^j-1}\delta_j(f^{(2^j-1)}(z)).$$
% Obviously, $0\le\gamma_j(f,z)\le 2^{2^j}-1$. A natural question arises:
% 
% {\slshape Given a compatible and ergodic mapping
% $f\colon\mathbb Z_2\rightarrow\mathbb Z_2$ and a $2$-adic integer $z\in\mathbb
% Z_2$, what infinite string $\gamma_0=\gamma_0(f,z),\gamma_1=\gamma_1(f,z),
% \gamma_2=\gamma_2(f,z),\dots$ (where $\gamma_j\in\{0,1,\dots,2^{2^j}-1\}$
% for
% $j=0,1,2,\dots$) could be obtained?} 
% 
% The answer is:  {\slshape any one.} 
The answer is {\slshape yes}. Namely, the
following theorem holds.
\begin{thm}[\cite{anashin4}]
\label{thm:WP:AnyHalfPer}
Let $m> 1$ be a rational integer, and let $\Gamma=\{\gamma_1,\gamma_2,\dots\}$ 
be an arbitrary sequence over $\N_0$ %=\{0,1,2,\ldots\}$
such that $\gamma_j\in\{0,1,2,\ldots,2^{2^jm}-1\}$ for
all $j=1,2,\dots$. Then there exist  a finite sequence 
$\mathcal G=\{g_0,\ldots,g_{m-1}\}$
of compatible  measure preserving mappings of $\Z_2$ onto itself and a
$2$-adic integer $x_0=z\in\Z_2$ such that $\mathcal G$ satisfies conditions
of theorem \ref{thm:WP}, and the base-$2$ expansion of $\gamma_j$ agrees
with the first $2^jm$ terms of the sequence $\delta_j(\mathcal Z)$ %satisfies \eqref{eq:num:coord}
for all $j=1,2,\dots$,
where the recurrence sequence $\mathcal Z=\{x_0,x_1,\ldots\in\Z_2\}$ is
defined by the recurrence relation $x_{i+1}=g_{i\bmod m}(x_i)$, $(i=0,1,2,\dots)$.
In case $m=1$ the assertion holds for an arbitrary $\Gamma=\{\gamma_0,\gamma_1,\dots\}$,
where $\gamma_j\in\{0,1,2,\ldots,2^{2^j}-1\}$, $j=0,1,2,\dots$.
\end{thm}
\begin{proof} We will prove the theorem only for $m=1$ (i.e., for $T$-functions)
by  two reasons.
First, in this case use of methods of 2-adic analysis becomes more transparent,
and second, the proof for $m>1$ is much more technical and complicated (an
interested reader is referred to \cite{anashin4}). 

Speaking informally, we fill a table with countable infinite number of rows
and columns in such a way that the first $2^j$ entries of the $j$\textsuperscript{th}
column represent $\gamma_j$ in its base-2 expansion, and the other entries
of this column are obtained from these by applying recursive relation of Proposition \ref{note:halfper-odd}; that is, the next $2^j$ entries are bitwise negation
of the first $2^j$ entries, the third $2^j$ entries are bitwise negation
of the second $2^j$ entries, etc.
Then we read each $i$\textsuperscript{th} row of the table  as a  2-adic canonical
representation of 2-adic integer which we denote via $z_i$. Thus we define
a set $Z=\{z_0,z_1,\ldots\}$ of 2-adic integers.

We shall prove that $Z$ is a dense subset in $\mathbb Z_2$, and then
define $f$ on $Z$ in such a way that $f$ is compatible and ergodic on $Z$.
This will imply the assertion of the theorem. 

Proceeding along this way we claim that $Z\bmod 2^k = \mathbb Z/2^k\Z$ for all $k=1,2,3,\ldots$,
i.e., a natural ring homomorphism $\bmod\, 2^k\colon z\mapsto z\bmod 2^k$ maps
$Z$ onto the residue ring $\mathbb Z/2^k\Z$. Indeed, this trivially holds
for $k=1$. Assuming our claim holds for $k< m$ we prove it for $k=m$.
Given arbitrary $t\in\{0,1,\ldots,2^{m}-1\}$ there exists $z_i\in Z$ such
that $z_i\equiv t\pmod{2^{m-1}}$. If $z_i\not\equiv t\pmod{2^{m}}$ then
$\delta_{m-1}(z_i)\equiv\delta_{m-1}(t)+1\pmod 2$ and thus
$\delta_{m-1}(z_{i+2^{m-1}})\equiv\delta_{m-1}(t)\pmod 2$. However, 
$z_{i+2^{m-1}}\equiv z_i\pmod {2^{m-1}}$. Hence
$z_{i+2^{m-1}}\equiv t\pmod {2^m}$. 

A similar argument shows that for each $k\in\mathbb N$ 
the sequence $\{z_i\bmod 2^k\colon i=0,1,2,\ldots\}$
is purely periodic with period length $2^k$, and each $t\in\{0,1,\ldots,2^{k}-1\}$
occurs at the period exactly once (in particular, all elements of $Z$ are
pairwise distinct 2-adic integers). Moreover, $i\equiv i^{\prime}\pmod{2^k}$
iff $z_{i}\equiv z_{i^{\prime}}\pmod{2^k}$. Consequently, $Z$ is dense
in $\mathbb Z_2$ since for each $t\in\mathbb Z_2$ and each $k\in\mathbb
N$ there exists $z_i\in Z$ such that $\|z_i-t\|_2\le 2^{-k}$. Moreover, if
we define $f(z_i)=z_{i+1}$ for all $i=0,1,2,\ldots$ then 
$\|f(z_i)-f(z_{i^{\prime}})\|_2=\|z_{i+1}-z_{i^{\prime}+1}\|_2=
\|(i+1)-(i^{\prime}+1)\|_2=\|i-i^{\prime}\|_2=\|z_i-z_{i^{\prime}}\|_2$.
Hence, $f$ is well defined and compatible on $Z$; it follows that the continuation
of $f$ to the whole space $\mathbb Z_2$ is compatible. Yet $f$ is transitive
modulo $2^k$ for each $k\in\mathbb N$, so its continuation is ergodic.
\end{proof}  
\begin{note}[Representation by T-functions]
\label{note:wr=tf}
%\textcolor{red}{???Is it difficult to prove this???}
% The following follows from \cite[Theorems 5.9 and 5.10]{anashin4}. For the
% wreath product $g=h\rightthreetimes f_i$ of claim~1 of Proposition~\ref{prop:wr}, there exists a single-cycle T-function $G\colon\Z/2^{t+n}\rightarrow\Z/2^{t+n}$ such that $g(i,x)=G(i+2^tx)$ $\forall i\in\Z/2^t\Z$ and $\forall x\in \Z/2^n\Z$.
Suppose $m=2^k$ under conditions of Theorem \ref{thm:WP:AnyHalfPer}. Then, considering the sequence $\delta_j(\mathcal Z)$, %from %$h\rightthreetimes
%f_i$ 
one deals with the %most significant 
$(j+m)$-th coordinate sequence of a
single-cycle T-function.
\end{note}
\subsection{Linear complexity} The latter is an important cryptographic
measure of
complexity of a binary sequence; being a number of cells of the
shortest linear feedback shift register (LFSR) that outputs the given sequence%
\footnote{i.e., degree of the minimal polynomial over $Z/2\Z$ of given
sequence} 
%$\mathcal Z_j$},
it estimates dimensions of a linear system an attacker must solve to
obtain initial state. %An expectation of the linear complexity of a random finite sequence is
%a half of its length. 
%With the use of this theorem and \ref{note:halfper-odd} one may prove the
%following
\begin{thm}[\cite{anashin4}]
\label{thm:lincomp:sharp} For $\mathcal Z$ and $m$ of theorem \ref{thm:WP} let
$\mathcal Z_j=\delta_j(\mathcal Z)$, $j>0$, be
the $j$\textsuperscript{th}
coordinate sequence. 
% of a wreath product of automata {\rm(described by any of
% \ref{WP-even}, \ref{WP-even-trunc}, \ref{WP-odd}, \ref{le:WP-odd},
% and \ref{thm:WP}: thus $\mathcal Z_j$ is a purely periodic binary sequence
% of period length $2^{j+1}\ell$, where $\ell=2^m$ for wreath products described
% by \ref{WP-even} or \ref{WP-even-trunc}, and $\ell=m$ otherwise\rm)}. 
Represent $m=2^kr$, where $r$
is odd.
Then length of the shortest period of $\mathcal Z_j$ is $2^{k+j+1}s$ for 
some $s\in\{1,2,\dots,r\}$, and both extreme cases $s=1$ and $s=r$ 
occur: For every sequence $s_1,s_2,\ldots$ over a set $\{1,r\}$
there exists a sequence $\mathcal Z$ of theorem \ref{thm:WP}
%wreath product of automata 
such that length of the shortest period  of
%of the  $j$\textsuperscript{th} coordinate sequence 
$\mathcal Z_j$ is 
$2^{k+j+1}s_j$, $(j=1,2,\ldots)$.
Moreover, linear complexity $\lambda_2(\mathcal Z_j)$ 
of the sequence $\mathcal
Z_j$ 
satisfies the following inequality:
$$2^{k+j}+1\le \lambda_2(\mathcal Z_j)\le 2^{k+j}r+1.$$
Both these bounds are sharp: 
For every sequence $t_1,t_2,\ldots$ over a set $\{1,r\}$ 
there exists a sequence $\mathcal Z$ of theorem \ref{thm:WP}
%wreath product of automata 
such that linear complexity of
%of the $j$\textsuperscript{th} coordinate sequence 
$\mathcal Z_j$ is exactly
$ 2^{k+j}t_j+1$, $(j=1,2,\ldots)$.
\end{thm}
\begin{note*} The linear complexity %$\lambda_2(\mathcal S_j)$
of the $j$-th coordinate sequence of a $T$-function %$\mathcal S_j$ 
%over
%$GF(2)$ 
is exactly $2^j+1$, i.e., approximately half of the length of the period of the sequence. {Note} that %\textcolor{yellow}
{the expectation of the linear complexity $\lambda_{2}(\mathcal C)$
%: Let $R=\mathbb Z/2$, let $\mathcal Z$ be 
of a random sequence $\mathcal C$
of length $L$ %. Then 
%the expectation of $\lambda_{\mathbb Z/2}(\mathcal Z)$ 
is $\frac{L}{2}$.} %Thus, the coordinate sequences {are rather good} with respect
%to their linear complexities.
\end{note*}
% \begin{note*} Somewhat similar estimates hold for $2$-adic span (see definition
% in \cite{Kl-Gor}),
% one more cryptographic
% measure of complexity of a sequence. We have to omit exact statements due
% to space limitations.
% \end{note*}
Whereas the linear complexity of a binary sequence $\mathcal X$ is the length of the shortest LFSR %linear feedback shift register  
that produces $\mathcal X$,
the {\it $\ell$-error linear complexity} is the length of the  shortest LFSR  that produces a sequence with almost the same (with the exception of not more than $\ell$ terms)  period  as that of $\mathcal
X$; that is, the two periods coincide everywhere  but at $t\le\ell$
places.
Obviously, a random sequence of length $L$
coincides with a sequence that has a period of length $L$ approximately at
$\frac{L}{2}$ places. That is, the $\ell$-error linear complexity makes sense
only for $\ell<\frac{L}{2}$. The following proposition holds.
\begin{prop}%[$\ell$-error linear complexity of coordinate sequences]
\label{prop:coord}
Let $\mathcal Z$ %$\mathcal V=\mathcal V(z_0,x_0)$ 
be a sequence of Theorem \ref{thm:WP}, and let $m=2^s>1$. %claim 1 of
%Proposition \ref{prop:wr}. 
Then for $\ell$  less than the half of the length of the shortest
period of the $j$-th coordinate sequence $\delta_j(\mathcal Z)%=\{\delta_j(x_i)\}_{i=0}^\infty
$,
the $\ell$-error linear complexity of $\delta_j(\mathcal Z)$ exceeds $2^{j+m-1}$, the half of the length of its
shortest period.
\end{prop}
\begin{proof} 
In view of Note~\ref{note:wr=tf} it suffices to prove the statement for the coordinate sequences of a $T$-function %$G(i+2^tx)=g(i,x)$, $y=i+2^tx\in\Z/2^{t+n}\Z$
only. According to Proposition \ref{note:halfper-odd}, the $j$-th coordinate sequence $\mathcal Y=\{x_i\colon i=0,1,2,\ldots\}=\delta_j(\mathcal Z)$ %$\mathcal Y_k=\{\delta_k(y^{(j)})\}_{j=1}^{\infty}$ of a single-cycle T-function
%$G$ 
of a $T$-function is a periodic sequence with the length of the shortest period $2^{j+1}$,
%Moreover, it 
which satisfies the 
relation 
\begin{equation}
\label{eq:inv}
\delta_j(x_{i+2^j})\equiv
\delta_j(x_i)+1\pmod 2,
%\delta_k(y^{(m+2^k)})=\delta_k(y^{(m)})\oplus1~\forall m\in\{1,2,3,\dots\}\text{}. 
\end{equation}
for all $i=0,1,2,\ldots$
Since $2^{j+1}$ is the length of a period of a (binary) sequence $\mathcal Y$,  $$w(X)=X^{{2^{j+1}}}+1=(X+1)^{2^{j+1}}$$ is a
characteristic polynomial (over a field $\Z/2\Z$ of two elements) of the sequence
$\mathcal Y$.% see \cite[Proof of Lemma 5.4]{anashin4}.

Let $\mathcal Q=\{q_{i}\colon i=0,1,2,\ldots\}$ be a binary sequence produced by a LFSR with $d$ cells such that $\mathcal Q$ has a period of length $2^{j+1}$,
and $x_{i}=q_i$ for all $i\in\{0,1,2,\ldots,2^{j+1}-1\}$ with the exception of $\ell$
indexes
$j=j_1,\ldots,j_\ell\in\{0,1,\ldots,2^{j+1}-1\}$. Since $2^{j+1}$ is the
length of a period of $\mathcal Q$, the minimal polynomial $\mu(X)$ of the
sequence $\mathcal Q$ (which
is of degree $d$ then) must be a multiple of the polynomial $X^{2^{j+1}}+1=
(X+1)^{2^{j+1}}$ over the field $\Z/2\Z$. Hence, $\mu(X)=(X+1)^d$, and $d\le 2^{j+1}$.  

On the other hand, if $\ell <2^j$, then in view of \eqref{eq:inv} the length
of the shortest period of the sequence $\mathcal Q$ cannot be less than $2^{j+1}$.
Hence, $d\ge 2^j+1$, since otherwise $\mu(x)$ is a multiple of $(X+1)^{2^j}=X^{2^j}+1$;
yet the latter would imply 
that $\mathcal Q$ has a period of length $2^j$.
 
\end{proof}
We can consider linear complexity of a sequence with terms from an arbitrary
commutative ring, not necessarily from the field of two elements.
\begin{defn} 
%\hypertarget{lincomp}{{\bf Definition.}}
Let $\mathcal Z=\{z_i\}$ be a sequence over a commutative ring $R$. The {\it linear complexity}
$\lambda_R(\mathcal Z)$
of $\mathcal Z$ over $R$ is the smallest $r\in\mathbb N_0$ such that there
exist
$c, c_0,c_1, \ldots,c_{r-1}\in R$ (not all equal to $0$) such that for
all $i=0,1,2,\ldots$ holds
\begin{equation}
\label{eq:lin}
%$$
c+\sum_{j=0}^{r-1} c_j\cdot z_{i+j}=0.
%$$
%}
\end{equation}   
\end{defn}
%\underline
%\begin{boxitpara}{box 0.9 setgray fill}
%\fromSlide{3}{%
%{Geometrically this equation %(\ref{eq:lin}) 
%means the following}: 
For instance, if $R=\mathbb Z/p^n\Z$; then 
geometrically %this 
equation (\ref{eq:lin}) 
means  that
all the points 
$(\frac{z_{i}}{p^n},\frac{z_{i+1}}{p^n},\ldots,\frac{z_{i+r-1}}{p^n})$, 
$i=0,1,2,\ldots$, of a unit $r$-dimensional Euclidean hypercube fall into parallel hyperplanes. For instance, with the use of linear complexity over the
residue ring $\Z/2^k\Z$ we can study distribution of $r$-tuples of the sequence
produced by an ergodic $T$-function modulo $2^k$. We already know that this
sequence, being considered as the sequence of elements over $\Z/2^k\Z$ is strictly uniformly distributed: Every element from $\Z/2^k\Z$ occurs
at the period exactly once. But what about distribution of consecutive pairs
of elements? Triples? etc. It varies...
%\end{boxitpara}
% \newslide
% In fact, linearity tests turn out to be ones of the most effective. For
% instance, linear congruential generators do not pass these since their
% linear
% complexity over the corresponding residue ring is only 2; that is, distribution
% of pairs in produced sequences is rather poor: All the points that correspond
% to pairs of consequtive numbers fall  into  a small
% number of parallel straight lines in a unit square.
% \begin{boxitpara}{box
% .9 setgray fill}
% Moreover, a number of sequences produced by $T$-functions with a single cycle
% property demonstrate poor distribution of pairs even after truncation of
% low order bits.
% \end{boxitpara}
% \begin{exmp} A $T$-function $x+x^2\vee C$ has a single cycle property whenever
% $C\equiv 5\pmod 8$, or $C\equiv 7\pmod 8$ (Klimov and Shamir, 2002)
% 
% However, the distribution of pairs of the sequence produced by this $T$-function
% varies from satisfactory (when there
% are few 1's in more significant bit positions) to poor (when there are
% more 1's).
% \end{exmp}
%\end{slide}
%}
%\overlays{6}{%
%\begin{slide}[Wipe]{Linear complexity}
%\onlySlide*{1}{%
%In fact, linearity tests turn out to be ones of the most effective.

\begin{figure}
\begin{minipage}[b]{.40\linewidth}
\centering\epsfig{file=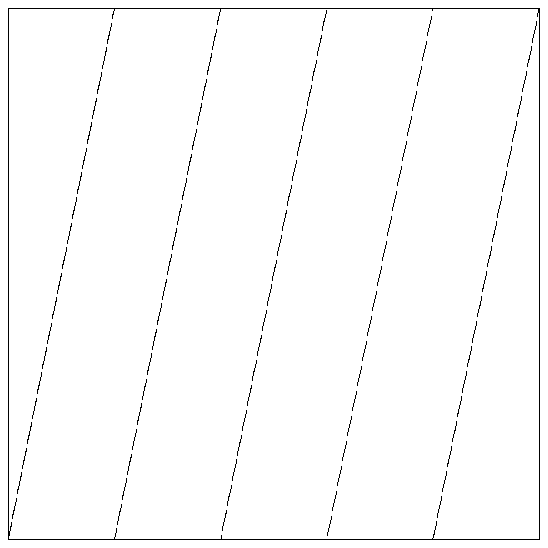,width=\linewidth}
\caption{\qquad\footnotesize{Linear congruential generator $3+5x$}}\label{fig:Lin}
\end{minipage}\hfill
\begin{minipage}[b]{.40\linewidth}
\centering\epsfig{file=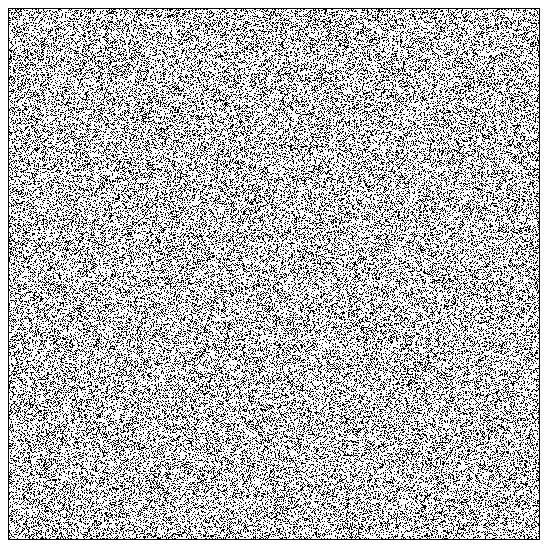,width=\linewidth}
\caption{\qquad\footnotesize{Polynomial generator of degree 8}}\label{fig:Mah}
\end{minipage}
\end{figure}

For
example, despite every transitive linear congruential generator $x_{i+1}=a+b\cdot x_i\pmod{2^k}$
produces a strictly uniformly distributed sequence over $\Z/2^k\Z$, linear
% distribtionLinear
complexity over $\mathbb Z/2^k\Z$ of this generator % linear congruential generators 
is only 2; hence, distribution
of pairs in produced sequences is rather poor:
All the points that correspond
to pairs of consecutive numbers fall  into  a small
number of %\href{E:/Mydir/Talks/Visit%DS/Talk//Lin.vrg}
{parallel straight lines} in a unit square, and this picture {\sl does not
depend on $k$}, see Figure \ref{fig:Lin}. 
\begin{figure}
\begin{minipage}[b]{.40\linewidth}
\centering\epsfig{file=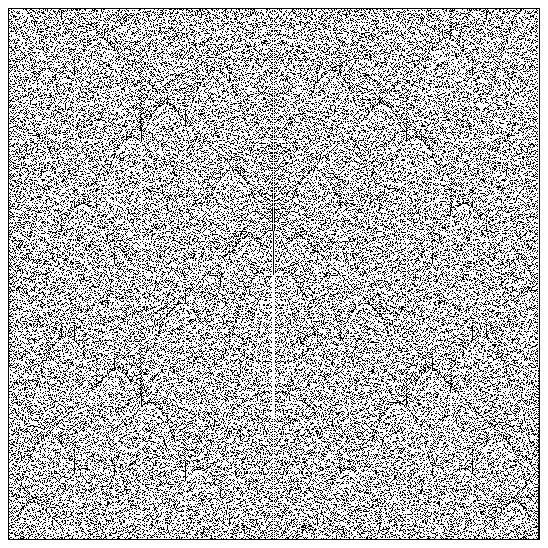,width=\linewidth}
\caption{\qquad\footnotesize{Klimov-Shamir generator with $C=101$}}\label{fig:KlSh}
\end{minipage}\hfill
\begin{minipage}[b]{.40\linewidth}
\centering\epsfig{file=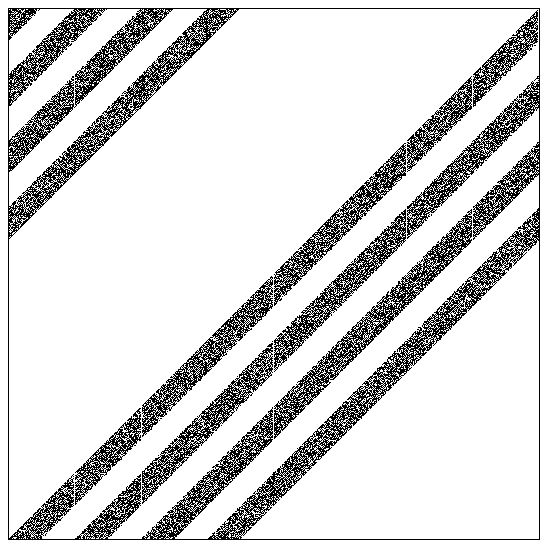,width=\linewidth}
\caption{\qquad\footnotesize{Same, with $C=10010000101010111$}}\label{fig:KlSh_bad}
\end{minipage}

\end{figure}

% All $T$-functions with a single cycle
% property  produce uniformly distributed sequences. However, some of these $T$-functions produce bad sequences, which have a number of linear dependencies modulo $p^n$,
% and poor distribution of pairs %even after truncation of
% %low order bits.

%\begin{exmp} 
Another example: The already mentioned $T$-function {$x+x^2\vee C$} of Klimov and Shamir has a single cycle property whenever
$C\equiv 5\pmod 8$, or $C\equiv 7\pmod 8$, see \ref{KlSh-3}. %(Klimov and Shamir, 2002)
However, distribution of pairs of the sequence produced by this $T$-function
varies from %\href{KlSh.vrg}
{satisfactory} (when there
are few 1's in more significant bit positions, see Figure \ref{fig:KlSh}) 
to %\href{KlSh_bad.vrg}
{poor} (when there are
more 1's in these positions, see Figure \ref{fig:KlSh_bad}).
%\end{exmp}

This is not easy to find a $T$-function that guarantees good distribution
of pairs.
For instance, this problem is not completely solved even for  quadratic generators with 
a single
cycle property, despite a number of works in the area (see  e.g. \cite{Emm,Eich}
and a survey \cite{EHHW}).

%\begin{boxitpara}{box 0.9 setgray fill}
However, we can prove that with respect to the linear complexity over residue
ring the {sequence %\hyperlink{concat}{$\mathcal X$} 
$\mathcal X_n=\{f^i(x_0)\bmod p^n\}$ over $\mathbb Z/p^n\Z$,
generated by} %those $T$-functions that are
{\sl compatible ergodic polynomial 
%\href{Mahler_8.vrg}
{$f(x)\in\Q[x]$} of degree $\ge 2$,
is `asymptotically good'} (cf. Figure \ref{fig:Mah} for distribution of pairs
for a polynomial generator of degree 8). Namely, the following theorem holds:

\begin{thm}[\cite{anashin3}]
$\lim_{n\to\infty}\lambda_{\mathbb Z/p^n\Z}(\mathcal X_n)=\infty$.
%\begin{note*}
%Within conditions of the theorem,
Moreover, 
$\lambda_{\mathbb Z/p^n\Z}(\mathcal X_n)$ tends to $\infty$ not
slower than $\log n$.
\end{thm}
We note, however, that in most real life ciphers the use of polynomials of higher degrees (say,
of degrees higher than 2) is too time-costly; so the search for good functions continues!
\subsection{The 2-adic span}
%With the use of theorem \ref{halfper} it is possible to estimate 
There are two other
measures of complexity of a binary sequence, which were introduced
in \cite{Kl-Gor}: namely, 
{\it $2$-adic complexity} and {\it $2$-adic span}. 
Whereas linear complexity (which is also known
as a {\it linear span}) is the number of cells in a linear
feedback shift register outputting a sequence $\mathcal S$ over
$\mathbb Z/2$, the $2$-adic span is the number of cells in both memory and register
of a feedback with carry shift register (FCSR) that outputs $\mathcal S$,
and
the
$2$-adic complexity estimates the number of cells in the register of this
FCSR. To be more exact, the $2$-adic complexity $\Phi_2(\mathcal S)$ of the (eventually) periodic
sequence $\mathcal S=\{s_0,s_1,s_2,\ldots\}$ over $\mathbb Z/2$ is $\log_2(\Phi(u,v))$, 
where $\Phi(u,v)=\max\{|u|,|v|\}$ and $\frac{u}{v}\in\mathbb Q$ 
is the irreducible fraction
such that its $2$-adic expansion agrees with $\mathcal S$, that is, 
$\frac{u}{v}=s_0+s_12+s_22^2+\dots\in\mathbb Z_2$. The number of cells in the register
of FCSR producing $\mathcal S$ is then $\lceil\log_2(\Phi(u,v))\rceil$, 
the least rational integer not smaller than $\log_2(\Phi(u,v))$.
% The $2$-adic span $\Lambda_2(\mathcal S)$ and the $2$-adic complexity 
% $\Phi_2(\mathcal S)$ of the sequence $\mathcal S$ are related by the
% following inequality (see Proposition 9.3 of \cite{Kl-Gor}):
% $$|(\Lambda_2(\mathcal S)-2)-\Phi_2(\mathcal S)|\le\log_2(\Phi_2(\mathcal S)).$$
Thus, we only need to estimate $\Phi_2(\mathcal S)$.
\begin{thm}[\cite{anashin4}]
\label{2-comp}
Let $\mathcal S_j=\{s_0,s_1,s_2,\dots\}$ be the $j$\textsuperscript{th} coordinate 
sequence of an ergodic $T$-function. 
%Then $\Phi_2(\mathcal S_j)>2^j-1$.
Then the $2$-adic complexity $\Phi_2(\mathcal S_j)$ of $\mathcal S_j$ is $$\log_2\Bigg(\frac{2^{2^j}+1}{\gcd(2^{2^j}+1,\gamma+1)}\Bigg),$$
% , that is, the number of
% cells in the register of FCSR, which outputs $\mathcal S$, is $\gcd (2^j+1,
% r+1)$, 
where $\gamma=s_0+s_12+s_22^2+\dots+s_{2^{j}-1}2^{2^{j}-1}$. 
\end{thm}
\begin{note*}
We note that $\gamma$ is a non-negative
rational integer, $0\le \gamma\le 2^{2^{j}}-1$; also we note that for each $\gamma$
of this range there exists an ergodic mapping such that the first half
of the period of the $j$\textsuperscript{th} 
coordinate sequence of the corresponding output is a base-$2$ expansion
of $\gamma$ (see Theorem \ref{thm:WP:AnyHalfPer}). Thus,
to find all possible values
of 2-adic complexity
of the $j$\textsuperscript{th} coordinate sequence one has to decompose
the $j$\textsuperscript{th} Fermat number $2^{2^j}+1$. It is known that
the $j$\textsuperscript{th} Fermat number is prime for $0\le j\le 4$ and that
it
is composite for $5\le j\le 23$. For each Fermat number outside 
this range it is not known whether
it is prime or composite.
The complete decomposition of $j$\textsuperscript{th} Fermat number is not known
for $j>11$. Assuming for some $j\ge 2$ the $j$\textsuperscript{th} Fermat number 
is composite, 
all its factors are of the form $t2^{j+2}+1$, see e.g. \cite{Brent} for
further references. So, {\it the following bounds for $2$-adic
complexity $\Phi_2(\mathcal S_j)$ of the $j$\textsuperscript{th} coordinate sequence
$\mathcal S_j$ hold:
$$ j+3\le\lceil\Phi_2(\mathcal S_j)\rceil\le 2^j+1,$$
%cells in the register of FCSR, which outputs the $j$\textsuperscript{th}
%coordinate sequence, is $j+3$, 
yet to prove whether the lower bound is sharp for a certain $j>11$, or whether  
%could for some sequence $\mathcal S_j$ the number 
$\lceil\Phi_2(\mathcal S_j)\rceil$ could be actually less
than $2^j+1$ for $j>23$ is as difficult as to decompose the $j$\textsuperscript{th} Fermat number
or, respectively, to determine whether the $j$\textsuperscript{th}
Fermat number 
%$2^{2^j}+1$ 
is prime or composite.}   
\end{note*}
\begin{proof}[Proof of theorem \ref{2-comp}]
We only have to express $s_0+s_12+s_22^2+\dots$ as an irreducible fraction. Denote
$\gamma=s_0+s_12+s_22^2+\dots+s_{2^{j}-1}2^{2^{j}-1}$. Then
using 
the second identity of \eqref{eq:id} we in view of \ref{note:halfper-odd} obtain that
$s_0+s_12+s_22^2+\dots+s_{2^{j+1}-1}2^{2^{j+1}-1}=\gamma+2^{2^j}(2^{2^j}-\gamma-1)=
\gamma^\prime$
and hence $s_0+s_12+s_22^2+\dots=
\gamma^\prime+\gamma^\prime 2^{2^{j+1}}+\gamma^\prime 2^{2\cdot 2^{j+1}}+
\gamma^\prime 2^{3\cdot
2^{j+1}}+\dots=\frac{\gamma+1}{2^{2^j}+1}-1$. 
% Thus, $\Phi(|r-\nobreak 2^{2^j}|,|1+2^{2^j}|)=
% 1+2^{2^j}$. 
This completes the proof in view of the definition of $2$-adic
complexity of a sequence. 
\end{proof}
\begin{note*}
%\label{note:WP:2comp}
Similar estimates of $\Phi_2(\delta_{n-1}(\mathcal S))$ could be obtained for coordinate sequences of wreath products.
% sequence
% $\mathcal S\in\{
% \mathcal W_n, \mathcal Y_n, \mathcal Z\}$ 
% of  \ref{WP-odd}, \ref{le:WP-odd},
% and \ref{thm:WP}, respectively (for $\mathcal S\in\{\mathcal U_n, \mathcal X_n\}$
% of \ref{WP-even} and \ref{WP-even-trunc} this estimate is 
% already given by \ref{2-comp} in view of \ref{WP-even-trunc}). 
In view
of \ref{note:halfper-odd} the argument of the proof of \ref{2-comp}
gives that the representation of the binary sequence $\delta_{n-1}(\mathcal S)$
as a $2$-adic integer is $\frac{\gamma+1}{2^{2^{n-1}m}+1}-1$, so we have
%%%%$\frac{\gamma}{1+2^{2^{n-1}m}}+1+\frac{1}{2^{2^{n-1}m}-1}$,
only to study a fraction 
%$\frac{2^{2^{n-1}m}+1}{\gcd(2^{2^j}+1,\gamma+1)}$
$\frac{\gamma+1}{2^{2^{n-1}m}+1}$, 
where
$\gamma=s_0+s_12+s_22^2+\dots+s_{2^{n-1}m-1}2^{2^{n-1}m-1}$, and $m$
is of statements of %\ref{WP-odd}, 
\ref{le:WP-odd},
and of \ref{thm:WP}. 
%Thus, $\Phi_2(\delta_{n-1}(\mathcal S))>2^{n-1}m-1$,
%$n=1,2,\ldots$.
Representing $m=2^km_1$ with $m_1>1$ odd, we can factorize
$2^{2^{n-1}m}+1=(2^{2^{n-1+k}}+1)(2^{2^{n-1+k}(m_1-1)}-2^{2^{n-1+k}(m_1-2)}
+\cdots-2^{2^{n-1+k}}+1)$, but the problem does not become much easier because
of the first multiplier. We omit further details.
\end{note*}

\section{Schemes}
In this section we are going to give some ideas how stream ciphers could
be designed on the basis of the theory discussed above.  We must now combine
state update and output functions into an automaton that produces a sequence
that {\sl might} be cryptographically secure.
\subsection{Improving lower order bits}
The drawback of the sequence produced by a  $T$-function
$f\colon\mathbb Z/2^n\Z\rightarrow\mathbb Z/2^n\Z$
with the single cycle property is that %\hypertarget{LSB}{\textcolor{yellow}
{\sl the less significant is the bit,
the shorter is the period of the sequence it outputs} (see \ref{note:halfper-odd}); that is:
%\begin{boxitpara}{box 0.9 setgray fill}
Despite the length of the period of the sequence 
$$\mathcal S=\{ u_0=u, u_1=f(u_0),u_2=f(u_1),\ldots\}%, z_{i+1}=F(z_i)=F^i(z_0),\ldots
$$
of $n$-bit words is $2^n$, the length of the period of the $j$\textsuperscript{th}
bit sequence (i.e., %which is called 
the $j$\textsuperscript{th} {coordinate sequence})
$$\mathcal S_j=\{\delta_j (u_0), \delta_j(u_1),\delta_j(u_2),\ldots, 
\delta_j(u_{i+1}),\ldots\}$$
is only $2^{j+1}$, $(j=0,1,\ldots,k-1)$.

From \ref{thm:lincomp:sharp} it follows also that the less is $j$, %the shorter is a period (and 
the smaller is 
linear
complexity of the
coordinate sequence. %$\mathcal Z_j$. %of the sequence  $\mathcal Z$ of \ref{thm:WP} This is what we call a {\it less significant bit effect} of a $T$-function.
Obviously, in applications we must get rid of this effect.

Thus, designing a PRNG (see Fig. \ref{fig:PRNG}) we must understand what output function $F$ one should use: $F$ must add security, $F$ must be balanced (for not
to spoil the uniform distribution), and $F$ must cure the very unpleasant
%\hyperlink{LSB}
{low order bits
effect} of $T$-functions.
%}
%\onlySlide*{2}{%
%\textcolor{yellow}

{One way (that of Corollary \ref{cor:distr}) is to truncate low order bits.
But this obviously will reduce the performance of the generator ... Are there other ways?} Since the %\hyperlink{LSB}
{low order bits effect} is an inherent property
of $T$-functions, one should include in output function  some %\hyperlink{instr}
{basic
chip operations} other than $T$-functions. Thus, output function will not be a $T$-function
any more. Could one construct the output function this way, yet not `spoil' good properties of the
sequence of states? 

A solution is given at Figure \ref{fig:PRNG-rev}: We include into
a composition only one mapping $\pi$ which permute bit order of the state
(which is an $n$-bit word),
sending the most significant bit (that is, $(n-1)$-th bit) to the least significant
bit position. An important example of such a permutation $\pi$ is a word
{\sl
rotation}, $\chi_{n-1}\chi_{n-2}\cdots\chi_1\chi_0\mapsto\chi_{n-2}\chi_{n-3}\cdots\chi_1\chi_0\chi_{n-1}$,
which is also a standard instruction in most processors.

The following could be proved regarding the output sequence of the so constructed
counter-dependent generator:
%}
\begin{figure}
\begin{quote}\psset{unit=0.4cm}
 \begin{pspicture}(1,0)(24,16)
%\pscustom[fillstyle=slopes,
%slopecolors=0 1 1 .9  17 .5 1 .5  23 0 0.5 0.5  3]{
%\psccurve(0,2.5)(12,3.5)(20,4)(23,2)(17,2.5)}
%\psaxes(0,0)(0,12)(25,0)
\pscircle[linewidth=2pt](12,14){1}
%\pscircle[linewidth=2pt](12,10){1}
\pscircle[linewidth=2pt](12,7){1}
%\psline(10,14)(10,10)
%\psline{->}(10,10)(5,10)
%\psline{->}(4,11)(4,13)
  \psline{->}(12,11)(12,13)
  \psline{->}(22,10)(16,10)
  %\psline{->}(8,14)(11.5,14)
  \psline(22,14)(22,10)
  \psline(13,14)(22,14)
  \psline{<-}(12,8)(12,9)
  \psline{->}(12,2)(12,1)
  \psline{<-}(12,5)(12,6)
  %\psframe[linewidth=2pt](0,13)(8,15)
  \psframe[linewidth=2pt](8,9)(16,11)
  \pspolygon[linewidth=2pt](8,5)(16,5)(12,2)
  \uput{0}[180](12.4,6.9){{$\pi$}}
  \uput{0}[90](12,9.6){$x_i$}
  %\uput{0}[90](4,9.7){$L$}
  \uput{0}[90](12,13.5){$f_i$}
  \uput{1}[90](12.1,2.2){$H_i$}
  \uput{1}[0](15.3,10.7){\scriptsize{$x_{i+1}=f_i(x_i)$}}
  \uput{1}[0](-1.9,7.5){\scriptsize{$\pi$ permutes bits so that}}
  \uput{1}[0](-1.9,6.5){\scriptsize{$\delta_0(\pi(x_i))=\delta_{n-1}(x_i)$;}}
  \uput{1}[0](-1.9,5.5){{\scriptsize{i.e., $\pi$ sends the most}}}
  \uput{1}[0](-1.9,4.5){{\scriptsize{significant bit of $x_i$}}}
  \uput{1}[0](-1.9,3.5){{\scriptsize{to the least significant}}}
  \uput{1}[0](-1.9,2.5){{\scriptsize{bit position!}}}
  %\uput{1}[0](12.5,7.5){\red{\scriptsize{$\pi$ sends the most significant bit}}}
  %\uput{1}[0](12.5,6.5){\red{\scriptsize{to the least significant position!}}}
  %\uput{1}[0](-1,-0.5){\scriptsize{$\pi$ is a permutation of bits such %that $\delta_0(\pi(x))=\delta_{n-1}(x)$}}
  %\footnotesize{$L(c)=2\cdot c\oplus \overline u\cdot\delta_{k-1}(c)$;
  %$\overline u$ agrees to coefficients of the polynomial $u$}}
  %\uput{1}[0](11.6,12){$h_i(x_i)$}
  %\uput{1}[0](3.3,12){$c_{i+1}=L(c_i)$}
  %\uput{1}[0](8.5,14.7){$c_i$}
  %\uput{1}[0](2.6,14){$c_i$}
  \uput{1}[0](15.2,11.6){\scriptsize{\sf state update}}
  \uput{1}[0](12,1.5){\scriptsize{$y_i=H_i(\pi(x_i))$}}
  \uput{1}[180](12,1.5){\scriptsize{\sf output}}
%\psgrid(0,0)(-1,-1)(3,2)
 \end{pspicture}
\end{quote}
\caption{PRNG with a bit order reverse permutation}
\label{fig:PRNG-rev}
\end{figure}
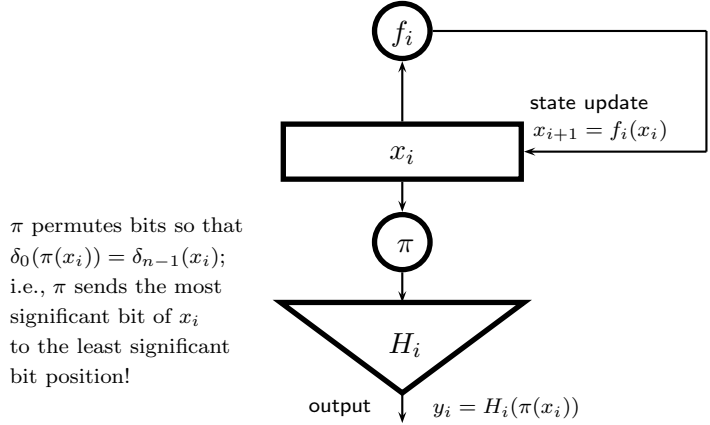

%have smaller period lengths
%and smaller linear complexities.
%This could be improved by truncation of less significant bits (see \ref{cor:distr}) %or, if necessary,
%with the use of clock output functions of special kind:
\begin{prop}[\cite{anashin4}]
\label{prop:reverse}
Let $H_i\colon\Z_2\rightarrow\Z_2$ $(i=0,1,2,\ldots,m-1)$
be compatible and ergodic mappings. %Define
%$F_i\colon\Z/2^n\>\Z/2^n$ by
For $x\in\{0,1,\ldots,2^n-1\}$ let $$F_i(x)=(H_i(\pi(x)))\bmod 2^n,$$
where $\pi$ is a permutation of bits of  $x\in\Z/2^n$ such that $\delta_0(\pi(x))=\delta_{n-1}(x)$. 
%{\rm(see
%Section \ref{Prelm} for the definition of the latter)}. 
Consider a sequence $\mathcal F$ of \ref{cor:WP}.
%=\{F_{i\bmod m}(x_i)\colon i=0,1,2\ldots\}$, 
%where $x_{i+1}=f_{i\bmod m}(x_i)\bmod 2^n$
Then  the shortest period of the $j$\textsuperscript{th}
coordinate sequence $\mathcal F_j=\delta_j(\mathcal F)$ $(j=0,1,2,\dots,n-1)$ 
%of the sequence $\mathcal F$ of \ref{cor:WP},
is of length $2^nk_j$ for a suitable $1\le k_j\le m$. Moreover, linear complexity of
the sequence $\mathcal F_j$ exceeds $2^{n-1}$.

%$\Psi_2(\mathcal F_j)>2^{n-1}$. 
% 
% Moreover, the same holds if $m=1$,
% %{\rm (}and whence $\ell=1${\rm )}, 
% i.e., when $\mathcal F$ is an output
% sequence of the automaton ${\mathfrak A}=\langle N,M,\bar f,F,u_0\rangle $,
% where $N=M=\Z/2^n$, $\bar f=f\bmod 2^n$, $f$ and $h$ are compatible
% and ergodic mappings of $\Z_2$ onto itself, 
% $F(x)=(h(\pi(x)))\bmod 2^n$, $x\in\{0,1,\ldots,2^n-1\}${\rm :} The exact period length of the $j$\textsuperscript{th}
% coordinate sequence $\delta_j(\mathcal F)$ is $2^n$ for all $j=0,1,2,\dots,n-1$.  
\end{prop}
% \begin{note*} In particular, $\mathcal F$ is a purely periodic sequence over
% $\Z/2^n$ of period
% length exactly $2^nm$, and  each element of $\Z/2^n$ occurs at the
% period exactly $m$ times (see \ref{prop:Auto}).
% \end{note*}
\subsection{The ABC stream cipher}
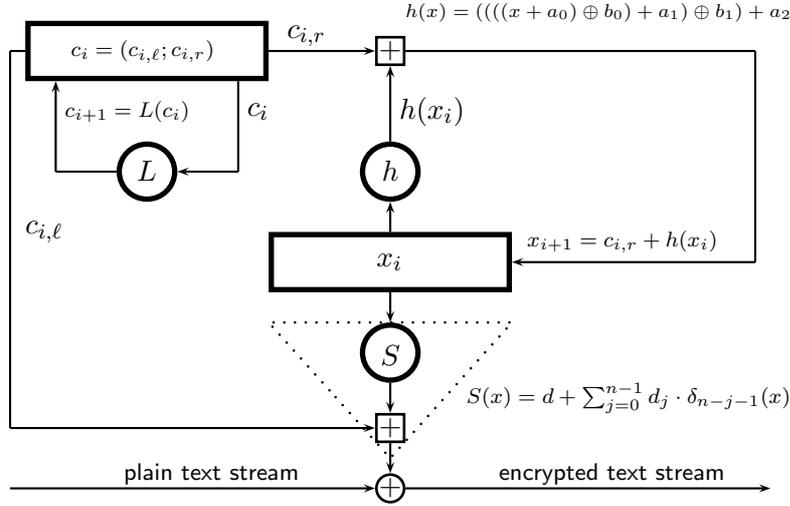
\begin{figure}
\begin{quote}\psset{unit=0.4cm}
 \begin{pspicture}(1,0)(24,16)
% \pscustom[linecolor=white]
%\pscustom[fillstyle=slopes,
%slopecolors=0 1 1 .9  17 .5 1 .5  23 0 0.5 0.5  3]{
%\psccurve(0,2.5)(12,3.5)(20,4)(23,2)(17,2.5)}
%\psaxes(0,0)(0,12)(25,0)
%\pscircle[linewidth=1pt](12,14){.5}
\psframe[linewidth=1pt](11.5,13.5)(12.5,14.5)
\psframe[linewidth=1pt](11.5,1)(12.5,2)
\pscircle[linewidth=2pt](12,10){1}
\pscircle[linewidth=2pt](4,10){1}
\pscircle[linewidth=2pt](12,4){1}
\pscircle[linewidth=1pt](12,-0.5){0.5}
%\psline(10,14)(10,10)
%\psline{->}(10,10)(5,10)
%\psline{->}(4,11)(4,13)
  \psline{->}(12,11)(12,13.5)
  \psline{->}(24,7)(16,7)
  \psline{->}(8,14)(11.5,14)
  \psline(24,14)(24,7)
  \psline(12.5,14)(24,14)
  \psline{->}(12,8)(12,9)
  \psline{->}(12,3)(12,2)
  \psline{->}(12,1)(12,0)
  \psline(7,10)(7,13)
  \psline{->}(7,10)(5,10)
  \psline(3,10)(1,10)
  \psline{->}(1,10)(1,13)
  \psline(0,14)(-0.5,14)%Were dotted
  \psline(-0.5,14)(-0.5,1.5)%Were dotted
  \psline{->}(-0.5,1.5)(11.5,1.5)%Were dotted
  \psline{->}(-0.5,-0.5)(11.5,-0.5)
  \psline{->}(12.5,-0.5)(24.5,-0.5)
  \psline{<-}(12,5)(12,6)
  \psframe[linewidth=2pt](0,13)(8,15)
  \psframe[linewidth=2pt](8,6)(16,8)
 \pspolygon[linestyle=dotted,linewidth=1pt](8,5)(16,5)(12,0.5)
  \uput{0}[180](12.4,7){$x_i$}
  \uput{0}[90](12,9.7){$h$}
  \uput{0}[90](4,9.7){$L$}
  %\uput{0}[90](5,15.7){{\large\bf A}}
  %\uput{0}[90](17,15.7){{\large\bf B}}
  %\uput{0}[90](17,3.7){{\large\bf C}}
  \uput{0}[90](12,13.7){+}
  \uput{0}[90](12,1.2){+}
  \uput{0}[90](12,-0.8){+}
%\onlySlide*{1}{
\uput{1}[90](12,2.6){%\hypertarget{rev}
{$S$}}
%\fromSlide*{2}{\uput{1}[90](12,2.5){\hypertarget{rev}{$\hat S$}}
  \uput{1}[0](15.5,7.7){{\scriptsize $x_{i+1}=c_{i,r}+h(x_i)$}}
  %\uput{1}[0](-1,0){\footnotesize{$L(c)=2\cdot c\oplus \overline u\cdot\delta_{n-1}(c)$;
 % $\overline u$ agrees to coefficients of the polynomial $u$}}
%\onlySlide*{1}{
\uput{1}[0](11.5,15.3){%\hyperlink{ex:5-1}
{{\tiny $h(x)=((((x+a_0)\oplus b_0)+a_1)\oplus b_1)+a_2$}}
}
%\fromSlide*{2}{\uput{1}[0](11.5,15.3){\hyperlink{ex:4}{{\tiny $h(x)=a+b\cdot(x\oplus
%a_1)$}}}
%}
  \uput{1}[0](11.3,12){{$h(x_i)$}}
  \uput{1}[0](0.3,12){{\scriptsize $c_{i+1}=L(c_i)$}}
  \uput{1}[0](6.3,12){$c_{i}$}
  \uput{1}[0](0.5,14){{\scriptsize$c_{i}=(c_{i,\ell};c_{i,r})$}}
 % \uput{1}[0](16.3,12){$f_{i}(x)=c_i+h_i(x)$}
  %\uput{1}[0](8.5,14.7){$c_i$}
  \uput{1}[0](-1,8){$c_{i,\ell}$}
  \uput{1}[0](7.6,14.5){$c_{i,r}$}
  %\uput{1}[0](15.6,8.6){\footnotesize{\sf state update}}
  \uput{1}[0](13.5,2.5)%{\hyperlink{ex:5}
  {{\scriptsize$S(x)=d+\sum_{j=0}^{n-1}d_{j}\cdot\delta_{n-j-1}(x)$}}
  %\uput{1}[0](13.5,3.9){%\hyperlink{ex:5}
%\fromSlide*{2}{{{\scriptsize$\hat S(x)=%S(x)+
%\circlearrowright(S(x))$}}}}
  %\uput{1}[180](12,1.5){\footnotesize{\sf output}}
  \uput{1}[180](10,0){\footnotesize{\sf plain text stream}}
  \uput{1}[180](24,0){\footnotesize{\sf encrypted text stream}}
%\psgrid(0,0)(-1,-2)(28,16)
 \end{pspicture}
\end{quote}
\label{fig:ABC}
\caption{The ABC stream cipher template. Here $L$ is a linear transformation,
$\boxplus$ and $+$ stand for integer addition, and $\oplus$ stands for $\XOR$.}
\end{figure}

With the use of the above considerations %it is possible to develop 
a %very
fast software-oriented stream cipher ABC is being developed now, %within a eSTREAM project,
see \cite{abc-v2}. In this subsection we outline underlying ideas of the
design to demonstrate their relations with the theory developed above. To
make  these ideas more transparent, we consider the ABC `template' (see Figure
\ref{fig:ABC}) rather than the actual design; the later has some differences
from the template due to necessity to withstand certain attacks. However,
we do not discuss these differences here since our aim is to illustrate the
2-adic techniques in stream cipher design rather than to give a comprehensive cryptographical analysis of a particular algorithm.

The main goal of the design was to achieve high performance and to {\sl prove}
some important properties of the key stream, e.g. long period and uniform
distribution.  

The high performance is achieved by a very restricted set of instructions
that are used: Actually, only fastest instructions, such as $+$, $\XOR$ and
shifts are allowed. That's why the clock state update function $f_i$ (c.f.
Figure \ref{fig:PRNG-rev}) is of the
form $h_i(x)=c_{i,r}+((((x+a_0)\oplus b_0)+a_1)\oplus b_1)+a_2$. 

Now recall Example
\ref{XOR} and Example 3 of \ref{WP-even}. Note that $L$ is a linear transformation
that is produced by a linear feedback shift register of a maximum period
length; $c_{i,r}$ is a right-hand part of the outputted word, so the sequence
$\{c_{i,r}\colon i=0,1,2,\ldots\}$ is a LFSR sequence with a maximum period
length. Thus, the state sequence $\{x_i\}$ has a maximum period length, and
is strictly uniformly distributed. 

After producing a uniformly distributed sequence of states,  we need to improve period lengths of output sequence. In ABC we do it with the use of Proposition
\ref{prop:reverse}, that is,  by a circuit described by Figure
\ref{fig:PRNG-rev}. 

Actually, in ABC  we take $\pi$ to be a bit order reverse permutation,
$$\delta_j(\pi(x))=\delta_{n-j-1}(x),$$ for all $x\in\Z/2^n\Z$. However, this
permutation is rather slow in software since one has to work with bits rather
than with words. Yet we use a trick to avoid this undesirable reduce
of performance. The trick is based on the use of special output function
$S(x)=d+\sum_{j=0}^{n-1}d_{j}\cdot\delta_{n-j-1}(x)$, which is a composition
of two functions, of a permutation $\pi$, and of the function 
$F(x)=d+d_0\cdot\delta_0(x)+d_1\cdot\delta_1(x)+\cdots$. Thus, to apply Proposition
\ref{prop:reverse}, we must know when $F$ is ergodic. 

The following Proposition could be proved:
\begin{prop}[\cite{anashin1}]
\label{prop:erg_sum}
The function $F(x)=d+d_0\cdot\delta_0(x)+d_1\cdot\delta_1(x)+\cdots$
is compatible and ergodic if and only if $\|d\|_2=1$, $d_0\equiv 1\pmod 4$,
and $\|d_j\|_2=2^{-j}$ for $j=1,2,\ldots$
\end{prop}
Now we just take clock output functions $H_i$ (c.f. Figure \ref{fig:PRNG-rev})
of the form $H_i(x)=c_{i,\ell}+F(x)$, where $c_{i,\ell}$ is the left-hand
part of the word produced by LSFR $L$. Thus, the circuit at Figure \ref{fig:ABC}
is a special case of the circuit at Figure \ref{fig:PRNG-rev}.  We note,
once
again, that compare to the template, the real-life stream cipher ABC has some important differences, yet however use of the above mentioned  ideas enable us to prove
crucial cryptographic properties of the cipher, long period, uniform distribution
and high linear complexity of output sequence, see \cite{abc-v2} for details.

\newpage
\bibliographystyle{plain}
\bibliography{lecture_notes}

%%+Bibliography
%\begin{thebibliography}{99}
%\bibitem{Label1} ...
%\bibitem{Label2} ...
%\end{thebibliography}
%%-Bibliography

\end{document}